\documentclass[11pt,dvipsnames]{article}
\usepackage{caption}
\usepackage{scalerel}
\usepackage{amsthm,latexsym,amsxtra,graphicx,appendix,epstopdf,feynmf,hyperref,setspace,fix-cm}
\usepackage[normalem]{ulem}
\usepackage{makeidx}
\usepackage{fullpage}
\usepackage{amsmath}
\usepackage{amssymb}
\usepackage{setspace}
\usepackage{bbm}
\usepackage{dsfont}
\usepackage{epsfig}
\usepackage[font=footnotesize,labelfont=bf,justification=centerlast,width=.94\textwidth]{caption}
\usepackage{cite}
 \usepackage{multirow}
\usepackage{array,booktabs}
\usepackage{subfigure}
\usepackage{mathtools}
\usepackage{color}
\usepackage[makeroom]{cancel}
\usepackage{tikz}
\usepackage{pgfplots}
\usepackage{bm}

\usepackage{tensor}
\usepackage{simplewick}
\usepackage{frcursive}
\usepackage{pgfornament,multicol}
\usepackage{caption}
\usepackage{hyperref}

\usepackage[skip=0.2cm, indent=0pt]{parskip}

\hypersetup{
    colorlinks,%
    citecolor=blue,%
    filecolor=blue,%
    linkcolor=blue,%
    urlcolor=blue
}
\usepackage{float}
\usepackage{slashed}
\usepackage{tikz}
\usetikzlibrary{decorations.pathmorphing}

\usepackage{tcolorbox}
\usepackage{tabularx}
\usepackage{array}
\usepackage{colortbl}
\tcbuselibrary{skins}

\tcbset{tab2/.style={enhanced,fonttitle=\bfseries,fontupper=\normalsize\sffamily,
colback=Goldenrod!20!white,colframe=Blue!50!black,colbacktitle=Goldenrod!50!white,
coltitle=black,center title}}

\def\r{\rightarrow}

\def\b{\beta}
\def\d{\delta}

\newcommand{\bi}{\begin{itemize}}
\newcommand{\ei}{\end{itemize}}
\newcommand{\bea}{\begin{eqnarray}}
\newcommand{\eea}{\end{eqnarray}}

\def\={\, = \,}
\def\rh{r_{\text{h}}}
\def\rb{r_{\text{b}}}
\def\frakr{\mathfrak{r}}

\def\bomega{\bm{\omega}}
\def\r{\mathfrak{r}}
\def\k{\mathcal{K}}
\def\b{\tilde\beta}

\def\r{\mathfrak{r}}
\def\frakr{\mathfrak{r}}
\def\bomega{{\boldsymbol{\omega}}}

\def\XXint#1#2#3{{\setbox0=\hbox{$#1{#2#3}{\int}$}
     \vcenter{\hbox{$#2#3$}}\kern-.5\wd0}}

\def\={\, = \,}

\newcommand{\beq}{\begin{equation}}

\newcommand{\eeq}{\end{equation}}

\usepackage{tikz}
\usetikzlibrary{positioning}
\usetikzlibrary{intersections}
\usetikzlibrary{fadings} 
\usetikzlibrary{arrows.meta} 
\usetikzlibrary{arrows}

\tikzfading[name=fade out,
inner color=transparent!0,
outer color=transparent!100]

\definecolor{cherryblossompink}{rgb}{1.0, 0.72, 0.77}
\definecolor{lightblue}{rgb}{0.68, 0.85, 0.9}

\usetikzlibrary{decorations.pathmorphing}
\usetikzlibrary{decorations.pathreplacing,decorations.markings}

\usetikzlibrary{backgrounds,automata}

\newsavebox\CBox
\newcommand\hcancel[2][0.5pt]{%
  \ifmmode\sbox\CBox{$#2$}\else\sbox\CBox{#2}\fi%
  \makebox[0pt][l]{\usebox\CBox}%  
  \rule[0.75\ht\CBox-#1/2]{\wd\CBox}{#1}}

\numberwithin{equation}{section}
\allowdisplaybreaks  % allow page breaks in displayed eqs

\begin{document}

\vspace*{2.5cm}
\begin{center}
{ \huge {Conformal boundaries near extremal black holes} }\\

\vspace*{1cm}
\end{center}

\renewcommand{\thefootnote}{\fnsymbol{footnote}}

\begin{center}
Dami\'an A. Galante, Chawakorn Maneerat\footnotemark[2] \footnotetext[2]{corresponding author}, and Andrew Svesko
\end{center}
\begin{center}
{
\footnotesize
\vspace{0.2cm}
Department of Mathematics, King's College London, Strand, London WC2R 2LS, UK}
\end{center}
\begin{center}
{\textsf{\footnotesize{
damian.galante@kcl.ac.uk, chawakorn.maneerat@kcl.ac.uk, andrew.svesko@kcl.ac.uk}} } 
\end{center}

\renewcommand{\thefootnote}{\arabic{footnote}}

\vspace*{0.5cm}

%\end{center}
\vspace*{1.5cm}
\begin{abstract}
\noindent We examine four dimensional, near-extremal black hole solutions in the presence of a finite boundary obeying conformal boundary conditions, where the conformal class of the induced metric and the trace of the extrinsic curvature are fixed. Working in Euclidean signature and at fixed charge, we find the near-extremal regime is dominated by a double-scaling limit which reveals new scaling laws for the quasi-local conformal entropy at low temperatures. Upon spherical dimensional reduction, we obtain the effective two-dimensional dilaton-gravity theory that describes the near-extremal regime. In contrast to Dirichlet boundaries, for conformal boundaries a linear dilaton potential is not sufficient to capture the leading correction away from extremality and higher orders are needed. We also examine near-Nariai solutions and the circular reduction of pure three-dimensional gravity in (Anti-) de Sitter space. In the latter, provided the boundary is placed near the conformal boundary of three-dimensional Anti-de Sitter space, the dynamics of the spherically symmetric boundary mode is governed by a Liouville equation that descends from a (minus) Schwarzian effective action.

\end{abstract}

\newpage

\setcounter{tocdepth}{2}
\tableofcontents

\newpage

\section{Introduction} \label{sec:intro}

Two-dimensional theories of gravity play an important role in understanding quantum aspects of spacetime. The reason is two-fold. On one hand, the appearance of quantum mechanical (random) models that reproduce some features of two-dimensional theories of gravity, and on the other, the relation of these lower dimensional models to universal aspects of near-extremal black holes in higher dimensions, including leading 1-loop quantum corrections. For a review, see, e.g., \cite{Turiaci:2024cad}.

In two dimensions the Einstein-Hilbert action is purely topological, denoted $I_{\text{top}}$, and thus has trivial dynamics. The simplest two-dimensional models of gravity with non-trivial dynamics include a dilaton (scalar) field $\Phi$ in addition to the metric $g_{\mu\nu}$. Up to local field redefinitions, the most general (Euclidean) action for dilaton-gravity models with at most two derivatives, is given by \cite{Cavaglia:1998xj,Grumiller:2007ju,Anninos:2017hhn, Witten:2020ert} 
\begin{equation}
    I_E = I_{\text{top}} - \frac{1}{16\pi G_2} \int d^2x \sqrt{g} \left( \Phi R + U(\Phi) \right) + I_{\text{bdy}} \,. \label{action_int}
\end{equation}
In the case of manifolds with boundaries, boundary conditions need to be specified through the one-dimensional boundary action $I_{\text{bdy}}$, yielding a well-defined variational problem.
Each theory is characterized by a smooth function $U(\Phi)$, the dilaton potential.  

A paradigmatic example is Jackiw-Teitelboim (JT) gravity \cite{Jackiw:1984je,Teitelboim:1983ux}, that has $U(\Phi)=2\Phi/L^2$. Classical solutions are locally Euclidean Anti-de Sitter (AdS) space (with characteristic length $L$) in two dimensions. Famously, when the boundary is close to the conformal boundary of AdS$_2$, this theory captures the linear-in-temperature dependence of the entropy of higher-dimensional near-extremal black holes \cite{Maldacena:2016upp}. Notably, this behavior can be reproduced by quantum mechanical models, such as the Sachdev-Ye-Kitaev (SYK) model at low energies \cite{Sachdev:1992fk,Kit_SYK} and double-scaled matrix models \cite{Saad:2019lba,Johnson:2019eik}.

So far, much effort has been devoted to understanding how to deform the dilaton potential away from the linear (with positive slope) regime, from the microscopic theory. This includes deformations to SYK models \cite{Jiang:2019pam, Anninos:2020cwo, Anninos:2022qgy, Chapman:2024pdw} or matrix model descriptions \cite{Maxfield:2020ale, Witten:2020wvy, Turiaci:2020fjj, Eberhardt:2023rzz,Kruthoff:2024gxc}. One motivation for all such deformed models is that they might provide insights into microscopic theories of spacetimes undergoing accelerated expansion, like de Sitter (dS) space \cite{Anninos:2017hhn, Anninos:2018svg, Anninos:2022hqo, Chapman:2021eyy}.\footnote{The theory of pure dS JT gravity with $U(\Phi)=-2\Phi/L^2$ is interesting in its own right, cf. \cite{Maldacena:2019cbz, Cotler:2019nbi, Nanda:2023wne, Cotler:2024xzz}. Alternatively, one could also couple the two-dimensional theory of gravity with a positive cosmological constant to quantum fields, see \cite{Anninos:2024iwf}, and references therein.} See also  \cite{Blommaert:2024ydx, Berkooz:2024lgq, Blommaert:2024whf, Collier:2025pbm} for developments regarding periodic dilaton potentials.
Most of these works assume the boundary obeys Dirichlet boundary conditions (where both the induced metric and dilaton at the boundary are fixed). 
Alternative boundary conditions have been considered in e.g., \cite{Goel:2020yxl, Godet:2020xpk}.

In this work, we initiate a systematic study of two-dimensional dilaton-gravity theories \eqref{action_int} with \emph{finite} boundaries whose field content obey a broader set of boundary conditions. Specifically, we consider two-dimensional Euclidean manifolds with a finite-size boundary, where we keep the following quantities fixed,
\beq 
\left\{ \Phi^{-\alpha}\sqrt{h} \, , \, \Phi^{-\alpha}\left(\Phi K+\alpha \, n^{\mu}\nabla_{\mu}\Phi\right) \right\} \,.
\label{eq:confbcsind_intro}
\eeq
Here, $h$ denotes the determinant of the induced metric, $n^{\mu}$ is the outward-pointing unit vector and $K$ is the trace of the extrinsic curvature at the boundary. The parameter $\alpha$ is a new real parameter, part of the definition of the theory. These boundary conditions can be used for any sign of the cosmological constant, and in particular, the boundaries need not to be close to the conformal boundary of AdS$_2$.

Including finite timelike boundaries in gravity by now has a long history. Notably,
York \cite{York:1986it} and collaborators \cite{Whiting:1988qr,Braden:1990hw,Brown:1992bq,Brown:1994gs} introduced finite timelike boundaries obeying Dirichlet boundary conditions to address puzzles of (asymptotically flat) black hole thermodynamics, initiating the program of  
quasi-local gravitational thermodynamics. There has since been renewed interest in finite timelike boundaries, especially in regards to addressing fundamental aspects of de Sitter physics. In particular, the insertion of boundaries inside the de Sitter static patch is a promising way toward dS holography from a more local perspective \cite{Anninos:2011af, Anninos:2012qw, Anninos:2022ujl, Galante:2023uyf}. Finite boundaries also offer a method to microscopically account for the Gibbons-Hawking entropy of the dS static patch \cite{Shyam:2021ciy,Coleman:2021nor,Silverstein:2022dfj,Silverstein:2024xnr}.  In fact, including a finite boundary allows one to define thermal ensembles in spacetimes with cosmological horizons in the first place \cite{Banihashemi:2022jys}, and clarify the interpretation of the thermodynamic quantities in the first law for cosmological horizons \cite{Banihashemi:2022htw}. The study of finite Dirichlet boundaries is now well trodden for theories of gravity in two \cite{Gross:2019ach,Svesko:2022txo, Anninos:2022hqo,Batra:2024qju,Aguilar-Gutierrez:2024nst}, three \cite{Shyam:2021ciy,Coleman:2021nor, Silverstein:2022dfj,Batra:2024kjl}, and four spacetime dimensions \cite{Hayward:1990zm, Wang:2001gt, Draper:2022ofa, Banihashemi:2022jys, Banihashemi:2022htw, Silverstein:2024xnr}. Generic Dirichlet boundary data (in four dimensions), though, leads to a boundary problem that is proven to be ill-posed both in Euclidean \cite{Anderson:2006lqb} and Lorentzian signature \cite{An:2021fcq}.\footnote{Generic boundary induced metrics will have both existence and uniqueness problems. Of course this does not mean that there are no solutions with finite Dirichlet boundaries. A special class of them are the spherical Dirichlet boundaries, which are proven to be well-posed in Euclidean \cite{anderson2010extension, Witten:2018lgb} and Lorentzian \cite{An:2025gvr} signature. A linearized analysis in Lorentzian signature was previously done in \cite{Anninos:2023epi}. In contrast, both the Minkowski and the Rindler corner with Dirichlet boundary conditions are shown to have non-uniqueness issues already at the linear level \cite{Anninos:2024xhc}.}

Here, we instead focus on finite boundaries obeying the seemingly strange boundary conditions (\ref{eq:confbcsind_intro}). Our motivation comes from studies of higher-dimensional theories of gravity with finite boundaries obeying conformal boundary conditions (CBCs), 
in which the conformal metric $[h_{\mu\nu}]$ and the trace of the extrinsic curvature $\k$ are fixed at the boundary, 
\begin{equation}
    \text{Conformal boundary conditions:} \, \, \{ [h_{\mu\nu}] , \k \} \,\, \text{fixed} \,.
\label{eq:CBChigherdintro}\end{equation}
The spherical reduction of these conformal boundary conditions gives \eqref{eq:confbcsind_intro} in the two-dimensional theory for specific values of $\alpha$ (that we discuss later), as first shown in \cite{Banihashemi:2025qqi, edgar}. 
As opposed to the generic Dirichlet (or Neumann) problem, conformal boundary conditions (\ref{eq:CBChigherdintro}) are proven to lead to a well-posed (elliptic) boundary value problem in Euclidean signature \cite{Anderson:2006lqb, Witten:2018lgb}. In Lorentzian signature, the well-posedness of the initial boundary value problem with conformal boundary conditions has been very recently proven at a linearized level about \textit{any} Einstein background \cite{An:2021fcq, An:2025rlw}.\footnote{More precisely, for the initial boundary value problem with conformal boundary conditions to be well-posed, additional data needs to be supplemented at the intersection between the timelike boundary and the initial Cauchy slice \cite{Anninos:2023epi, Anninos:2024wpy, Liu:2024ymn, Anninos:2024xhc}. Given this `corner' data, the problem becomes an \textit{initial boundary corner value problem} (IBCVP), which has been shown to be well-posed, at least locally in time and linearly about any Einstein background \cite{An:2025rlw}.} A systematic study of linearized dynamics with these boundary conditions about maximally symmetric backgrounds with vanishing, positive, and negative cosmological constant was done in \cite{Anninos:2023epi, Anninos:2024wpy, Anninos:2024xhc}. See also earlier work on conformal boundaries in the stretched horizon limit \cite{Bredberg:2011xw, Anninos:2011zn}. In \cite{An:2021fcq, Odak:2021axr}, the equivalent of the Brown-York stress tensor \cite{Brown:1992br}  has been derived for the case of conformal boundaries, which led to the study of thermal properties of these conformal boundary conditions.

In Euclidean signature with boundary topology $S^1 \times \Sigma_{d-2}$, fixing the conformal class of the metric and the trace of the extrinsic curvature defines a thermal partition function
\beq
\mathcal{Z}(\tilde{\beta},\k)\approx \sum_{g_{\mu\nu}^{\ast}}e^{-I_{d} [g_{\mu\nu}^{\ast}]}\;, \label{z_intro}
\eeq
that depends on the trace of the extrinsic curvature $\k$ at the boundary and the inverse conformal temperature $\tilde\beta \equiv \beta/\r$, where $\beta$ is the size of the $S^1$ and $\r$, the characteristic scale of $\Sigma_{d-2}$. We refer to this partition function as the conformal canonical partition function. Analogous to the Gibbons-Hawking prescription \cite{Gibbons:1976ue}, in the saddle-point approximation, the partition function is computed by summing over Euclidean on-shell actions $I_{d}$ for smooth Einstein metrics $g_{\mu\nu}^{\ast}$ satisfying the conformal boundary conditions.  A one-parameter family generalization of the boundary conditions (\ref{eq:CBChigherdintro}) and the subsequent gravitational thermodynamics has recently been explored in \cite{Liu:2024ymn}. 

The study of spherical black hole and cosmological solutions revealed an interesting structure for the partition function  (\ref{z_intro}) in the saddle-point approximation.\footnote{Note that in \cite{Liu:2024ymn}, fluctuations about the black hole saddle-point have been computed, finding one static and spherically symmetric unstable mode for the type of conformal boundary conditions under consideration in this work. In the Lorentzian linearized analysis \cite{Anninos:2023epi, Anninos:2024wpy, Anninos:2024xhc}, there were also modes with complex frequencies, associated to an exponential growth in the size of the worldtubes. It would be interesting to understand the physical meaning of these modes both in Euclidean and Lorentzian signature, and the non-linear fate of conformal boundaries at late times.} In all cases, the conformal entropy $\mathcal{S}_{\text{conf}}$, defined through \eqref{z_intro}, gives an area-law formula. For $d\geq4$, the conformal entropy can be written in a high-temperature expansion as \cite{Anninos:2023epi, Anninos:2024wpy, Anninos:2024xhc, Banihashemi:2024yye, Banihashemi:2025qqi} 
\begin{eqnarray} \label{eq: intro_entropy}
   \mathcal{S}_{\text{conf}} (\b, \k) = \frac{N_{\text{d.o.f.}}(\k)}{\tilde\beta^{d-2}} + \mathcal{O} \left( \b^{-(d-4)} \right)\,, 
   \end{eqnarray}
   with 
   \begin{eqnarray}
       N_{\text{d.o.f.}}(\k)=\begin{cases}
       \frac{\Omega_{d-2}\ell^{d-2}}{4G_{d}}\left(\frac{4\pi}{(d-1)^{2}}\right)^{ d-2}\left(\sqrt{\k^{2}\ell^{2}+(d-1)^{2}}-\k\ell\right)^{ d-2} \,  , & \,\Lambda>0 \,, \\
        \tfrac{\Omega_{d-2}(2\pi)^{d-2}}{4G_d \k^{d-2}} \, ,  &\, \Lambda=0 \,, \\
         \frac{\Omega_{d-2}\ell^{d-2}}{4G_{d}}\left(\frac{4\pi}{(d-1)^{2}}\right)^{ d-2}\left(\k\ell-\sqrt{\k^{2}\ell^{2}-(d-1)^{2}}\right)^{ d-2} \,, & \, \Lambda<0 \,, \\
   \end{cases} \label{eq: ndof}
\end{eqnarray}
where $G_d$ is Newton's constant in $d$-spacetime dimensions, $\Omega_{n}\equiv \frac{2\pi^{(n+1)/2}}{\Gamma[(n+1)/2]}$ is the volume of a unit $n$-sphere, and $\Lambda=\pm\frac{(d-1)(d-2)}{2\ell^{2}}$ is the $d$-dimensional cosmological constant with (A)dS length $\ell$. 
This leading term in the conformal entropy can be identified as the thermal entropy of a conformal field theory in $(d-1)$ dimensions, where $N_{\text{d.o.f.}}$ is identified with some count of degrees of freedom, like a central charge in odd $d$. Below are some interesting properties of $N_{\text{d.o.f.}}$:
\begin{itemize}
    \item In three spacetime dimensions the leading expression in \eqref{eq: intro_entropy} is actually the exact result at all temperatures, both for the BTZ black hole and the cosmological horizon \cite{Anninos:2024wpy}.
    \item $N_{\text{d.o.f.}}$ is a monotonically decreasing function from the boundary of AdS (for $\Lambda<0$) to the stretched black hole horizon, and from the worldline of dS ($\Lambda>0$) to the cosmological horizon.
    \item $N_{\text{d.o.f.}}$ (for $\Lambda>0$) is a positive function even when $\k \ell$ is negative.
\end{itemize}
Subleading corrections to \eqref{eq: intro_entropy} in the high-temperature expansion have been computed (in different spacetime dimensions and different spatial topologies) in \cite{Banihashemi:2024yye, Banihashemi:2025qqi}. 

A distinct feature of these results is that both in three and four dimensions, there exist patches of spacetime which contain cosmological horizons (dubbed cosmic patches), with positive specific heat. This is in stark contrast with the case of Dirichlet boundary conditions, where the specific heat is negative definite for all cosmic patches. Conformal patches with positive specific heat include patches of (near-)Nariai solutions \cite{Anninos:2024wpy}. Analogous to (near-)extremal black holes, these spacetimes have a near-horizon geometry that is dS$_2\times S^2$, which brings us back to the present paper. 

Here, we study effective two-dimensional theories that arise from dimensionally reducing the spherically symmetric sector of Einstein and Einstein-Maxwell theory in higher dimensions with conformal boundary conditions. 
The main results of this paper can be summarized as follows:
\begin{itemize}
    \item In four-spacetime dimensions, we consider the near-extremal limit of Reissner-Nordstr{\"o}m (RN) black holes at fixed charge in the presence of a static, spherically symmetric boundary obeying conformal boundary conditions. In Euclidean signature, the boundary data is given by $\b$, the inverse conformal temperature; $\k$, the trace of the extrinsic curvature, and $Q$, the charge of the black hole. For general $\b$ and $\k$, the extremal limit is obtained by requiring 
    \beq \b = \b_{\text{ex}} (\k, Q)= \frac{2\pi}{\sqrt{\k^2 Q^2 -1}} \,. \eeq 
    The near zero-temperature extremal limit is obtained by taking a double-scaled limit where both $\b \to \infty$ and $\k Q\to 1$. (This is unlike the Dirichlet case, where there is only one parameter controlling the near-extremal limit). In this regime, the linear-in-temperature behavior of the entropy gets enhanced and we find new universal scaling regimes, all within the near-extremal limit. These are summarized as shown in Table \ref{table1}. 

\begin{center}
\begin{tcolorbox}[tab2,tabularx={c||c||c},title= Near-zero temperature extremal limits,boxrule=2pt, width=11cm, code={\setstretch{1.25}}]
   $\boldsymbol{\mathcal{K}Q}$ & \textbf{Scaling of $\mathcal{S}- \mathcal{S}^{\text{ex}}_{\text{conf}}$} & \textbf{Regime of validity} \\ 
   \hline \hline
     \multirow{2}{*}{ $\mathcal{K}Q<1$} & $\sqrt{1-\k Q}\tilde{\beta}^{-1}$ & $\tilde{\beta}^{-1}\ll\sqrt{1-\k Q} \ll 1$ \\ 
     & $\tilde{\beta}^{-2}$ & $\sqrt{1-\k Q}\ll \tilde{\beta}^{-1}\ll 1$ \\
   \hline \hline
     {} & $(\mathcal{K}Q-1)|\delta\tilde{\beta}|$ & $|\delta\tilde{\beta}|\ll\sqrt{\mathcal{K}Q-1} \ll1$ \\
   $\mathcal{K}Q>1$ & $(\mathcal{K}Q-1)^{5/4} |\delta\tilde{\beta}|^{1/2}$ & $\sqrt{\mathcal{K}Q-1}\ll|\delta\tilde{\beta}|\ll \frac{1}{\sqrt{\k Q-1}}$ \\
   {} & $\tilde{\beta}^{-2}$ & $\frac{1}{\sqrt{\k Q-1}}\lesssim |\delta\tilde{\beta}|\lesssim \frac{\sqrt{2}\pi}{\sqrt{\k Q-1}}$ %\vspace{0.3mm} \\
\end{tcolorbox}
\end{center}
\begin{minipage}[c]{0.3 \textwidth}
\captionof{table}{Near-extremal scaling regimes. Here $\d\b \equiv \b - \b_{\text{ex}}$ and $\mathcal{S}^{\text{ex}}_{\text{conf}} = \tfrac{\pi Q^2}{ G_4}$ is the entropy of the extremal black hole.}
\label{table1}
%\end{center}
\end{minipage}

    These scaling regimes complement the high-temperature result \eqref{eq: intro_entropy} and are derived in Section \ref{sec:CBCsthermalensem}, where we also provide a summary of results regarding conformal thermodynamics in three and four dimensions (including the Nariai and near-Nariai thermodynamics).
    \item In Section \ref{sec:spherered}, we show that by dimensionally reducing the spherically symmetric sector of Einstein-Maxwell theory, we obtain effective two dimensional actions of the form (\ref{action_int}), while the conformal boundary conditions reduce to (\ref{eq:confbcsind_intro}). In Section \ref{sec:confthermo2D}, we solve this class of theories for generic $U(\Phi)$ and $\alpha$, and compute their gravitational thermodynamics. We find the entropy obeys an area-like formula, while the specific heat at fixed $\kappa$ (the reduced CBC of fixed $\mathcal{K}$) is 
\begin{equation}
    C_\kappa = \left(\frac{ f''(\rh)}{f'(\rh)} + \frac{ \rb f'(\rh)((2\alpha-1)f'(\rb)-\rb f''(\rb))}{4\alpha^2f(\rb)^2+\rb^2 f'(\rb)^2-2\rb f(\rb)(f'(\rb)+\rb f''(\rb))}\right)^{-1}\frac{\tilde{\Phi}}{4G_2}  \,,
\end{equation}
where $f(r) = \int_{r_h}^r U(\Phi) d\Phi$, $\tilde\Phi$ is a normalization of the dilaton field, $\rb$ is the position of the two-dimensional boundary and $\rh$, the position of the event horizon. In order to express the specific heat in terms of boundary variables, $\b$ and $\kappa$, one would further need to invert \eqref{eqn: beta generic potential} and \eqref{eqn: kappa generic potential}. This is a completely generic formula valid for any dilaton potential and any value of $\alpha$, independent of whether the theory has a higher-dimensional pedigree or not. We then match the higher dimensional solutions for (A)dS$_3$, near-Nariai and near-extremal RN black holes in four dimensions, finding agreement between both.
    \item Notably, we find the JT action with reduced conformal boundary conditions is not enough to capture the thermodynamic behavior of neither near-Nariai nor near-extremal black hole solutions. In fact, we show the thermodynamics derived from the JT action are the same as the extremal thermodynamics (up to a constant shift in the conformal energy). We explicitly show higher order terms in the dilaton potential are needed to reproduce the higher dimensional near-extremal thermodynamics.
    \item Most of our work focuses on \textit{static} spherically symmetric boundaries. But given conformal boundary conditions do not fix the conformal factor of the induced metric at the boundary, it is possible to have \textit{non-static} boundaries as well, where the conformal factor has dynamics. The equation of motion for such non-static boundaries in higher dimensions is derived in Section \ref{sec:CBCsthermalensem}. In Section \ref{sec: schwarzian}, we initiate an investigation into finding an effective action for the spherically symmetric non-static boundary mode from the two-dimensional perspective. We focus on the particular case where the higher dimensional theory is three-dimensional with a negative cosmological constant. Using the same methodology used to derive the Schwarzian action in near-AdS$_2$ \cite{Maldacena:2016upp}, we find that for the case of conformal boundary conditions, and when the finite boundary is close enough to the conformal boundary of AdS$_3$, i.e., $\k = 2 + \epsilon^2 \kappa_2$, the effective boundary action is given by
    \begin{equation}
    I_{\text{bdy}} =-\frac{1}{8\pi G_{2}}\int_{\partial \mathfrak{m}}du\left[S(t(u),u) - \frac{\kappa_2}{2} \Gamma'(u)^{2} - S\left(\Gamma(u),u\right) \right]\;.  
    \end{equation}
    Here $\Gamma (u) = \int^u \Phi_r(\tilde{u})$, $\Phi_r(u)$ is the renormalized value of the dilaton at the boundary, and $t(u)$ is a parametrization of the boundary curve. The first Schwarzian term, $S(t(u),u)$, is the same one that appears with Dirichlet boundary conditions, while the second Schwarzian is a consequence of introducing the conformal boundary. Interestingly, the same relative minus sign appears when coupling the Schwarzian to one-dimensional quantum gravity \cite{Anninos:2021ydw}. The equation of motion for $\Gamma(u)$ that stems from this action yields a one-dimensional Liouville equation of motion for the conformal factor of the metric. As the boundary is pushed towards the interior of the spacetime, bulk and boundary degrees of freedom are coupled to each other.   
\end{itemize}

Finally, for this article to be self-contained, we include a number of appendices. Appendix \ref{app:varprobCBC} derives the form of the appropriate boundary term in the action to have a well-posed variational problem for boundaries obeying CBCs. In Appendix \ref{app:reductionscoords} we carry out the spherical reduction of Einstein-Maxwell theory. Appendix \ref{app: brane dynamics} presents details of the dynamics of non-static boundaries. In Appendix \ref{app: highT} we derive the conformal thermodynamics in the high temperature limit from the two-dimensional perspective.

\section{Conformal boundary conditions and thermal ensembles} \label{sec:CBCsthermalensem}

We start by setting up the problem of Euclidean Einstein gravity with a cosmological constant for manifolds with boundaries obeying conformal boundary conditions in Section \ref{sec2: gravity}. We explore solutions that are spherically symmetric both at a static and non-static level. We then interpret these solutions in terms of thermal ensembles, using the Gibbons-Hawking prescription. We provide a review of results for $d=3$ and $d=4$ in this context. Then, in Section \ref{sec2: maxwell}, we extend our study to Einstein-Maxwell theory. In particular, we focus on the near-extremal limit of charged black hole solutions, which will play an important role in the two-dimensional picture.

\subsection{Pure gravity} \label{sec2: gravity}

We consider Einstein gravity with a cosmological constant, on a manifold $\mathcal{M}$ in $d$ spacetime dimensions. 
 The (Euclidean) action is given by
\beq I_{d}=-\frac{1}{16\pi G_{d}}\int_{\mathcal{M}} \hspace{-2mm}d^{d}X\sqrt{G}[\mathcal{R}-2\Lambda_{d}]-\frac{\Theta_d}{8\pi  G_{d}}\int _{\Gamma}d^{d-1}Y\sqrt{H}\mathcal{K} %+ I_{\text{matter}}
\;,\label{eq:einhilbv1_Euc}\eeq
where $G_d$ is the Newton's constant in $d$ dimensions and  $\Gamma=\partial\mathcal{M}$ denotes the boundary of $\mathcal{M}$. Here  $G_{MN}$ is the metric endowed on $\mathcal{M}$, $\mathcal{R}$ is the Ricci scalar, and $H_{MN}$ is the induced metric on the codimension-1 boundary $\Gamma$. The trace of the extrinsic curvature at the boundary is defined as $\mathcal{K} = G^{MN} \mathcal{K}_{MN}$, where $\mathcal{K}_{MN} =\tfrac{1}{2}\mathcal{L}_{n^M} G_{MN}$ for unit outward-pointing normal vector $n^{M}$. 

The second term in \eqref{eq:einhilbv1_Euc} 
makes the variational problem well-posed under a class of boundary conditions (see Appendix \ref{app:varprobCBC} for a derivation). 
For example,
\[ \Theta_{d} =  \begin{cases} 
      1\,, & \text{Dirichlet:} \;\; H_{MN} \;\; \text{fixed on} \;\; \partial\mathcal{M}\\
      -\frac{(d-4)}{2}\,, & \text{Neumann:} \;\; \pi^{MN} \;\; \text{fixed on} \;\; \partial\mathcal{M}\\
      \frac{1}{(d-1)}\,, & \text{Conformal:} \;\; ([H_{MN}],\mathcal{K}) \;\; \text{fixed on} \;\; \partial\mathcal{M}
   \end{cases}
\] 
where $\pi^{MN}$ is the conjugate momentum to $H_{MN}$ and 
$[H_{MN}]=H^{-1/(d-1)}H_{MN}$ is the conformal metric. 
Our primary focus will be on conformal boundary conditions, such that we fix $\Theta_d = \tfrac{1}{d-1}$, however, there exists a one-parameter extension leading to 
 more general conformal-like boundary conditions for any $\Theta_d$
 (see Appendix \ref{app:varprobCBC}) \cite{Liu:2024ymn}. Note it is possible to compute the analog of the quasi-local Brown-York stress tensor stemming from the Lorentzian version of \eqref{eq:einhilbv1_Euc} for any $\Theta_d$. If we require this stress tensor to be traceless, this uniquely imposes $\Theta_d = \tfrac{1}{d-1}$.

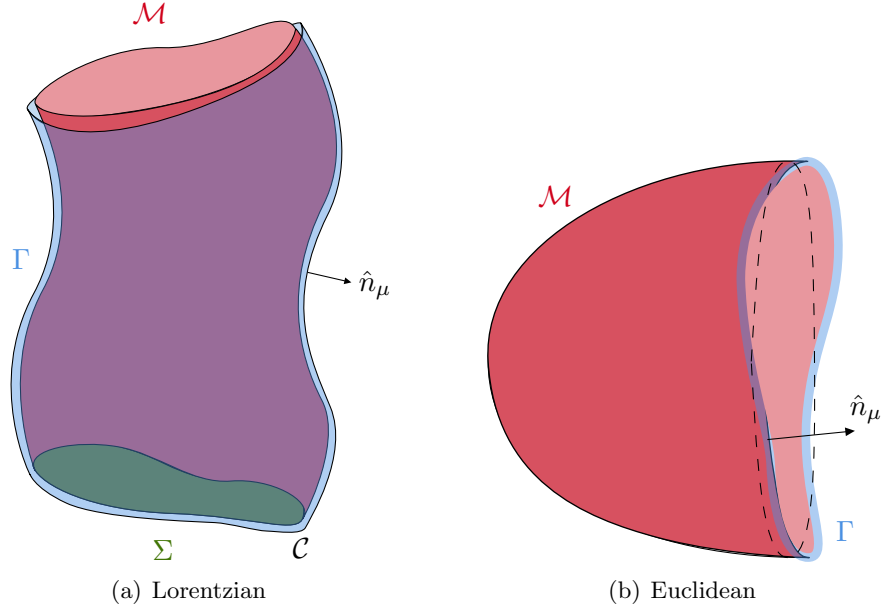
\begin{figure}[t!]
        \centering
         \subfigure[Lorentzian]{
               %\centering
%
%
%\tikzset{every picture/.style={line width=0.75pt}} %set default line width to 0.75pt        
%
\begin{tikzpicture}[x=0.75pt,y=0.75pt,yscale=-1,xscale=1]
%uncomment if require: \path (0,340); %set diagram left start at 0, and has height of 340

%Shape: Polygon Curved [id:ds35817102512356125] 
\draw  [fill={rgb, 255:red, 208; green, 2; blue, 27 }  ,fill opacity=0.46 ] (271.2,260.8) .. controls (267.34,252.12) and (285.93,247.16) .. (291.75,248.25) .. controls (297.56,249.35) and (308.47,244.98) .. (335.93,258.25) .. controls (363.38,271.53) and (365.38,264.07) .. (383.38,267.53) .. controls (401.38,270.98) and (408.35,275.48) .. (406.4,285.2) .. controls (404.45,294.93) and (382.45,285.17) .. (340.95,283.67) .. controls (299.45,282.17) and (275.06,269.48) .. (271.2,260.8) -- cycle ;
%Shape: Polygon Curved [id:ds36127878005637015] 
\draw  [fill={rgb, 255:red, 65; green, 117; blue, 5 }  ,fill opacity=0.75 ] (271.2,260.8) .. controls (267.34,252.12) and (285.93,247.16) .. (291.75,248.25) .. controls (297.56,249.35) and (308.47,244.98) .. (335.93,258.25) .. controls (363.38,271.53) and (365.38,264.07) .. (383.38,267.53) .. controls (401.38,270.98) and (408.35,275.48) .. (406.4,285.2) .. controls (404.45,294.93) and (382.45,285.17) .. (340.95,283.67) .. controls (299.45,282.17) and (275.06,269.48) .. (271.2,260.8) -- cycle ;
%Shape: Polygon Curved [id:ds4301343481165173] 
\draw  [fill={rgb, 255:red, 151; green, 190; blue, 232 }  ,fill opacity=0.57 ] (268,80.82) .. controls (265.14,76.71) and (301.14,53.63) .. (333.14,54.43) .. controls (365.14,55.23) and (402.72,35.57) .. (405.11,37.69) .. controls (407.5,39.81) and (408.06,59.34) .. (380.86,71.29) .. controls (353.66,83.23) and (343.43,85.86) .. (310.57,90.43) .. controls (277.71,95) and (270.86,84.93) .. (268,80.82) -- cycle ;
%Shape: Polygon Curved [id:ds5928842365427999] 
\draw  [fill={rgb, 255:red, 232; green, 151; blue, 158 }  ,fill opacity=1 ] (272.4,78.8) .. controls (266.43,72.83) and (304,49.2) .. (336,50) .. controls (368,50.8) and (395.71,28.07) .. (402.4,40.4) .. controls (409.09,52.73) and (374.53,72.13) .. (331.42,80.13) .. controls (288.31,88.13) and (278.37,84.77) .. (272.4,78.8) -- cycle ;
%Shape: Polygon Curved [id:ds23374855632726188] 
\draw  [fill={rgb, 255:red, 215; green, 81; blue, 96 }  ,fill opacity=1 ] (402.6,40.6) .. controls (402.86,39.21) and (437,87.8) .. (416.6,128.6) .. controls (396.2,169.4) and (403.8,203.8) .. (415.8,227) .. controls (427.8,250.2) and (402.29,293.71) .. (406.6,285.4) .. controls (410.91,277.09) and (390.2,265.8) .. (370.2,267.4) .. controls (350.2,269) and (334.2,253.8) .. (313.4,250.2) .. controls (292.6,246.6) and (267.72,251.39) .. (271.4,261) .. controls (275.08,270.61) and (251,217.4) .. (275.8,171.4) .. controls (300.6,125.4) and (268.67,74.27) .. (272.6,79) .. controls (276.53,83.73) and (299.06,89.27) .. (349.4,76.2) .. controls (399.74,63.13) and (402.34,41.99) .. (402.6,40.6) -- cycle ;
%Shape: Polygon Curved [id:ds21807976999890055] 
\draw  [fill={rgb, 255:red, 74; green, 144; blue, 226 }  ,fill opacity=0.44 ] (405.11,37.69) .. controls (403.75,35.06) and (440.89,89.36) .. (418.67,133.8) .. controls (396.44,178.24) and (410.13,206.33) .. (420,230.24) .. controls (429.87,254.16) and (409.56,286.82) .. (407.56,290.69) .. controls (405.56,294.56) and (394,292.69) .. (388,292.47) .. controls (382,292.24) and (370,288.69) .. (362,288.24) .. controls (354,287.8) and (304.89,286.47) .. (287.56,278.47) .. controls (270.22,270.47) and (268.85,262.69) .. (269.33,264.24) .. controls (269.82,265.8) and (246.4,219.22) .. (271.2,173.22) .. controls (296,127.22) and (266.29,77.64) .. (268,80.82) .. controls (269.71,84) and (286.76,104.18) .. (349.56,81.8) .. controls (412.35,59.42) and (406.47,40.32) .. (405.11,37.69) -- cycle ;
%Straight Lines [id:da9717336214356753] 
\draw    (408.57,162.7) -- (427.65,167.08) ;
\draw [shift={(430.57,167.75)}, rotate = 192.93] [fill={rgb, 255:red, 0; green, 0; blue, 0 }  ][line width=0.08]  [draw opacity=0] (3.57,-1.72) -- (0,0) -- (3.57,1.72) -- cycle    ;

% Text Node
\draw (259,149.9) node [anchor=north west][inner sep=0.75pt]  [color={rgb, 255:red, 74; green, 144; blue, 226 }  ,opacity=1 ]  {$\Gamma $};
% Text Node
\draw (329.5,295.9) node [anchor=north west][inner sep=0.75pt]  [color={rgb, 255:red, 65; green, 117; blue, 5 }  ,opacity=1 ]  {$\Sigma $};
% Text Node
\draw (319.5,26.4) node [anchor=north west][inner sep=0.75pt]  [color={rgb, 255:red, 208; green, 2; blue, 27 }  ,opacity=1 ]  {$\mathcal{M}$};
% Text Node
\draw (433,160.6) node [anchor=north west][inner sep=0.75pt]    {$\hat{n}_\mu$};
\draw (400,295) node [anchor=north west][inner sep=0.75pt]    {$\mathcal{C}$};
\end{tikzpicture} \label{fig: lorentzian tube}} \quad \quad 
                 \subfigure[Euclidean]{

%\tikzset{every picture/.style={line width=0.75pt}} %set default line width to 0.75pt        

\begin{tikzpicture}[x=0.75pt,y=0.75pt,yscale=-0.85,xscale=0.9]

%Curve Lines [id:da6371966294830084] 
\draw    (380.96,265.76) .. controls (278.52,266.28) and (203.24,218.06) .. (203.24,150.02) .. controls (203.24,81.98) and (280.52,31.95) .. (379.25,31.95) ;
%Curve Lines [id:da6507110486867125] 
\draw [fill={rgb, 255:red, 215; green, 81; blue, 96 }  ,fill opacity=1 ][line width=0.5]    (381.75,31.95) .. controls (345.98,33.2) and (353.57,140.64) .. (360.24,198.59) .. controls (366.91,256.54) and (375.6,265.62) .. (381.8,265.76) .. controls (388,265.9) and (339.81,265.72) .. (298.12,254.04) .. controls (256.43,242.37) and (207.17,220.27) .. (203.24,150.02) .. controls (199.31,79.77) and (289.36,28.45) .. (381.04,31.95) ;
%Curve Lines [id:da0807925976795929] 
\draw [color={rgb, 255:red, 74; green, 144; blue, 226 }  ,draw opacity=0.45 ][fill={rgb, 255:red, 232; green, 151; blue, 158 }  ,fill opacity=1 ][line width=3.25]    (344.16,101.42) .. controls (344.63,49.01) and (374.08,32.52) .. (380.08,31.95) .. controls (386.09,31.38) and (397.05,36.62) .. (398.48,77.6) .. controls (399.9,118.57) and (384.18,136.2) .. (380.85,182.9) .. controls (377.51,229.59) and (396.32,265.51) .. (381.8,265.76) .. controls (367.28,266.01) and (361.79,234.36) .. (360.83,211.49) .. controls (359.88,188.62) and (345.59,150.5) .. (344.16,101.42) -- cycle ;
%Curve Lines [id:da33085117861724467] 
\draw  [dash pattern={on 4.5pt off 4.5pt}]  (349.99,150.02) .. controls (352.66,105.33) and (355.35,33.48) .. (369.83,31.95) .. controls (384.3,30.43) and (384.78,85.22) .. (385.25,150.02) .. controls (385.73,214.82) and (384.78,267.23) .. (369.83,265.76) .. controls (354.88,264.28) and (352.66,197.38) .. (349.99,150.02) -- cycle ;
%Straight Lines [id:da14921082796465257] 
\draw    (358.96,196.16) -- (404.82,191.32) ;
\draw [shift={(407.8,191)}, rotate = 173.97] [fill={rgb, 255:red, 0; green, 0; blue, 0 }  ][line width=0.08]  [draw opacity=0] (5.36,-2.57) -- (0,0) -- (5.36,2.57) -- cycle    ;

% Text Node
\draw (229.71,45.16) node [anchor=north west][inner sep=0.75pt]  [color={rgb, 255:red, 208; green, 2; blue, 27 }  ,opacity=1 ]  {$\mathcal{M}$};
% Text Node
\draw (396,242.83) node [anchor=north west][inner sep=0.75pt]  [color={rgb, 255:red, 74; green, 144; blue, 226 }  ,opacity=1 ]  {$\Gamma $};
% Text Node
\draw (403.5,167.95) node [anchor=north west][inner sep=0.75pt]  [color={rgb, 255:red, 0; green, 0; blue, 0 }  ,opacity=1 ]  {$\hat{n}_{\mu }$};

%\draw (254.5,149.9) node [anchor=north west][inner sep=0.75pt]  [color={rgb, 255:red, 74; green, 144; blue, 226 }  ,opacity=1 ]  {$\Gamma $};
% Text Node
%\draw (329.5,295.9) node [anchor=north west][inner sep=0.75pt]  [color={rgb, 255:red, 65; green, 117; blue, 5 }  ,opacity=1 ]  {$\Sigma $};
% Text Node
%\draw (319.5,26.4) node [anchor=north west][inner sep=0.75pt]  [color={rgb, 255:red, 208; green, 2; blue, 27 }  ,opacity=1 ]  {$\mathcal{M}$};

\end{tikzpicture} \label{fig: euclidean tube}}                           
                \caption{Illustrations of manifolds with finite boundaries. In the Lorentzian problem, we supplement boundary conditions at $\Gamma$ with initial data at Cauchy surface $\Sigma$ (and the corner $\mathcal{C} = \Sigma \cap \Gamma$). In Euclidean signature, we look for smooth bulk solutions that satisfy the boundary conditions at $\Gamma$.} \label{fig: tubes}
\end{figure}

\noindent \textbf{Spherically symmetric boundary conditions.}
In Euclidean signature, it is usually common to consider boundaries $\Gamma$ whose topology is $S^1 \times S^{d-2}$.\footnote{See \cite{Banihashemi:2025qqi} for recent results with $S^1\times \mathbb{R}^{d-2}$ and $S^1\times \mathbb{H}^{d-2}$ conformal boundaries. See also \cite{Capoferri:2024sgo} for a linearized analysis with boundaries of the form $[-T,T]\times\Sigma$, for finite Euclidean time $T$ in four dimensions.} See Figure \ref{fig: tubes}. We then impose conformal boundary conditions so that the induced metric and the trace of the extrinsic curvature at $\Gamma$ are
\beq ds^{2}|_{\Gamma}=e^{2\bm{\omega} (Y^M)}(du^{2}+\mathfrak{r}^{2}d\Omega_{d-2}^{2})\;,\qquad \mathcal{K}|_{\Gamma}=\text{constant}\;.\label{eq:cbcsintrov2}\eeq
Here $\bm{\omega} (Y^M)$ is some unspecified function of the boundary coordinates, $\mathfrak{r}$ is a positive parameter characterizing the size of the sphere $S^{d-2}$, and $u$ is a Euclidean time coordinate with periodicity $u\sim u+\beta$ for some constant $\beta$. Given that only the conformal class of the induced metric is specified, only the dimensionless ratio $\tilde{\beta} \equiv \tfrac{\beta}{\frakr}$ is geometrically meaningful.

\noindent \textbf{Regular static solutions.} An ansatz for the metric that is spherically symmetric and static (in the $\tau$-coordinate) is given by
\begin{equation} \label{bulk_metric_static}
    ds^2 = f(r)d\tau^2+\frac{dr^2}{f(r)} + r^2 d\Omega^2_{d-2} \,  , \qquad r \in (r_+,\mathfrak{r}e^{\omega_{s}}) \, ,
\end{equation}
where $r_+$ is the largest positive root
of $f(r)$ and $\omega_{s}$ is a constant which will be later fixed by the boundary conditions. The boundary is located at $r=\frakr e^{\omega_{s}}$, at which the induced metric agrees with \eqref{eq:cbcsintrov2} for a constant $\bomega = \omega_{s}$ and a rescaled Euclidean time, $\tau\to\tfrac{e^{\omega_{s}}}{\sqrt{f(\r e^{\omega_{s}})}}u$. 

Requiring that the boundary has constant trace of the extrinsic curvature $\mathcal{K}$ yields\footnote{If $f(r)$ has another positive root $r_-<r_+$, it is also possible to consider the region with $r \in (\mathfrak{r} e^{\omega_{s}},r_{-})$. In that case, \eqref{eqn: K static d-dim} and \eqref{eqn: beta tilde d-dim} remain the same, but $\sigma = \text{sign} (f'(r_-))$.}
\begin{equation}\label{eqn: K static d-dim}
    \left.\mathcal{K}\right|_{\Gamma} = \sigma\left(\frac{f'(\frakr e^{\omega_{s}})}{2\sqrt{f(\frakr e^{\omega_{s}})}} + \frac{d-2}{e^{\omega_{s}}\frakr}\sqrt{f(\frakr e^{\omega_{s}})}\right) \, ,
\end{equation}
where $f'(x) \equiv \partial_x f(x)$, and $\sigma = \text{sign} (f'(r_+))$. Further assuming that $f(r)$ has a single pole at $r=r_+$, regularity of the solution requires an appropriate periodicity of the coordinate $\tau$. As a result, we obtain
\begin{equation}\label{eqn: beta tilde d-dim}
    \tilde{\beta}= \frac{4\pi\sqrt{f(\frakr e^{\omega_{s}})}}{\frakr e^{\omega_{s}}|f'(r_+)|} \, .
\end{equation}
In principle, for a given $f(\r e^{\omega_{s}})$, \eqref{eqn: K static d-dim}-\eqref{eqn: beta tilde d-dim} could be inverted to obtain $\omega_{s}$ as a function of the boundary data $\mathcal{K}$ and $\b$.

\noindent \textbf{Regular non-static solutions.} At fixed constant $\mathcal{K}$, there are also solutions that are still spherically symmetric but not static.\footnote{ Note that this does not imply any violation in Birkhoff's theorem. Indeed, Birkhoff's theorem restricts only the geometry  exterior to the gravitating mass. As an embedded surface, the boundary is allowed to be dynamical.} These can be found by computing the trace of the extrinsic curvature for an induced metric with $u$-dependent $\bomega$. Specifically, using the metric \eqref{bulk_metric_static}, we now set the boundary location to be $r= \r e^{\bomega(u)}$, which depends on the boundary coordinate $u$. Imposing that the induced metric is \eqref{eq:cbcsintrov2} yields an equation determining the coordinate $\tau$ at the boundary as a function of $u$,
\begin{equation}\label{eqn: tdu non-static d-dim}
    \left.\partial_u\tau\right|_{\Gamma} = \frac{e^{\bomega}\sqrt{f(\r e^{\bomega})-\r^2 \partial_u\bomega^2}}{f(\r e^{\bomega})} \, .
\end{equation}
Integrating this equation over the $S^1$, we obtain a relation between the periodicity of $\tau$ and $u$. The condition that the boundary has trace of the extrinsic curvature $\mathcal{K}$ then gives \cite{Liu:2024ymn}
\begin{equation}\label{eqn: K non-static d-dim}
    \r^2\partial_u^2\bomega = (d-2)\left(f(\r e^{\bomega})-\r^2 (\partial_u\bomega)^2\right)+\frac{1}{2}\r e^{\bomega}f'(\r e^{\bomega}) -\sigma  \mathcal{K}\r e^{\bomega}\sqrt{f(\r e^{\bomega})-\r^2 (\partial_u\bomega)^2} \,.
\end{equation}
For a given $f(r)$ and constant $\k$,\footnote{Note that this equation also holds for a time-dependent $\k= \k(u)$.} this is a non-linear, second-order differential equation for the conformal factor at the boundary, $\bomega(u)$.  In principle, to completely specify the solution, this equation needs to be supplemented by two extra pieces of boundary data. This might seem in contradiction with the well-posedness of conformal boundary conditions. While true in Lorentzian signature (where additional initial data at the intersection between the timelike boundary and the initial Cauchy slice must be provided \cite{An:2025rlw}), this is not the case in Euclidean signature. As we will see below at the linearized level, the reason is that the periodicity of $u$ imposes constraints on the allowed solutions. 

It is straightforward to check that, for a constant $\bomega=\omega_{s}$, \eqref{eqn: K non-static d-dim} reduces to \eqref{eqn: K static d-dim}, for the static solution. Linearizing \eqref{eqn: K non-static d-dim} about the static solution, we obtain
\begin{equation}\label{eqn: K non-static lin d-dim}
    \r^2 \partial_u^2\delta\bomega(u) = -\Omega^2 \delta\bomega(u) \quad , \quad \Omega^2 \equiv - \sigma\frac{\sqrt{f(\r e^{\omega_{s}})}\partial_{\omega_{s}} \mathcal{K}}{\r e^{\omega_{s}}} \,,
    \end{equation}
where $\delta \bomega(u) \equiv \bomega(u) - \omega_{s}$, and $\partial_{\omega_{s}} \mathcal{K}$ is understood as taking $\omega$-derivative of \eqref{eqn: K static d-dim}. For the class of Euclidean geometries analyzed in this paper, $\sigma \mathcal{K}$ is a monotonically decreasing function of $\omega_{s}$, which makes $\Omega^2>0$. This leads to a general solution to \eqref{eqn: K non-static lin d-dim} which exhibits oscillatory behavior in the $u$-coordinate, i.e., 
\begin{equation}\label{eqn: lin non-static sol}
    \delta \bomega(u) = \upsilon \cos\left(\Omega \frac{u - u_0}{\r}\right) \, ,
\end{equation}
where $u_0$ and $\upsilon$ are constants of integration. The $u_{0}$ simply corresponds to a shift in the boundary time, while it might seem that $\upsilon$ is a free undetermined parameter, generating a continuum of non-unique solutions. As we show in Appendix \ref{app: brane dynamics}, however, the amplitude $\upsilon$ (and the frequency $\Omega$) are in fact completely determined by the boundary data $(\tilde{\beta}, \k)$, so there is no violation of uniqueness. Furthermore, the linearized Weyl factor must obey $\delta\bomega(u+\beta)=\delta\bomega(u)$, which leads to
\begin{equation}\label{eqn: winding condition}
    \Omega \tilde{\beta} = 2 \pi n \, , \qquad n \in \mathbb{N} \, .
\end{equation}
As $\Omega$ is a function of $\k$ and $\tilde{\beta}$, this condition implies this linearized solution can only exist for finely tuned pairs of boundary data $(\tilde{\beta},\mathcal{K})$. Finally, note the analogous solutions in Lorentzian signature will exhibit exponential behavior in boundary time.

Returning to the non-linear case, if we consider $f(r) = 1 + \tfrac{2 \Lambda_d r^2}{(d-1)(d-2)}$, then, for maximally symmetric spacetimes, \eqref{eqn: K non-static lin d-dim} simplifies to \cite{Anninos:2024wpy, Liu:2024ymn, Anninos:2024xhc}
\begin{equation}
    \r^2\partial_u^2\bomega = (d-2)\left(1-\r^2(\partial_u\bomega)^2\right)-\frac{2\Lambda_d\r^2 e^{2\bomega}}{d-2} -\sigma \mathcal{K}\r e^{\bomega}\sqrt{1-\frac{2\Lambda_d \r^2 e^{2\bomega}}{(d-1)(d-2)}-\r^2 (\partial_u\bomega)^2} \,.
\end{equation}

Consider now a general $f(r)$ in a theory with a negative cosmological constant $\Lambda_d = -\tfrac{(d-1)(d-2)}{2\ell^2}$. In this case, it is possible to consider the limit when the boundary is very large, i.e., when it approaches the conformal boundary of asymptotically AdS spacetimes. When this happens, $e^{\bomega} \gg 1$, and $f(\r e^{\bomega}) \approx \tfrac{\r^2 e^{2\bomega}}{\ell^2} +1+\mathcal{O}(e^{-\bomega})$. In particular, this implies all terms containing $u$-derivatives of $\bomega(u)$ in \eqref{eqn: K non-static d-dim} are suppressed. The only term that remains imposes $\mathcal{K}=\tfrac{d-1}{\ell}$, in agreement with the near AdS boundary limit \cite{Anninos:2024xhc}. Going to next order yields an equation that captures the dynamics of $\bomega$. In particular, let $\bomega \to \bomega - \log(\epsilon)$ and $\mathcal{K} = \tfrac{d-1}{\ell}+\epsilon^2 \delta \mathcal{K}$. By taking $\epsilon\to 0$, \eqref{eqn: K non-static d-dim} becomes (to leading order),
\begin{equation}
    \r^2 \partial_u^2\bomega = - \delta \mathcal{K}  \frac{\r^2e^{2\bomega}}{\ell} - \frac{d-3}{2}\left(\r^2 (\partial_u\bomega)^2-1\right) \, .
\label{eq:dynamicsofomega}\end{equation}
This is still hard to solve analytically for generic spacetime dimension. In the particular case of $d=3$, however, the differential equation \eqref{eq:dynamicsofomega} becomes the one-dimensional Liouville equation,
\begin{eqnarray}\label{eqn: 0th non-static eqn}
    \partial_u^2\bomega+ \frac{\delta\mathcal{K}}{\ell} e^{2\bomega}= 0\,,
\end{eqnarray}
where $\tfrac{\delta\k}{\ell}$ plays the role of the Liouville coupling.  We will come back to this equation from the two-dimensional perspective in Section \ref{sec: schwarzian}.

Let us expand \eqref{eqn: K non-static d-dim} further to first sub-leading order in $\epsilon$. For simplicity, we take the bulk metric to be the BTZ black hole, namely $d=3$ and $f(r)=\tfrac{r^2-r_{\text{bh}}^2}{\ell^2}$. Now we consider $\bomega \to \bomega_0 +\epsilon^2\,\bomega_1 - \log{\epsilon}$ while keeping $\mathcal{K} = \tfrac{2}{\ell}+\epsilon^2 \delta \mathcal{K}$. 
The leading order imposes that $\bomega_0$ satisfies \eqref{eqn: 0th non-static eqn}. At order $\epsilon^2$ we obtain 
\begin{equation} \label{eq: brane dynamics second order}
    \partial_u^2\bomega_1 +\frac{2 \delta\mathcal{K}}{\ell} \bomega_1 e^{2\bomega_0}=   \frac{\ell^2}{4  e^{2\bomega_0}} \left((\partial_u \bomega_0)^2+\frac{r_{\text{bh}}^2}{\ell^2\r^2}\right)\left((\partial_u\bomega_0)^2+\frac{r_{\text{bh}}^2}{\ell^2\r^2}+\frac{2\delta \k}{\ell}e^{2\bomega_0}\right)\, .
\end{equation}
While at leading order the boundary dynamics is decoupled from the bulk, this equation shows this no longer holds at next order, where the boundary equation for $\bomega_1(u)$ starts receiving bulk contributions from, for instance, the black hole horizon. Note that even for $r_{\text{bh}}=0$, the RHS of \eqref{eq: brane dynamics second order} results in a complicated source for $\bomega_1(u)$.

\subsubsection{Conformal canonical thermodynamics} \label{sec: cct}

Following the Gibbons-Hawking prescription \cite{Gibbons:1976ue}, at leading order in a semi-classical saddle-point approximation, the conformal canonical thermal partition function is 
\beq \mathcal{Z}(\tilde{\beta},\mathcal{K})\approx \sum_{g_{\mu\nu}^{\ast}}e^{-I_{d} [g_{\mu\nu}^{\ast}]}\;,\eeq
where $I_{d} [g_{\mu\nu}^{\ast}]$ corresponds to evaluating the on-shell Euclidean action on $g_{\mu\nu}^{\ast}$, Euclidean smooth metrics obeying the classical Einstein field equations and conformal boundary conditions (\ref{eq:cbcsintrov2}). In principle, both static and non-static solutions contribute to this partition function. Here, we will only consider spherically symmetric and static solutions, though it would be interesting to understand the contributions from non-static saddles as well. We leave this for future work.

Via the partition function, the conformal energy, entropy, and specific heat at fixed $\mathcal{K}$ are defined, respectively, 
\beq E_{\text{conf}}\equiv -\partial_{\tilde{\beta}}\big|_{\mathcal{K}}\log\mathcal{Z}\;,\quad \mathcal{S}_{\text{conf}}\equiv (1-\tilde{\beta}\partial_{\tilde{\beta}})\big|_{\mathcal{K}}\log \mathcal{Z}\;,\quad C_{\mathcal{K}}\equiv \tilde{\beta}^{2}\partial^{2}_{\tilde{\beta}}\big|_{\mathcal{K}}\log \mathcal{Z}\;.\label{eq:confESC}\eeq
Below we provide a brief summary of new and old results regarding conformal thermodynamics.

%\vspace{2mm}

\noindent \textbf{(A)dS$_{3}$ conformal thermodynamics.} Let us briefly review the results for conformal thermodynamics 
in (A)dS$_{3}$ \cite{Anninos:2024wpy}. We first consider the case of a positive cosmological constant with $\Lambda_3 = + \ell^{-2}$. The metric is now given by \eqref{bulk_metric_static} with
\begin{equation}
    f(r)=\frac{r_{\text{c}}^{2}-r^{2}}{\ell^{2}} \,.
\end{equation}
The cosmological horizon is located at $r=r_{\text{c}}$. The boundary at $r=\r e^{\omega_s}<r_{\text{c}}$, separates the static patch of dS$_3$ into two regions:
(i) a pole patch, the domain with $r\in[0,\r e^{\omega_s}]$ that does not contain a horizon, 
and (ii) a cosmic patch with $r\in[\r e^{\omega_s},r_{\text{c}}]$, thus containing the cosmological horizon. For the cosmic patch, the canonical conformal energy, entropy and specific heat are\footnote{Here we report the energy coming from the unregulated action. In \cite{Anninos:2024wpy}, the thermodynamic quantities are derived from a regulated action, $I_{E,reg}=I_{E}^{\text{cosmo}}-I_{E}^{\text{pole}}$, where $I_{E}^{\text{pole}}$ is the on-shell Euclidean action for the pole patch. The pole patch action is linear in inverse conformal temperature $\tilde{\beta}$ 
such that it does not contribute to the entropy.}
\beq
\begin{cases}
 E_{\text{conf}}=\frac{2\pi^{2}\ell}{8G_{3}\tilde{\beta}^{2}}\left[\sqrt{\mathcal{K}^{2}\ell^{2}+4}-\mathcal{K}\ell\right]\;,\\
 \mathcal{S}_{\text{conf}}=\frac{\text{Area}(r_{\text{c}})}{4G_{3}}=\frac{\pi^{2}\ell}{2G_{3}\tilde{\beta}}\left[\sqrt{\mathcal{K}^{2}\ell^{2}+4}-\mathcal{K}\ell\right]\;, \\
 C_\k = \frac{\pi^{2}\ell}{2G_{3}\tilde{\beta}}\left[\sqrt{\mathcal{K}^{2}\ell^{2}+4}-\mathcal{K}\ell\right] \,.
\end{cases}
\label{eq:dS3confthermo}\eeq
Note the specific heat is positive for all allowed values of canonical thermodynamic data  $(\tilde{\beta},\mathcal{K})$.  
The conformal entropy may be suggestively massaged into a microcanonical Cardy-like form\footnote{If one works with the regulated on-shell action by subtracting off the pole patch contribution, then $\mathcal{S}_{\text{conf}}=2\pi\sqrt{\frac{\mathfrak{c}_{\text{dS}}}{3}\left(E_{\text{conf}}-\frac{\mathfrak{c}_{\text{dS}}}{12}\right)}$ \cite{Anninos:2024wpy}.}
\beq \mathcal{S}_{\text{conf}}=2\pi\sqrt{\frac{\mathfrak{c}_{\text{dS}}}{3}E_{\text{conf}}}\;,\quad \mathfrak{c}_{\text{dS}}\equiv \frac{3\ell}{4G_{3}}\left(\sqrt{\mathcal{K}^{2}\ell^{2}+4}-\mathcal{K}\ell\right)\;,\label{eq:cardyformdS3}\eeq
for ``central charge'' $\mathfrak{c}_{\text{dS}}$ \cite{Anninos:2024wpy}, which relates to $N_{\text{d.o.f.}}$ in \eqref{eq: ndof} for $d=3$ via
$
    N_{\text{d.o.f.}} = \frac{2 \pi^2\mathfrak{c}_{\text{dS}}}{3}\,.
$

The AdS$_3$ results for a finite boundary obeying conformal boundary conditions in a BTZ background follow from analytic continuation of the dS results, such that $\ell \to - i \ell_{\text{AdS}}$. In particular, the ``central charge'' becomes 
\begin{equation}\label{eq:cardyformAdS3}
    \mathfrak{c}_{\text{AdS}}=\frac{3\ell_{\text{AdS}}}{4G_{3}}\left(\mathcal{K}\ell_{\text{AdS}}-\sqrt{\mathcal{K}^{2}\ell_{\text{AdS}}^{2}-4}\right) \,,
\end{equation}
which is a monotonically decreasing function from the Brown-Henneaux central charge ($c=\tfrac{3\ell_{\text{AdS}}}{2G_{3}}$) when $\k \ell_{\text{AdS}}= 2$ (the conformal boundary of AdS$_3$) to zero, as $\k \ell_{\text{AdS}} \to \infty$, near the BTZ black hole horizon. See \cite{edgar} for a more complete treatment of this case.

%\vspace{2mm}

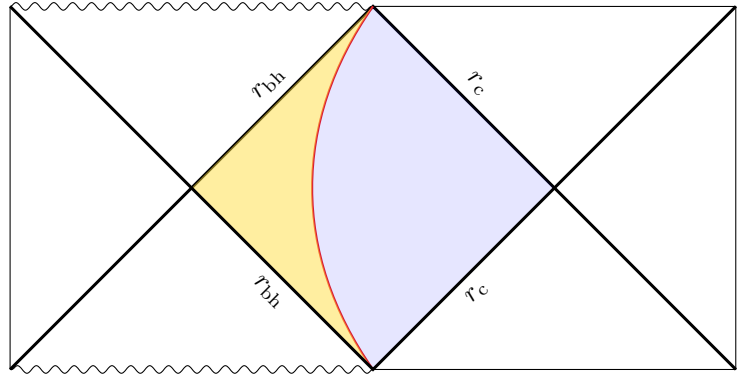
\begin{figure}[t]
\centering
\begin{tikzpicture}[scale=1.2]
    \draw (0,0) coordinate (a) -- (0,4) coordinate (b); 
    \draw[decorate, decoration={snake, amplitude=0.5mm, segment length=2.5mm}] (b) -- (4,4) coordinate (c) ;
    \draw (c) -- (8,4) coordinate (d) -- (8,0) coordinate (e) -- (4,0) coordinate (f);
    \draw[decorate, decoration={snake, amplitude=0.5mm, segment length=2.5mm}] (f) -- (a); 
    \draw [line width = .4mm] (a) -- (c) node[pos=.75, above, sloped] {{\small $r_{\text{bh}}$}} -- (e) node[pos=.25, above, sloped] {{\small $r_{\text{c}}$}}; 
    \draw[thick, Red, name path=rB] (c) to[bend right=35] (f);
    \path (c) to[bend right=35] node[pos=.65,above,sloped] {} (f);
    \fill[fill=Goldenrod, fill opacity=0.5] (c) to[bend right=35] (f) -- (2,2) -- (c);
    \fill[fill=blue, fill opacity=0.1] (c) to[bend right=35] (f) -- (6,2) -- (c);
    \draw [line width = .4mm] (b) -- (f) node[pos=.75, below, sloped] {{\small $r_{\text{bh}}$}} -- (d) node[pos=.25, below, sloped] {{\small $r_{\text{c}}$}};
   % \draw[dashed] (c) -- (f) node[pos=.5, below, sloped] {{\small $\rho = \beta/2$}};
\end{tikzpicture}
\caption{Penrose diagram of Schwarzschild-de Sitter black hole with finite timelike boundaries. The cosmic patch (right shaded blue region) is the region bounded between the timelike boundary  (thick red arc) and the cosmological horizon $r_{\text{c}}$.  The black hole patch  patch (left gold shaded region) is the region between the boundary and the black hole horizon $r_{\text{bh}}$.
}
\label{fig:dS2bdryB}  
\end{figure}

\noindent \textbf{Schwarzschild-dS$_{4}$ conformal thermodynamics.}  Next let us review Einstein gravity in $d=4$, with a positive cosmological constant $\Lambda_4 = + \tfrac{3}{\ell^2}$  \cite{Anninos:2024wpy}. Regular static solutions include pole and cosmic patches, as in $d=3$, but now there are also black hole patches, i.e., the region between the finite boundary and a Schwarzschild-dS black hole horizon (see Figure \ref{fig:dS2bdryB} for a Lorentzian depiction).

In particular, solving the Einstein equation in this case gives metric \eqref{bulk_metric_static} with
\begin{equation}
    f(r) = 1 - \frac{2 \mu}{r} - \frac{ r^2}{\ell^2} \,,
\end{equation}
where the black hole mass parameter $\mu \geq 0$ is related to the size of the cosmological horizon $r_c$ via
\begin{eqnarray}
    \mu = \frac{r_{\text{c}}}{2} \left( 1 - \frac{r_{\text{c}}^2}{\ell^2} \right) \,.
\end{eqnarray}
For $\mu =0$, we obtain the empty de Sitter solution, with cosmological horizon at $r_{\text{c}} = \ell$, while for $\mu >0$, the cosmological horizon reduces its size and a black hole horizon appears at $r = r_{\text{bh}}$. The size of the cosmological and the black hole horizon are related by,
\begin{equation}
    r_{\text{bh}} = \frac{1}{2} \left( \sqrt{4\ell^2 - 3 r_{\text{c}}^2} - r_{\text{c}} \right) \,.
\end{equation}
Both horizons coincide when $r_{\text{c}} = r_{\text{bh}} = \tfrac{\ell}{\sqrt{3}}\equiv r_{\text{N}}$, namely, the Nariai limit. This limit will play an important role in the two-dimensional description as the near horizon geometry is locally dS$_2 \times S^2$. The phase space of static spherically symmetric solutions as a function of both $\tilde{\beta}$ and $\k$ is more rich than its three dimensional counterpart, and the details can be found in \cite{Anninos:2024wpy}. Here we focus on solutions that are (near)-Nariai, which will be relevant for the lower dimensional study.

\noindent \textbf{Near-Nariai.} Consider the conformal thermodynamics of (near-) Nariai solutions. In the exact limit, the black hole and cosmological
horizons coincide such that the inverse conformal temperature is tuned to $\tilde{\beta} = \tilde{\beta}_\text{N}$ and obeys,
\begin{equation}
    \tilde{\beta}_\text{N} = \frac{2\pi}{\sqrt{\k^2 r_{\text{N}}^2 +1}} \,.
\label{eq:Naritemp4D}\end{equation}

Unlike the Dirichlet case, note that here there are Nariai solutions for all values of $\k \ell \in \mathbb{R}$. To compute thermodynamic quantities we need to go slightly away from the Nariai temperature, obtaining
\begin{equation}
\begin{cases}
    E_{\text{conf}} = \frac{\k r_{\text{N}}^3}{3G_4} \,, \\
    \mathcal{S}_{\text{conf}} = \frac{\pi r_{\text{N}}^2}{ G_4} \,, \\
    C_\k = C_\k^{(0)} \equiv \frac{6\pi r_{\text{N}}^2
    \left(1+\k^2 r_{\text{N}}^2\right)^{3/2}}{\left(9 \k r_{\text{N}}+5 \k^3 r_{\text{N}}^3 - 2 (1+\k^2r_{\text{N}}^2)^{3/2}\right)G_4} \,.
\end{cases}
\label{eq:Nariaiconfthermo4D}\end{equation}
The entropy is the Bekenstein-Hawking area-entropy for (half) the Nariai solution, $\mathcal{S}_{\text{N}}=\frac{4\pi r_{\text{N}}^{2}}{4G_{4}}$. Note again that there are Nariai patches with positive specific heat for $\mathcal{K}\ell \gtrsim  0.405$, in contrast to the Dirichlet case \cite{Svesko:2022txo,Banihashemi:2022jys}. 

Finally, one can perturb the solution slightly away from the Nariai geometry. In terms of the boundary data, this amounts to taking temperature infinitesimally away from the Nariai temperature so that $\tilde{\beta} = \tilde{\beta}_\text{N} + \delta \tilde{\beta}$, for small real $\delta \tilde{\beta}$. In that case, the thermodynamic quantities are modified to
\begin{equation}
\begin{cases}
    E_{\text{conf}} = \frac{\k r_{\text{N}}^3}{3G_4} - C_\k^{(0)}\frac{\delta\tilde{\beta}}{\tilde{\beta}_{\text{N}}^2} \,, \\
    \mathcal{S}_{\text{conf}} = \frac{\pi r_{\text{N}}^2}{ G_4}- C_\k^{(0)}\frac{\delta\tilde{\beta}}{\tilde{\beta}_{\text{N}}} \,, \\
    C_\k = C_\k^{(0)} - \frac{3r_{\text{N}}^2\left(1+\k^2r_{\text{N}}^2\right)^2\left(-74+21\k^2r_{\text{N}}^2+39\k^4r_{\text{N}}^4+8\k^6r_{\text{N}}^6+2\left(9\k r_{\text{N}}+5\k^3 r_{\text{N}}^3\right)\left(1+\k^2 r_{\text{N}}^2\right)^{3/2}\right)}{\left(9\k r_{\text{N}}+5\k^3 r_{\text{N}}^3-2\left(1+\k^2 r_{\text{N}}^2\right)^{3/2}\right)^3G_4} \delta\tilde{\beta} \,.
\end{cases}
\label{eq:nearnariaithermo4D}\end{equation}
We also see that the conformal entropy is equal to the Nariai entropy plus near-Nariai corrections proportional to the deviation $\delta\tilde{\beta}$.

\subsection{Einstein-Maxwell} \label{sec2: maxwell}

As noted in the introduction, two-dimensional dilaton gravity captures the horizon thermodynamics of the near-horizon limit of near-extremal black holes. Near-extremal black holes must be charged (or rotating), the simplest being solutions to Einstein-Maxwell theory (with or without cosmological constant). 
Here we elaborate on boundary conditions for the Maxwell field and the consequences for the thermodynamics of near-extremal black holes, as this will be relevant for comparison to the effective 2D model we uncover in Section \ref{sec:confthermo2D}.

Euclidean Einstein-Maxwell gravity is characterized by the Einstein-Hilbert action (\ref{eq:einhilbv1_Euc}) plus the Maxwell action describing a $U(1)$ gauge field $\mathcal{A}_{M}$,
\beq I_{\text{Max}}=\frac{1}{64\pi G_{d}\mu_{0}}\int_{\mathcal{M}}\hspace{-2mm} d^{d}X\sqrt{G}\mathcal{F}_{MN}^{2}\;,\label{eq:bulkmaxact}\eeq
where $\mathcal{F}_{MN}=\partial_{M}\mathcal{A}_{N}-\partial_{N}\mathcal{A}_{M}$ and $\mu_{0}$ is a dimensionless coupling constant.
The Maxwell action (\ref{eq:bulkmaxact}) has a well-posed variational problem if the gauge field obeys Dirichlet boundary conditions, $\delta\mathcal{A}_{M}|_{\Gamma}=0$, independent of the metric boundary conditions; the boundary term associated with the Einstein-Hilbert term remains unchanged. Alternatively, the Maxwell action is to be supplemented by the boundary term 
\beq I_{A}=-\frac{1}{16\pi G_{d}\mu_{0}}\oint_{\Gamma}d^{d-1}Y\sqrt{H}n_{M}\mathcal{F}^{MN}\mathcal{A}_{N}\;,\label{eq:IAbdrytermmain}\eeq
if $\delta(\sqrt{H}n_{M}\mathcal{F}^{MN})|_{\Gamma}=0$ instead (and any metric boundary conditions). Whether the boundary term (\ref{eq:IAbdrytermmain}) is included influences the type of thermal ensemble one considers.
Namely, we follow standard terminology and refer to fixed $\mathcal{A}_{M}$ as the `grand canonical ensemble', and fixed $\sqrt{H}n_{M}\mathcal{F}^{MN}$ as the `canonical ensemble'. In the former, the electrostatic potential is kept fixed, while the electric charge is held fixed in the latter. In this article we only focus on the canonical (fixed charge) ensemble. See \cite{Banihashemi:2025qqi} for an analysis of the conformal grand canonical ensemble in arbitrary dimensions.

\subsubsection{Conformal canonical thermodynamics} 

Consider asymptotically flat charged black hole solutions of four-dimensional Einstein-Maxwell theory in the conformal canonical ensemble, i.e., fixed charge. For concreteness, we consider Euclidean four-dimensional solutions \eqref{bulk_metric_static} that are static, spherically symmetric and regular with
\begin{equation}\label{eqn: f(r) charged bh}
    f(r) = 1+\frac{Q^2}{r^2} - \frac{Q^2+r_{\text{bh}}^2}{r_{\text{bh}} r} \,,
\end{equation}
for charge parameter $Q$. In the above we have set $\mu_0=1/4$ for simplicity, which corresponds to $Q$ having units of length.
There are two positive real roots to the blackening factor, characterizing the locations of the inner ($r_{\text{in}}$) and outer ($r_{\text{bh}}$) black hole horizons with $r_{\text{in}}\leq r_{\text{bh}}$, and $r_{\text{bh}}\geq|Q|$. In particular, these horizons are related via
\begin{equation}
    r_{\text{in}} r_{\text{bh}} = Q^2 \, .
\end{equation}
Here we have cast the black hole mass in terms of $Q$ and $r_{\text{bh}}$. The charged black hole becomes extremal when $r_{\text{in}}=r_{\text{bh}}=|Q|$. 

\begin{figure}[h!]
\centering
\begin{tikzpicture}[scale=0.9]
	\pgfmathsetmacro\myunit{4} 
           \draw [dashed, white]	(0,0)			coordinate (a)
		--++(90:\myunit)	coordinate (b);
	\draw [dashed, white] (b) --++(0:\myunit)		coordinate (c);
							
	\draw[dashed, white] (c) --++(-90:\myunit)	coordinate (d);
         % \draw [line width = .4mm] (b)  --  node[pos=.5, above, sloped] {${\color{black} r=r_{-}}$} (-2,2) -- (a);
           %\draw [line width = .4mm] (a)  --  node[pos=.5, below, sloped] {${\color{black} r=r_{-}}$} (-2,2) -- (b);
            \draw [line width = .4mm] (a)  --  node[pos=.5, above, sloped] {} (-2,-2) -- (0,-4) coordinate (n3);
             \draw [line width = .4mm] (n3)  --  node[pos=.5, below, sloped] {} (-2,-2) -- (a);
	\draw [line width = .4mm] (b)  --  node[pos=.5, above, sloped] {${\color{black} r_{\text{in}}}$} (2,2) -- (d);
 \draw [line width = .4mm] (d)  --  node[pos=.5, below, sloped] {${\color{black} r_{\text{in}}}$} (2,2) -- (b);
           \draw [line width = .4mm] (c) --  node[pos=.5, above, sloped] {${\color{black} r_{\text{in}}}$} (2,2) -- (a);
            \draw [line width = .4mm] (a) --  node[pos=.5, below, sloped] {${\color{black} r_{\text{in}}}$} (2,2) -- (c);
            %\draw [line width = .4mm] (c) -- node[pos=.5, above, sloped] {${\color{black} r=r_{-}}$} (6,2) -- (d);
             % \draw [line width = .4mm] (d) -- node[pos=.5, below, sloped] {${\color{black} r=r_{-}}$} (6,2) -- (c);
    \draw[dashed,white] (-2,0) coordinate (e) -- (-2,-4) coordinate (n1);
   % \draw [line width = .4mm] (a) to [out=15, in=165] node[pos=.5, above] {$r=-\infty$} (d);
    %\draw [line width = .4mm] (a) to [out=-15, in=-165] node[pos=.5, below] {$r=\infty$} (d);
    \draw [dashed,white]  (e) -- (-2,4) coordinate (f);
    \draw [white] (f) -- (b);   %node[pos=.5, above] {$r=4\infty$}; 
    %\draw [line width = .4mm]  (b) to [out=-15, in=-165] node[pos=.5, below] {$r=-\infty$} (c);
    \draw [white] (c) -- (6,4) coordinate (g);   %node[pos=.5, above] {$r=3\infty$}; 
    \draw [dashed,white] (g) -- (6,0) coordinate (h);
     \draw[dashed,white] (h) -- (6,-4) coordinate (n2);
    %\draw (h) -- (d)  node[pos=.5, below] {$r=2\infty$}; 
    \draw [decorate, decoration={snake, amplitude=0.5mm, segment length=2.5mm}] (c) -- (d)   node[pos=.5, above, sloped] {}; %rightsingularity
    \draw[thick, Red, name path=rB] (c) to[bend right=35] (d);
 \fill[fill=blue, fill opacity=0.1] (c) to[bend right=35] (d)  -- (2,2) -- (c);
     \draw [decorate, decoration={snake, amplitude=0.5mm, segment length=2.5mm}] (b) -- (a)   node[pos=.5, above, sloped] {}; %leftsingularity
     %\draw (n1) -- (n3);
     \draw [white] (n2) -- (4,-4) coordinate (n4);
      \draw [line width = .4mm] (n4)  --  node[pos=.5, below, sloped] {} (6,-2) -- (d);
             \draw [line width = .4mm] (d)  --  node[pos=.5, above, sloped] {} (6,-2) -- (n4);
              \draw [line width = .4mm] (a) --  node[pos=.5, above, sloped] {${\color{black} r_{\text{bh}}}$} (2,-2) -- (n4);
              \draw [line width = .4mm] (n4) --  node[pos=.5, below, sloped] {${\color{black} r_{\text{bh}}}$} (2,-2) -- (a);
               \draw [line width = .4mm] (n3)  --  node[pos=.5, below, sloped] {${\color{black} r_{\text{bh}}}$} (2,-2) -- (d);
                \draw [line width = .4mm] (d)  --  node[pos=.5, above, sloped] {${\color{black} r_{\text{bh}}}$} (2,-2) -- (n3);
                 \draw[thick, Red, name path=rB] (d) to[bend right=25] (n4);
\fill[fill=Goldenrod, fill opacity=0.5] (d) to[bend right=25] (n4)  -- (2,-2) -- (d);
               % \draw [line width = .4mm] (n3) to [out=15, in=165] node[pos=.5, above] {$r=\infty$} (n4);
\end{tikzpicture}
\caption{\small Penrose diagram of non-extremal Reissner-Nordstr{\"o}m black hole with finite timelike boundaries. An inner patch (upper shaded blue region) lies between the boundary (thick red arc) and the inner horizon $r_{\text{in}}$, while an outer patch (bottom gold shaded region) is between the boundary and the outer horizon $r_{\text{bh}}$.}
\label{fig:penRN}
\end{figure}
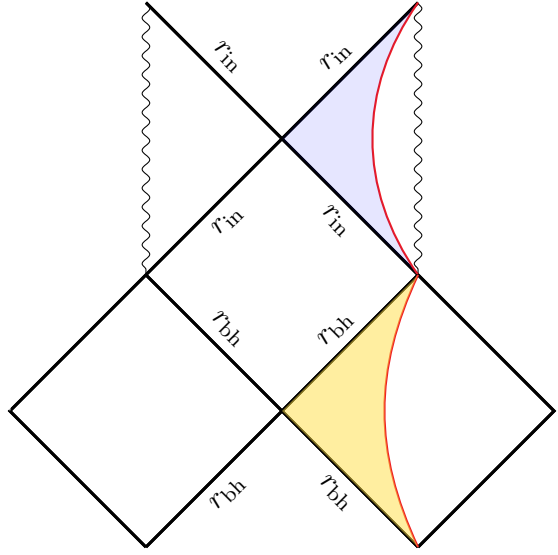

We consider a finite boundary located at $r=\r e^{\omega_s}$. For charged black holes, this boundary delimits two different  patches containing horizons: (i) outer patches, i.e., regions between the outer horizon and the boundary, namely $r\in \left[r_{\text{bh}},\r e^{\omega_s}\right]$; and (ii) inner patches, regions between the boundary and the inner horizon, $ \left[\r e^{\omega_s},r_{\text{in}}\right]$. Note that inner patches are allowed because the singularity at $r=0$ lies outside the patch. The equivalent patches in Lorentzian signature are depicted in Figure \ref{fig:penRN}. 

In addition to the metric, the gauge field 1-form is $A_MdX^M = \tfrac{iQ}{r_{\text{bh}}}\left(1-\tfrac{r_{\text{bh}}}{r}\right)d\tau$,
where we work in a gauge such that the electric potential $A_{\tau}$ vanishes at $r=r_{\text{bh}}$. Note that by pulling back the gauge field $A_M$ onto the boundary, we find that 
\begin{equation}
    \left.\sqrt{H} n_M F^{M u}\right|_{\Gamma} = i Q \sin(\theta)\, , \qquad \left.\sqrt{H} n_M F^{M \theta}\right|_{\Gamma} = \left.\sqrt{H} n_M F^{M \phi}\right|_{\Gamma} = 0 \,,
\end{equation}
from where it is clear that the black hole charge $Q$ is boundary data. The results presented in the remaining of this section will only depend on $|Q|$. We will assume $Q>0$ and write only $Q$ for simplicity.

As usual, we can compute thermodynamic quantities by evaluating the on-shell Euclidean action. 
The full phase space of static, spherically symmetric solutions in the fixed charge ensemble is a function of the thermodynamic data $(\tilde{\beta}, \k, Q)$. Given the trace of extrinsic curvature (\ref{eqn: K static d-dim}) and inverse conformal temperature (\ref{eqn: beta tilde d-dim}), the boundary data for the outer patches is
\begin{equation}
    \tilde{\beta}  = \frac{4\pi r_{\text{bh}}^3\sqrt{(\r e^{\omega_s}-r_{\text{bh}})(\r e^{\omega_s}-\frac{Q^2}{r_{\text{bh}}})}}{\r^2 e^{2\omega_s}(r_{\text{bh}}^2-Q^2)} \, , \quad \k = \frac{2Q^2r_{\text{bh}} + 4 \r^2 e^{2\omega_s}r_{\text{bh}}-3\r e^{\omega_s}(Q^2+r_{\text{bh}}^2)}{2\r^2 e^{2\omega_s}r_{\text{bh}}\sqrt{(\r e^{\omega_s}-r_{\text{bh}})(\r e^{\omega_s}-\frac{Q^2}{r_{\text{bh}}})}} \,. \label{eq: bdy data RN}
\end{equation}
Upon inverting these equations, we obtain the horizon and boundary radii as a function of the boundary data.
In the following, we focus on (near-)extremal black hole solutions.

\subsubsection{Near horizon limit of extremal black holes} 

Geometrically, the extremal black hole occurs when the inner and outer horizon radii coincide, at $r_{\text{bh}}=Q$. The near-horizon geometry of the extremal RN black hole is AdS$_{2}\times S^{2}$.  
Let us consider the case when the finite boundary is located inside the AdS$_{2}$ region.\footnote{There are also additional extremal black hole solution that have boundary data $\tilde{\beta}^{-1}=0$ and $\frac{\r e^{\omega_s}}{Q} = \frac{1+\sqrt{1- \k Q}}{\k Q}$. Note that in this case the boundary remains outside the AdS$_2$ with $0<\k Q<1$.} 

Concretely, we reparametrize the black hole horizon and the boundary radii as
\begin{equation}\label{eqn: near extr limit}
    r_{\text{bh}}= Q+ \epsilon \, , \qquad \r e^{\omega_s} = Q + \epsilon\, \rho \, ,
\end{equation}
for small parameter $\epsilon>0$ with dimensions of length and finite (dimensionless) parameter $\rho\geq1$ characterizing the relative deviation between the finite boundary and horizon.\footnote{By taking instead $\epsilon>0$ and $\rho<-1$, we can consider the inner patch region in the near-extremal and near-horizon limits. For a small $\epsilon$, the inner horizon radius is given by $r_\text{in}=Q-\epsilon$.} Taking $\epsilon \to 0$ with $\rho$ fixed then amounts to simultaneously taking the extremal limit while keeping the boundary inside the AdS$_2$ region.  
In this limit, the boundary data (\ref{eq: bdy data RN}) now leads to
\begin{equation}\label{eqn: extr RN temp}
    \tilde{\beta}=\tilde{\beta}_{\text{ex}} \equiv \frac{2\pi}{\sqrt{\k^2 Q^2 -1}} \, , \qquad \rho = \frac{ \k Q}{\sqrt{\k^2 Q^2-1}}\,,
\end{equation}
where the inverse extremal conformal temperature $\tilde{\beta}_{\text{ex}}$ is a function of $\k Q$.
Note that although the surface gravity of the extremal black hole is zero, the conformal (quasi-local) temperature is not necessarily vanishing. The trace of the extrinsic curvature is restricted to $\k Q>1$.

As $\k  Q\to 1^+$, the conformal temperature goes to zero, and the boundary is pushed to the conformal boundary of AdS$_2$ $\times$ $S^2$, where $\rho\to \infty$. In the thermodynamic phase space at fixed charge, this black hole lies at the point
\beq \text{Zero-temperature, extremal RN:}\;\; (\tilde{\beta}^{-1},\mathcal{K})=(0,Q^{-1})\;,\label{eq:extpoint}\eeq
where the thermodynamic quantities \eqref{eq: near-extremal RN thermo} become,
\begin{equation}
\begin{cases}
    E^{\text{ex}}_{\text{conf}} = \frac{ Q^2}{3G_4} \,, \\
    S^{\text{ex}}_{\text{conf}} = \frac{\pi Q^2}{ G_4} \,, \\
    C^{\text{ex}}_\k =0 \,.
\end{cases}
\label{eq:extconfthermo}\end{equation}
Near-extremal deviations away from this limit are subtle and depend on how we approach the double limit of $\tilde\beta^{-1} \to 0$ and $\k \to Q^{-1}$.  We first study deviations when $\beta_{\text{ex}}$ is finite in Section \ref{subsec: near_finite} and then carefully examine the near zero-temperature behavior in Section \ref{subsec: near_zero}.

\subsubsection{Near-extremal, finite temperature thermodynamics} \label{subsec: near_finite}
It is easier to perturb the extremal solution when the inverse temperature is finite. To obtain thermodynamics of the near-extremal black hole, we just perturb the inverse temperature away from its extremal value, $\tilde{\beta}= \tilde{\beta}_{\text{ex}}+\delta \tilde{\beta}$, keeping only the first $\delta \tilde{\beta}$ correction.  Consequently, the near-horizon geometry is no longer exactly AdS$_2$ $\times$ $S^2$ over the range $r \in \left[r_{\text{bh}} , \r e^{\omega_s}\right]$, where the correction to the AdS$_2$ metric is controlled by $\delta\tilde{\beta}$. For the outer patch, the inverse temperature is lower than the extremal value, $\delta\tilde{\beta}<0$. 
Inverting the boundary data (\ref{eq: bdy data RN}) together with the parametrization (\ref{eqn: near extr limit}) (with $0<\epsilon\ll1$) leads to
\begin{equation}
    \frac{r_{\text{bh}}}{Q} = 1 - \frac{\k^2 Q^2-1}{3 \k Q - 2 \sqrt{\k^2 Q^2-1}}\frac{\delta\tilde{\beta}}{2\pi} \, , \qquad \frac{\r e^{\omega_s}}{Q} = 1 - \frac{\k Q \sqrt{\k^2Q^2-1}}{3 \k Q -2\sqrt{\k^2Q^2-1}}\frac{\delta\tilde{\beta}}{2\pi}\,.
\end{equation}
The thermodynamic quantities at leading order in $\delta\tilde{\beta}$ are\footnote{We note that the same formulae hold for the inner patch thermodynamics. The only difference is that, for the inner patch, the near-extremal inverse temperature is larger than the extremal value, $\delta\tilde{\beta}>0$. This means also that the inner patches in the near-extremal and near-horizon limits are thermally stable when $\k Q>1$.}
\begin{equation}
\begin{cases}
    E_{\text{conf}} = \frac{\k Q^3}{3G_4} - C_\k\frac{\delta\tilde{\beta}}{\tilde{\beta}_{\text{ex}}^2} \,, \\
    \mathcal{S}_{\text{conf}} = \frac{\pi Q^2}{ G_4}- C_\k\frac{\delta\tilde{\beta}}{\tilde{\beta}_{\text{ex}}} \,, \\
    C_\k = \frac{2\pi Q^2 \sqrt{\k^2 Q^2-1}}{\left(3 \k Q - 2 \sqrt{\k^2 Q^2-1}\right)G_4}\;.
    %- \frac{Q^2\left(-11+13 \k^2Q^2-4\k^4Q^4+6 \k Q\left(\k^2 Q^2-1\right)^{3/2}\right)\delta\tilde{\beta}}{\left(3 \k Q - 2 \sqrt{\k^2Q^2-1}\right)^3G_4} \,.
\end{cases} \label{eq: near-extremal RN thermo}
\end{equation}
Similar to the Nariai case, the correction to the entropy is captured by linear deviations away from the extremal temperature, $\delta \tilde{\beta}$. Further taking the limit $\k Q\to1^{+}$ is subtle, as we now describe. 

\subsubsection{Near-extremal, near-zero temperature thermodynamics} \label{subsec: near_zero}

In order to find near-extremal solutions near the zero-temperature black hole, we need to specify how to take the two limits, $\mathcal{K}Q\to1$ and $\tilde{\beta}\to\infty$. It turns out that a meaningful way of doing so is by considering the double-scaling limit,
\begin{equation}
  \{   \mathcal{K}Q\to1, \tilde{\beta}\to\infty \} \quad \text{with} \quad \tilde\beta^2 (\k Q -1) \equiv \lambda \,\, \text{finite} \,. \label{eq: double scaled limit}
\end{equation}

Solving the boundary conditions in this limit gives to leading order,
\begin{equation}\label{eqn: rbh R near extr1}
    \frac{r_{\text{bh}}}{Q} = 1 + 2 \sqrt{2}\pi^2 \tilde{\beta}^{-2}\sqrt{1-\frac{\tilde{\beta}^2\left(\mathcal{K} Q-1\right)}{2\pi^2} }\, , \qquad  \frac{\frakr e^{\omega_s}}{Q} =1+  \sqrt{2}\pi \tilde{\beta}^{-1}\sqrt{1-\frac{\tilde{\beta}^2\left(\mathcal{K} Q-1\right)}{2\pi^2} } \, .
\end{equation}
From here, note that for $\mathcal{K}Q<1$ solutions exist for all inverse temperatures $\tilde{\beta}$. Instead, for $\mathcal{K}Q>1$, real solutions only exist for $\tilde\beta \leq \tilde\beta_{\text{ex}}$, that in this limit becomes $\b_{\text{ex}} = \tfrac{\sqrt{2}\pi}{\sqrt{\k Q -1}} + \mathcal{O}(\sqrt{\k Q -1})$.

Notice also that in this double-scaled limit,
\begin{equation}
    \frac{r_{\text{bh}}-Q}{\frakr e^{\omega_s}-Q} = 2\pi \tilde{\beta}^{-1} \,. 
\end{equation}
Since we are already in the limit of $\tilde{\beta}^{-1}\to0$, this implies that
\begin{equation}
    r_{\text{bh}}-Q \ll \frakr e^{\omega_s}-Q \ll Q \,,
\end{equation}
which means the finite boundary is taken to be near the conformal boundary of AdS$_2$ $\times$ $S^2$.

The thermodynamic quantities in this double-scaled limit are given by
\begin{equation}
\begin{cases}
    E_{\text{conf}} = \frac{\mathcal{K} \mathcal{Q}^3}{3G_4} + \frac{2Q^2\sqrt{2\pi^2-\tilde{\beta}^2\left(\mathcal{K} Q-1\right)}\left(4\pi^2+\tilde{\beta}^2(\mathcal{K} Q-1)\right)}{3\tilde{\beta}^3 G_4} \,, \\
    \mathcal{S}_{\text{conf}} = \frac{\pi Q^2}{G_4}+ \frac{4\pi^2 Q^2\sqrt{2\pi^2-\tilde{\beta}^2(\mathcal{K}Q-1)}}{\tilde{\beta}^2G_4} \,, \\
    C_{\mathcal{K}} = \frac{4\pi^2 Q^2(4\pi^2-\tilde{\beta}^2(\mathcal{K}Q-1))}{G_{4}\tilde{\beta}^2\sqrt{2\pi^2-\tilde{\beta}^2(\mathcal{K} Q-1)}}\,.
\end{cases} \label{eq: near-ex RN thermo b^2k fixed}
\end{equation}
The positive definiteness of the specific heat implies that the system is thermally stable.\footnote{For the inner patch, the specific heat takes the same form but with an overall minus sign, i.e. $C_{\mathcal{K}}^{(\text{inner})}=-C_{\mathcal{K}}$. This means that the inner patches near zero temperature and $\mathcal{K}Q\to1$ are thermally unstable.}  

Markedly different behaviors emerge depending on whether $\mathcal{K}Q>1$ or $\mathcal{K}Q<1$ as we approach the zero temperature extremal black hole (\ref{eq:extpoint}). Let us consider each of these in turn.

\noindent \fbox{$\mathcal{K}Q<1$}

When $\k Q<1$, there are two scaling regimes for the thermodynamic quantities (\ref{eq: near-ex RN thermo b^2k fixed}). The first scaling limit corresponds to having temperature as the lowest scale in the system, which translates into $\tilde{\beta}^{-1}\ll \sqrt{1-\mathcal{K}Q}\ll1$. Note that given that both $\tilde\beta^{-1}$ and $(\k Q-1)$ are small, we are still in the near-extremal region. In this case, the thermodynamic quantities become,
\begin{equation}
\tilde{\beta}^{-1}\ll \sqrt{1-\mathcal{K}Q}\ll1 \longrightarrow
\begin{cases}
    E_{\text{conf}} = \frac{\mathcal{K}Q^3}{3G_4} -\frac{2 (1-\mathcal{K}Q)^{3/2}Q^2}{3G_4} + \frac{2\pi^2 Q^2 \sqrt{1-\mathcal{K}Q}}{\tilde{\beta}^2G_4} \,, \\
    \mathcal{S}_{\text{conf}} = \frac{\pi Q^2}{G_4}  + \frac{4\pi^2Q^2 \sqrt{1-\mathcal{K}Q}}{\tilde{\beta}G_4}\,, \\
    C_{\mathcal{K}} = \frac{4\pi^2Q^2 \sqrt{1-\mathcal{K}Q}}{\tilde{\beta} G_4}\,,
\end{cases} \label{eq: near-ex RN thermo lin-T}
\end{equation}
which exhibits a positive and linear-in-temperature specific heat, reminiscent of the standard result for the thermodynamics of near-extremal black holes.

But given this is a double-scaled limit, we could also take $\sqrt{1-\mathcal{K}Q}\ll\tilde{\beta}^{-1}\ll1$, and still be in the near-extremal limit. In that case, thermodynamic quantities exhibit a different scaling,
   \begin{equation}
   \sqrt{1-\mathcal{K}Q}\ll\tilde{\beta}^{-1}\ll1 \longrightarrow
\begin{cases}
    E_{\text{conf}} = \frac{\mathcal{K}Q^3}{3G_4} + \frac{8\sqrt{2}\pi^3}{3\tilde{\beta}^3 G_4} \,, \\
    \mathcal{S}_{\text{conf}} = \frac{\pi Q^2}{G_4}  + \frac{4\sqrt{2}\pi^3}{\tilde{\beta}^2G_4}\,, \\
    C_{\mathcal{K}} = \frac{8\sqrt{2}\pi^3}{\tilde{\beta}^2G_4}\,.
\end{cases} \label{eq: near-ex RN thermo lin-T 2}
\end{equation}

The leading correction to the entropy away from extremality is quadratic in the conformal temperature. One could think of identifying this behavior with the scaling of the high temperature expansion \eqref{eq: intro_entropy}. However, this is just a coincidence of $d=4$. An analysis in higher dimensions reveals that while the near-extremal behavior scales as $\tilde\beta^{-2}$ in all spacetime dimensions, 
the high-temperature expansion goes like $\tilde\beta^{-(d-2)}$.

\begin{figure}[h!]
        \centering
         %\subfigure[$l = 2$]{
                \includegraphics[width = 0.85 \textwidth]{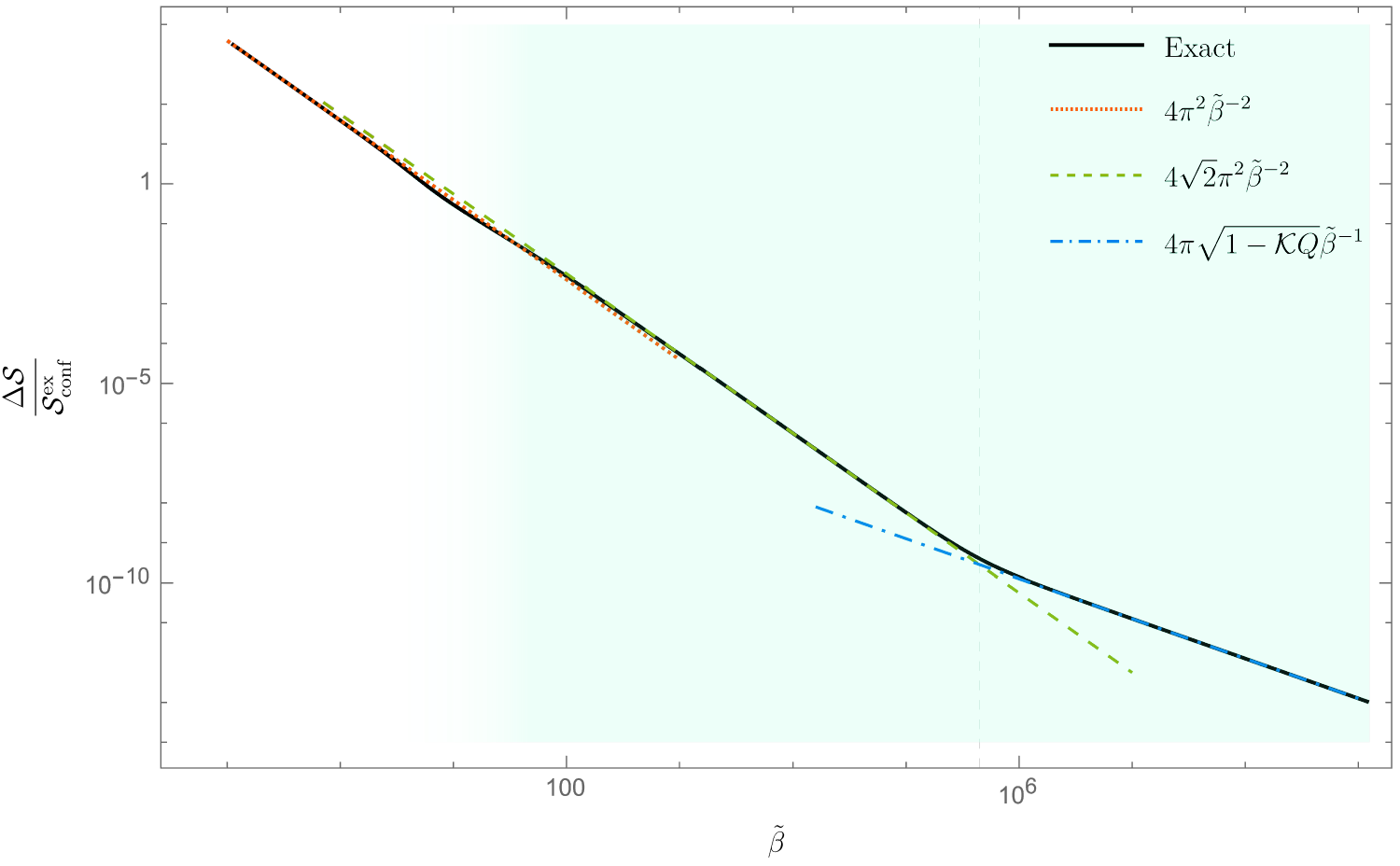}
                \caption{Entropy as a function of the conformal inverse temperature at a fixed value of $\k Q = 1 - 10^{-10}$. The shaded area corresponds to the near-extremal region. The black solid curve is the exact answer. In dotted orange, we show the high-temperature behavior that matches the exact answer in the non-extremal region. The green dashed line shows the first scaling in the near-extremal region, that matches the exact answer up until $\b \sim \sqrt{2}\pi(1-\k Q)^{-1/2}$, which is marked with a vertical grey dashed line. For larger inverse temperatures, the scaling regime is shown in a dotted-dashed blue line, where the entropy is linear in temperature, until reaching the exact extremal value when $\b\to \infty$.} \label{fig: kq<1}
\end{figure}

A nice way of depicting these behaviors is by looking at the quantity $\Delta\mathcal{S} \equiv \mathcal{S}_{\text{conf}}-\mathcal{S}^{\text{ex}}_{\text{conf}}$ as a function of the inverse conformal temperature for a fixed but small value of $\k Q -1$. The exact result together with all three scaling regimes for $\Delta\mathcal{S}$ when $\k Q -1<0$ can be found in Fig. \ref{fig: kq<1}.

\vspace{2cm}

\noindent \fbox{$\mathcal{K}Q>1$}

This case is a bit more intricate because the inverse conformal temperature is bounded, $\tilde\beta \lesssim \tilde\beta_{\text{ex}}$.\footnote{This is true for the outer patches analyzed in this section. We will comment on inner patches in Section \ref{sec:confthermo2D}.} It is then useful to introduce $\delta\tilde{\beta}\equiv \tilde{\beta}-\tilde{\beta}_{\text{ex}}\lesssim0$.

\begin{figure}[h!]
        \centering
         %\subfigure[$l = 2$]{
                \includegraphics[width = 0.85 \textwidth]{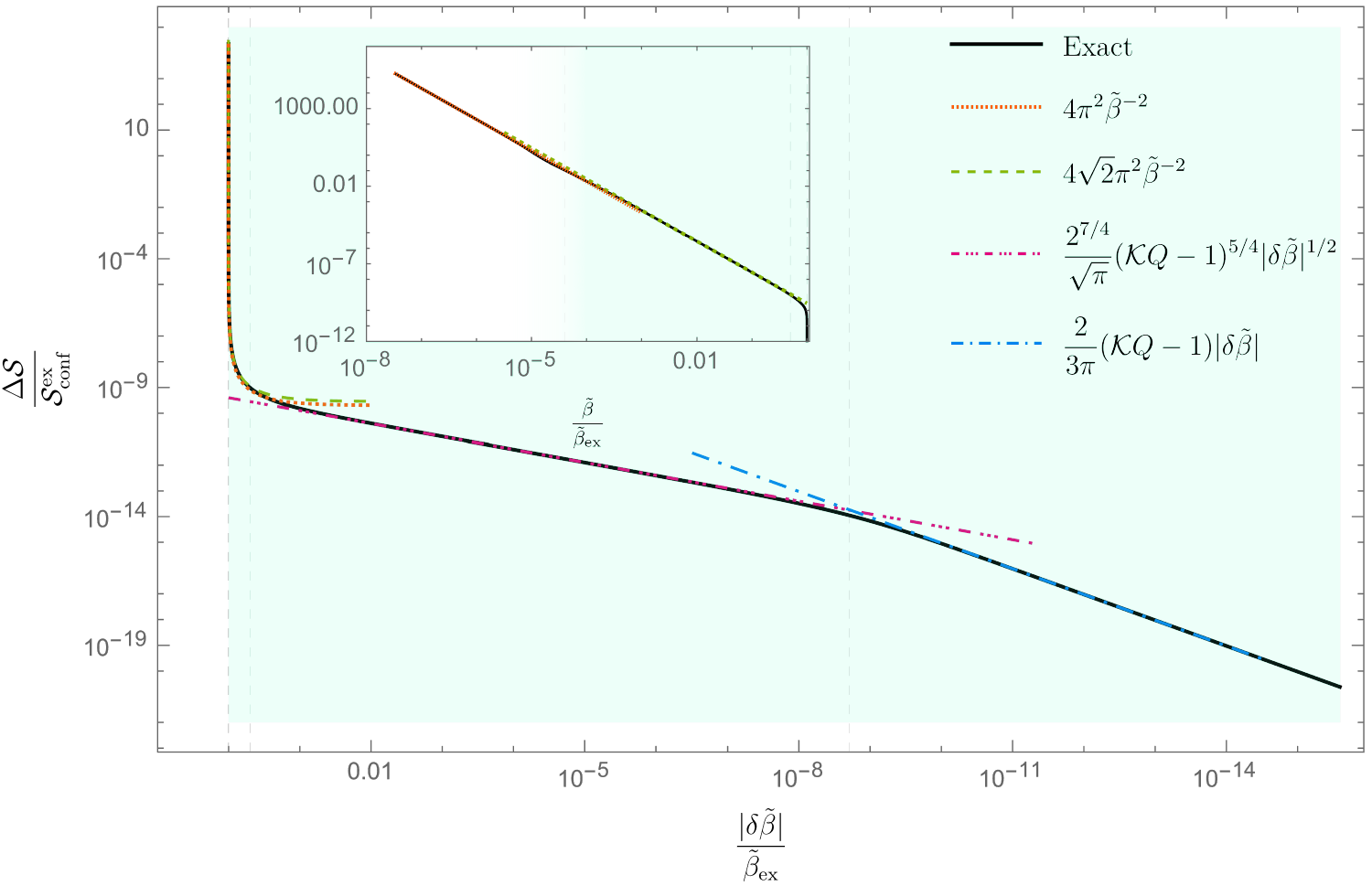}
                \caption{Entropy as a function of $|\delta\b|$ at a fixed value of $\k Q = 1 + 10^{-10}$. No outer patch black hole solutions exist for $\delta\b>0$. The shaded area corresponds to the near-extremal region. The black solid curve is the exact answer. We show the different scaling limits in dotted orange (high-temperature), dashed green, dashed-dotted pink and dashed-dotted light-blue (near-extremal). In the inset, we zoom (and plot as a function of inverse temperature) in the region between extremal and non-extremal solutions to distinguish between the high-temperature $\b^{-(d-2)}$-scaling and the near-extremal $\b^{-2}$-scaling. The extremal limit is reached as $|\delta\b| \to 0$.} \label{fig: kq>1}
\end{figure}

To describe the different scaling limits, it is convenient to start at high temperatures. As discussed, when $\tilde\beta \to 0$, $\mathcal{S}_{\text{conf}} \sim \tilde\beta^{-2}$, but this is far away from the near-extremal limit. We again get into the near-extremal regime by considering the double-scaled limit, \eqref{eq: double scaled limit}. At fixed $\k Q$, the higher temperature regime (already inside the near-extremal limit) is reached by taking $\lambda\to 0$. In terms of $\delta\b$, this roughly corresponds to $(\k Q-1)^{-1/2}\lesssim |\delta\tilde{\beta}|\lesssim \frac{\sqrt{2}\pi}{\sqrt{\k Q-1}}$. Note this regime is not parametrically separated, but there is still a scaling regime in this narrow window. In this case, we obtain
\begin{equation}
\frac{1}{\sqrt{\k Q-1}} \lesssim |\delta\tilde{\beta}|\lesssim \frac{\sqrt{2}\pi}{\sqrt{\k Q-1}} \longrightarrow 
\begin{cases}
    E_{\text{conf}} = \frac{\mathcal{K}Q^3}{3G_4}+ \frac{8\sqrt{2}\pi^3}{3\tilde{\beta}^3 G_4}\,, \\
    \mathcal{S}_{\text{conf}} = \frac{\pi Q^2}{G_4}+ \frac{4\sqrt{2}\pi^3}{\tilde{\beta}^2G_4} \,, \\
    C_{\mathcal{K}} = \frac{8\sqrt{2}\pi^3}{\tilde{\beta}^2G_4} \,,
\end{cases} \label{eq: near-ex RN kq>1 v1}
\end{equation}
matching (\ref{eq: near-ex RN thermo lin-T 2}). In principle, one could imagine taking the other limit of $\lambda\to\infty$, as in the previous case. However, given in this case, the inverse temoperature is bounded, we can at most take $\lambda \to 2\pi^2$. This limit in terms of $\delta\b$ gives $\sqrt{\mathcal{K}Q-1}\ll|\delta\tilde{\beta}|\ll \frac{1}{\sqrt{\k Q-1}}$, where the first limit is obtained by finding when the higher-order terms would become dominant. The thermodynamic quantities in this case scale non-analytically as,
\begin{equation}
\sqrt{\mathcal{K}Q-1}\ll|\delta\tilde{\beta}|\ll \frac{1}{\sqrt{\k Q-1}} \longrightarrow 
\begin{cases}
    E_{\text{conf}} = \frac{\mathcal{K}Q^3}{3G_4} + \frac{2^{5/4}(\mathcal{K}Q-1)^{7/4}Q^2|\delta\tilde{\beta}|^{1/2}} \,, \\
    \mathcal{S}_{\text{conf}} = \frac{\pi Q^2}{G_4}  + \frac{2^{7/4}(\mathcal{K}Q-1)^{5/4}\sqrt{\pi}Q^2|\delta\tilde{\beta}|^{1/2}}{G_4}\,, \\
    C_{\mathcal{K}} =  \frac{2^{5/4}(\mathcal{K}Q-1)^{3/4}Q^2\pi^{3/2}}{|\delta\tilde{\beta}|^{1/2}G_4}\,.
\end{cases} \label{eq: near-ex RN kq>1 v2}
\end{equation} 
Finally, one could consider the case $|\delta\tilde{\beta}|\ll\sqrt{\mathcal{K}Q-1}\ll1$. This limit is slightly away from the double-scaled limit with $\lambda$ finite. To find the scaling behavior of the thermodynamic quantities it is convenient to do an expansion in terms of small $\delta\b$ (instead of just $\b$), while keeping $\tfrac{\d\b}{\sqrt{\k Q-1}} $ finite. In that case, the boundary conditions to leading order give,
\begin{equation}
    \frac{r_{\text{bh}}}{Q} = 1 - \frac{(\k Q-1)\delta\tilde{\beta}}{\sqrt{-\frac{\pi \delta\tilde{\beta}}{2\sqrt{2}\sqrt{\k Q-1}}+\frac{9\pi^2}{4}}+\frac{3\pi}{2}} \, , \qquad \frac{\r e^{\omega_s}}{Q} = 1 - \frac{\sqrt{\k Q-1}\delta\tilde{\beta}}{\sqrt{-\frac{\pi \delta\tilde{\beta}}{\sqrt{2}\sqrt{\k Q-1}}+\frac{9\pi^2}{2}}+\frac{3\pi}{\sqrt{2}}}\,,
\end{equation}
which now can be used to take the desired limit. The thermodynamic quantities then become,
\begin{equation}
|\delta\tilde{\beta}|\ll\sqrt{\mathcal{K}Q-1} \ll1\longrightarrow
\begin{cases}
    E_{\text{conf}} = \frac{\mathcal{K}Q^3}{3G_4} - \frac{\sqrt{2}Q^2 (\k Q-1)^{3/2}\delta\tilde{\beta}}{3 \pi G_4} \,, \\
    \mathcal{S}_{\text{conf}} = \frac{\pi Q^2}{G_4}- \frac{2 Q^2 (\k Q-1)\delta\tilde{\beta}}{3 G_4}\,, \\
    C_{\mathcal{K}} =  \frac{2\sqrt{2}\pi Q^2 \sqrt{\k Q-1}}{3 G_4}\,.
\end{cases} \label{eq: near-ex RN kq>1 v3}
\end{equation}
These quantities coincide with the $\k Q \to 1^+$ limit of \eqref{eq: near-extremal RN thermo}.
The three scalings, together with the high temperature regime (at fixed $\k Q>1$) can be seen in Fig. \ref{fig: kq>1}.

\begin{center}
\pgfornament[height=5pt, color=black]{83}
\end{center}
\vspace{5pt}

To summarize this section, RN black holes exhibit a universal high-temperature regime where the entropy scales quadratic in the conformal temperature, see Eq. \eqref{eq: intro_entropy}. In the near-extremal regime, new scaling limits are usually expected. In the case of Dirichlet thermodynamics, there is only one way of taking the near-extremal limit, that corresponds to taking the temperature to zero. For conformal boundary conditions, we uncover a larger family of near-extremal limits, that depend on the way we simultaneously take the limits of $\b \to \infty$ and $\k Q \to 1$. Furthermore, this will depend on whether the limit is approached from $\k Q>1$ or $\k Q<1$. In short, all the scaling regimes in the different limits can be found in Table \ref{tab:NElimits}.

The thermodynamic behavior at fixed $\k Q <1$ shows features of a CFT$_{(d-1)}$ at high temperatures flowing to a near-CFT$_1$ with the characteristic linear-in-temperature behavior. This is reminiscent of what happens with holographic Fermi liquids, see e.g., \cite{Faulkner:2009wj,Faulkner:2010tq,Nickel:2010pr}, where the infrared behavior of the higher-dimensional theory is dominated by two-dimensional bulk physics. Interestingly, in \cite{Faulkner:2010tq,Nickel:2010pr} this was done by coupling dynamical boundary fields to a strongly coupled field theory. Examples of RG flows across spacetime dimensions in supersymmetric CFTs can be found in \cite{Bobev:2017uzs}. It would be desirable to have a microscopic understanding of these emergent scalings, including the intermediate $\b^{-2}$ regime and the $\k Q>1$ case.

\begin{center}
\begin{tcolorbox}[tab2,tabularx={c||c||c},title= Near-zero temperature extremal limits,boxrule=2pt, width=11cm, code={\setstretch{1.25}}]
   $\boldsymbol{\mathcal{K}Q}$ & \textbf{Scaling of $\Delta\mathcal{S}$} & \textbf{Regime of validity} \\ 
   \hline \hline
     \multirow{2}{*}{ $\mathcal{K}Q<1$} & $\sqrt{1-\k Q}\tilde{\beta}^{-1}$ & $\tilde{\beta}^{-1}\ll\sqrt{1-\k Q} \ll 1$ \\ 
     & $\tilde{\beta}^{-2}$ & $\sqrt{1-\k Q}\ll \tilde{\beta}^{-1}\ll 1$ \\
   \hline \hline
     {} & $(\mathcal{K}Q-1)|\delta\tilde{\beta}|$ & $|\delta\tilde{\beta}|\ll\sqrt{\mathcal{K}Q-1} \ll1$ \\
   $\mathcal{K}Q>1$ & $(\mathcal{K}Q-1)^{5/4} |\delta\tilde{\beta}|^{1/2}$ & $\sqrt{\mathcal{K}Q-1}\ll|\delta\tilde{\beta}|\ll \frac{1}{\sqrt{\k Q-1}}$ \\
   {} & $\tilde{\beta}^{-2}$ & $\frac{1}{\sqrt{\k Q-1}}\lesssim |\delta\tilde{\beta}|\lesssim \frac{\sqrt{2}\pi}{\sqrt{\k Q-1}}$ %\vspace{0.3mm} \\
\end{tcolorbox}
\end{center}
\noindent\begin{minipage}{0.3 \textwidth}
\begin{center}
\captionof{table}{\small Various near-extremal scaling of the conformal entropy as $\tilde{\beta}\to\infty$ and $\k Q \to 1$.}
\label{tab:NElimits}
\end{center}
\end{minipage}

For given $\{\b, \k, Q\}$, there probably also exist non-static, regular, solutions satisfying \eqref{eqn: K non-static d-dim}, that will also contribute to the conformal partition function and might have rich structure in the near-extremal limit, see Section \ref{sec: schwarzian}. In all, the conformal canonical ensemble  provides a plethora of different behaviors at low temperatures that are worth exploring.

\section{Effective 2D description of Einstein-Maxwell with CBCs} \label{sec:spherered}

In this section, we determine a general class of two-dimensional theories of dilaton gravity that effectively describe the spherically symmetric sector of higher-dimensional black holes obeying conformal boundary conditions. This is achieved by performing a spherical dimensional reduction of Einstein-Maxwell-$\Lambda$ with appropriate boundary terms ensuring a well-posed variational problem.

\subsection{Spherical reductions}

\subsection*{Spherical reduction of Einstein-Hilbert} 

Consider vacuum Einstein gravity in a $d$-dimensional spacetime $\mathcal{M}$ with cosmological constant $\Lambda_{d}$ characterized by action (\ref{eq:einhilbv1_Euc}). 
Assuming the $d$-dimensional metric $G_{MN}$ takes the form
\beq ds^{2}_{d}=G_{MN}dX^{M}dX^{N}=\gamma^{2}\Phi^{2\eta}(x)g_{\mu\nu}(x)dx^{\mu}dx^{\nu}+L^{2}_{d}\Phi^{2/(d-2)}(x)d\Omega_{d-2}^{2}\;,\label{eq:dimansatzmain}\eeq
where $M,N=0,1,...,d-1$; $\mu,\nu=0,1$, $\Phi(x)$ is the dimensionless dilaton, $\eta$ and $\gamma$ are dimensionless parameters, and $L_{d}$ is some bulk length scale. The Euclidean action (\ref{eq:einhilbv1_Euc}) then reduces to (see Appendix \ref{app:reductionscoords} for details)
\beq
\begin{split}
 I_{d}&=-\frac{1}{16\pi G_{2}}\int_{\mathfrak{m}}\hspace{-2mm}d^{2}x\sqrt{g}[\Phi R+U(\Phi)]-\frac{1}{8\pi G_{2}}\int_{\partial \mathfrak{m}}\hspace{-3mm}dy\sqrt{h}\left[\Theta_{d}\Phi K+\frac{(\Theta_{d}-1)(d-1)}{2(d-2)}n_{\mu}\nabla^{\mu}\Phi\right]\;,
\end{split}
\label{eq:redactnariaimain}\eeq
where $G_{2}^{-1}\equiv G_{d}^{-1}L_{d}^{(d-2)}\Omega_{d-2}$ is the (dimensionless) induced two-dimensional Newton's constant, $h_{\mu\nu}$ is the one-dimensional induced metric endowed on the boundary $\partial \mathfrak{m}$  of the two-dimensional manifold $\mathfrak{m}$, $n_{\mu}$ is the outward pointing unit normal to $\partial \mathfrak{m}$, and $K$ is the trace of the extrinsic curvature.   The dilaton potential is
\beq U(\Phi)=\gamma^{2}\left(\frac{(d-3)(d-2)}{L_{d}^{2}}\Phi^{-1/(d-2)}-2\Lambda_{d}\Phi^{1/(d-2)}\right)\;,\label{eq:dilapotgenred}\eeq
for $d$-dimensional cosmological constant $\Lambda_{d}=\pm(d-1)(d-2)/2\ell_{d}^{2}$.  Here the real parameter $\eta=-(d-3)/2(d-2)$ is chosen to eliminate possible kinetic terms in the bulk two-dimensional action but does not enter in $U(\Phi)$, and $\gamma$ is some real constant often chosen to attain a desired form of the potential, however, can be absorbed into $g_{\mu\nu}$. 

We highlight that while the presence of $\Theta_{d}$ at the level of the $d$-dimensional action (\ref{eq:einhilbv1_Euc})  is, naively, fairly innocuous, the resulting one-dimensional boundary term in the effective theory (\ref{eq:redactnariaimain}) is non-trivial. In particular, when the $d$-dimensional theory is assumed to obey Dirichlet boundary conditions $(\Theta_{d}=1)$, the one-dimensional boundary term simplifies to the familiar boundary term to JT gravity. Moreover, note the choice for $\Theta_{d}$ does not influence the form of the potential $U(\Phi)$. The potential does depend on the solution to the higher-dimensional theory, and, as we will show, is implicitly influenced by the choice of boundary conditions. From now on, we will focus on conformal boundary conditions, so we fix $\Theta_d = (d-1)^{-1}$.

\vspace{2mm}

\noindent \textbf{Reduced boundary conditions.} As with the action, the boundary conditions imposed on the field variables of the effective theory likewise have a higher-dimensional pedigree. For example, Dirichlet boundary conditions in the bulk, where $H_{MN}$ is held fixed, results in fixing both $h_{\mu\nu}$ and $\Phi$ in the two-dimensional dilaton theory. These are the standard boundary conditions assumed when studying JT gravity.\footnote{See \cite{Goel:2020yxl,Godet:2020xpk} for  JT gravity with alternative boundary conditions, though not from a dimensional reduction.} Alternatively, when the $d$-dimensional theory obeys conformal boundary conditions, where $([H_{MN}],\mathcal{K})$ are fixed, then the two-dimensional fields obey an induced set of boundary conditions, that for fixed $\k$ become (see Appendix \ref{app:reductionscoords} for details)
\beq \gamma\Phi^{-\alpha}\sqrt{h}\biggr|_{\partial \mathfrak{m}}\hspace{-2mm} = L_{d}\;,\quad \Phi^{-\alpha}\left(\Phi K+\alpha n^{\mu}\nabla_{\mu}\Phi\right)\biggr|_{\partial \mathfrak{m}}\hspace{-2mm} =\gamma \, \mathcal{K}\;, \label{eq:confbcsind}\eeq
where we set $\eta=-(d-3)/2(d-2)$ and introduced $\alpha\equiv(\eta+1)=(d-1)/2(d-2)$.\footnote{Without specifying $\eta$, the induced boundary conditions have a more general form, see Appendix \ref{app:reductionscoords}. } Independent of the higher-dimensional origin of the bulk two-dimensional action, well-posedness of the variational problem with boundary conditions (\ref{eq:confbcsind}) insists the bulk plus boundary theory is 
\beq
\begin{split}
 I&=-\frac{1}{16\pi G_{2}}\int_{\mathfrak{m}}\hspace{-2mm}d^{2}x\sqrt{g}[\Phi R+U(\Phi)]-\frac{1}{8\pi G_{2}}\int_{\partial \mathfrak{m}}\hspace{-3mm}dy\sqrt{h}\left[\frac{2\alpha-1}{2\alpha}\Phi K-\frac{1}{2}n_{\mu}\nabla^{\mu}\Phi\right]\;,
\end{split}
\label{eq:eff2Dactgenalpha}\eeq
which coincides with (\ref{eq:redactnariaimain}) for $\Theta_{d}=1/(d-1)$. In what follows, we will consider actions of the type (\ref{eq:eff2Dactgenalpha}) with solutions subject to boundary conditions (\ref{eq:confbcsind}).

\subsection*{Spherical reduction of Maxwell}

It is also worth considering models of dilaton gravity which capture charged near-extremal black holes subject to conformal boundary conditions. 
Restricting the metric ansatz (\ref{eq:dimansatzmain}) to $d=4$, with $\eta=-1/4$, the bulk Maxwell term (\ref{eq:bulkmaxact}) reduces to (assuming solutions with no magnetic charge)
\beq I_{\text{Max}}=\frac{L_{4}^{2}}{64\pi G_{2}\mu_{2}}\int_{\mathfrak{m}}d^{2}x\sqrt{g}\frac{\Phi^{3/2}}{\gamma^{2}}F_{\mu\nu}F^{\mu\nu}\;.\label{eq:redmaxact}\eeq
Here  $\mu_{2}\equiv \mu_{0}L_{4}^{2}$. To match with our conventions for the 4D solution (where $\mu_{0}=1/4$), we see $\mu_{2}$ is quadratic in length.\footnote{In our spherical dimensional reduction, the quantities $\tfrac{F_{2D}^2}{\mu_2}$ and $\tfrac{F_{4D}^2}{\mu_0}$ should have the same dimensions.} Further, $\mathcal{F}_{\mu\nu}\equiv F_{\mu\nu}=\partial_{\mu}A_{\nu}-\partial_{\nu}A_{\mu}$ is the (off-shell) two-dimensional Maxwell field strength.
The boundary term (\ref{eq:IAbdrytermmain}), meanwhile, reduces to (see Appendix \ref{app:reductionscoords} for details) 
\beq I_{A}=-\frac{L_{4}^{2}}{16\pi G_{2}\mu_{2}}\int_{\partial \mathfrak{m}} \hspace{-2mm} dy\sqrt{h}\,\frac{\Phi^{3/2}}{\gamma^{2}} n_{\mu}F^{\mu\nu}A_{\nu}\;.\label{eq:redbdrymax}
\eeq
Upon dimensional reduction, the bulk Dirichlet boundary conditions $\delta \mathcal{A}_{M}|_{\partial \mathcal{M}}=0$ trivially imply the Dirichlet boundary conditions of the two-dimensional gauge field
\beq \delta A_{\mu}|_{\partial \mathfrak{m}}=0\;.\label{eq:Dircgauge2D}\eeq
Meanwhile, the bulk boundary condition for fixed charge, $\delta (\sqrt{H}n_{M}\mathcal{F}^{MN})|_{\partial \mathcal{M}}=0$, yields
\beq \sqrt{h}n_{\mu}F^{\mu\nu}\Phi^{3/2}|_{\partial \mathfrak{m}}=\text{const}\;.\label{eq:fixedcharge2D}\eeq
These two boundary conditions accompany the reduced conformal boundary conditions (\ref{eq:confbcsind}).

\subsection{Equations of motion}

\subsection*{Equations of motion for Einstein-Hilbert}

For the pure two-dimensional dilaton-gravity theory (\ref{eq:redactnariaimain}), the metric and dilaton equations of motion are, respectively, 
\beq 
\begin{cases}
 E_{\mu\nu}\equiv-\frac{2}{\sqrt{g}}\frac{\delta I_{d}}{\delta g^{\mu\nu}}=\frac{1}{8\pi G_{2}}\left(\nabla_{\mu}\nabla_{\nu}\Phi-g_{\mu\nu}\Box\Phi+\frac{1}{2}g_{\mu\nu}U(\Phi)\right)=0\;,\\
 R+\partial_{\Phi}U(\Phi)=0\;.
\end{cases}
\label{eq:puregrav2Deoms}\eeq
The dilaton equation of motion fixes the metric, while the metric equations fix the form of the dilaton. 

\vspace{2mm}

\noindent \textbf{Near-Nariai solutions.} A special two-dimensional theory of dilaton gravity among the class (\ref{eq:redactnariaimain}) is de Sitter JT gravity. Such a theory follows from a spherical reduction of empty dS$_{3}$ (where $L_{d}=\ell_{3}$) or the (near-)Nariai solution in $d\geq4$. The latter follows from setting $L_{d}=r_{\text{N}}=\sqrt{\frac{d-3}{d-1}}\ell_{d}$, the Nariai radius, and choosing $\gamma=1/\sqrt{d-3}$.\footnote{In \cite{Svesko:2022txo}, one instead sets $\gamma=1/\sqrt{d-1}$ such that the coefficient in the dilaton potential is $(d-2)/\ell_{d}^{2}$.} The dilaton potential (\ref{eq:dilapotgenred}) simplifies to 
\beq U(\Phi)=\frac{(d-2)}{r_{\text{N}}^{2}}\left(\Phi^{-1/(d-2)}-\Phi^{1/(d-2)}\right)\label{eq:dilatonpotential2}\;,\eeq
obeying $U(\Phi=1)=0$ and $\partial_{\Phi}U|_{\Phi=1}=-\frac{2}{r_{\text{N}}^{2}}$. Expanding the reduced action (\ref{eq:redactnariaimain})  about $\Phi\approx \Phi_{0}+\phi$ for $\Phi_{0}=1$ and $\Phi_{0}\gg\phi$ gives, at leading order\footnote{Although here $\Phi_{0}=1$, we keep $\Phi_{0}$ to track its role.}
\beq 
\begin{split}
I_{\text{JT}}&=-\frac{1}{16\pi G_{2}}\int_{\mathfrak{m}}\hspace{-2mm}d^{2}x\sqrt{g}\left((\Phi_{0}+\phi)R- \frac{2}{r_{\text{N}}^{2}} \phi\right)-\frac{1}{8\pi G_{2}}\int_{\partial \mathfrak{m}}\hspace{-3mm}dy\sqrt{h}\left[\frac{(\Phi_{0}+\phi)}{(d-1)}K-\frac{1}{2}n_{\mu}\nabla^{\mu}\phi\right]\,.
\end{split}
\label{eq:JTactv2app}\eeq 
Naively, we might expect JT de Sitter to effectively capture small corrections to Nariai thermodynamics in the conformal ensemble, as is the case with Dirichlet boundary conditions. We will show below that this is not in fact the case--- another dilaton theory among the class (\ref{eq:redactnariaimain}) is needed.

\subsection*{Equations of motion for Einstein-Maxwell}

Varying the reduced Maxwell action (\ref{eq:redmaxact}) with respect to $g_{\mu\nu}$ yields
\beq T_{\mu\nu}^{\text{Max}}\equiv -\frac{2}{\sqrt{g}}\frac{\delta I_{\text{Max}}}{\delta g^{\mu\nu}}=\frac{4\ell_{4}^{2}}{64\pi G_{2}\mu_{2}}\left(F^{\alpha}_{\;\;\mu}F_{\alpha\nu}-\frac{1}{4}g_{\mu\nu}F^{2}\right)\frac{\Phi^{3/2}}{\gamma^{2}}\;,\label{eq:maxTmunu}\eeq
such that the metric equations of motion are $E_{\mu\nu}+T_{\mu\nu}^{\text{Max}}=0$. Meanwhile, the dilaton equation of motion (\ref{eq:puregrav2Deoms}) is modified to
\beq \frac{1}{16\pi G_{2}}(R+\partial_{\Phi}U)-\frac{\ell_{4}^{2}}{64\pi G_{2}\mu_{2}}\left(\frac{3}{2\gamma^{2}}\Phi^{1/2}F^{2}\right)=0\;.\label{eq:dileommax}\eeq
Finally, varying with respect to the two-dimensional gauge field $A_{\mu}$ gives
\beq \nabla_{\alpha}(\Phi^{3/2}F^{\alpha\beta})=0\;.\label{eq:Maxeom2d}\eeq
The general solution to the Maxwell equations is
\beq F_{\alpha\beta}=\frac{Q}{L_{4}^{2}}\Phi^{-3/2}\epsilon_{\alpha\beta}\;,\label{eq:Maxtensreed}\eeq
where $\epsilon_{\alpha\beta}$ is the Levi-Civita tensor, and $Q/L_{4}^{2}$ is an integration constant for  electric charge $Q$.\footnote{Recall $Q$ has dimensions of length, such that $F_{\alpha\beta}$ has dimensions of inverse length.} 
Substituting this solution (using $\epsilon_{\alpha\beta}^{2}=-2$), simplifies the metric and dilaton equations of motion. Specifically, the dilaton equation of motion (\ref{eq:dileommax}) becomes
\beq 
\frac{1}{16\pi G_{2}}(R+\partial_{\Phi}U)+\frac{3}{64\pi G_{2}\mu_{2}L_{4}^{2}}\frac{Q^{2}}{\gamma^{2}}\Phi^{-5/2} =0 \,,
\label{eq:dilaeomonshellA}\eeq
which can be written as $\frac{1}{16\pi G_{2}}(R+\partial_{\Phi}\tilde{U}) =0$ with
\beq \tilde{U}(\Phi)\equiv U(\Phi)-\frac{1}{2\mu_{2}L_{4}^{2}}\frac{Q^{2}}{\gamma^{2}}\Phi^{-3/2}\;.\label{eq:effdilapot}\eeq
Meanwhile, the Maxwell stress-tensor (\ref{eq:maxTmunu}) simplifies to 
\beq T_{\mu\nu}^{\text{Max}}=-\frac{1}{32\pi G_{2}\mu_{2}}\frac{Q^{2}}{\gamma^{2}}\frac{g_{\mu\nu}}{L_{4}^{2}}\Phi^{-3/2}\;,\eeq
such that the metric equations of motion $E_{\mu\nu}+T_{\mu\nu}^{\text{Max}}=0$ become
\beq \nabla_{\mu}\nabla_{\nu}\Phi-g_{\mu\nu}\Box\Phi+\frac{1}{2}g_{\mu\nu}\tilde{U}(\Phi)=0\;.\label{eq:meteomonshellA}\eeq
Since the equations of motion (\ref{eq:dilaeomonshellA}) and (\ref{eq:meteomonshellA}) are of the same form as (\ref{eq:puregrav2Deoms}) the general solution for the metric and dilaton will also have the same form.\footnote{Had we integrated out the gauge field $A_{\mu}$, results in an effective dilaton potential which does not coincide with (\ref{eq:effdilapot}).
In particular, the term proportional to the electric charge $Q^{2}$ comes in with the opposite sign, a consequence of exchanging a kinetic energy for a term in the effective dilaton potential. This observation was made in \cite{Brown:2018bms}, leading to a subtle departure from the AdS$_{2}$ JT model, however, can be resolved by rewriting the Maxwell boundary term as a two-dimensional bulk action.}

\vspace{2mm}

\noindent \textbf{Near-extremal black holes.} JT gravity in AdS$_{2}$ arises from, for example, a spherical reduction of near-extremal charged black holes. For concreteness, consider flat Reissner–Nordstr{\"o}m black hole in $d=4$. Setting $\gamma=1$ and $\mu_{2}=L_{4}^{2}/4$, the effective potential (\ref{eq:effdilapot}) is 
\beq \tilde{U}(\Phi)=\frac{2}{L_{4}^{2}}\Phi^{-1/2}\left(1-\frac{Q^{2}}{L_{4}^{2}}\Phi^{-1}\right)\;.\label{eq:dilaeffectpot}\eeq
Expanding $\Phi=\Phi_{0}+\phi$ for $\Phi_{0}=(Q/L_{4})^{2}$ with $\Phi_{0}\gg\phi$, the effective action reduces to AdS$_{2}$ JT gravity, 
\beq I\approx I_{\text{JT}}=-\frac{1}{16\pi G_{2}}\int_{\mathfrak{m}}d^{2}x\sqrt{g}\left((\Phi_{0}+\phi)R+\frac{2}{\ell_{2}^{2}}\phi+...\right)-\frac{1}{8\pi G_{2}}\int_{\partial \mathfrak{m}}\hspace{-3mm}dy\sqrt{h}\left[\frac{(\Phi_{0}+\phi)}{(d-1)}K-\frac{1}{2}n_{\mu}\nabla^{\mu}\phi\right]\;,\eeq
where the ellipsis denotes suppressed higher-order contributions in $\phi$. Here and what follows we will fix $L_{4}=Q$, such that $\Phi_{0}=1$ and the dilaton equation of motion fixes the background to be $\text{AdS}_{2}$ with length scale $\ell_{2}\equiv Q$. This action coincides (after proper identifications) with (E.40) in \cite{Banihashemi:2025qqi}. As in the near-Nariai case, in order to reproduce the thermodynamics of near-extremal black holes we will need to introduce higher order corrections in $\phi$.

\section{Conformal thermodynamics in 2D} \label{sec:confthermo2D}

Here we analyze the horizon thermodynamics of solutions to the effective two-dimensional dilaton gravity theories in the canonical conformal ensemble. We begin with remarks on a general class of two-dimensional dilaton-gravity models before specifying to specific theories which capture aspects of higher-dimensional horizon thermodynamics.

\subsection{General 2D dilaton gravity}

We are interested in studying the two-dimensional dilaton-gravity theory given by (Euclidean) action (\ref{eq:redactnariaimain}), that we recall here,
\begin{equation} \label{eqn: Euclidean action alpha=1}
    I_E = -\frac{1}{16 \pi G_2}\int_{\mathfrak{m}} \hspace{-2mm} d^2 x \sqrt{g} \left[\Phi R + L^{-2}U(\Phi)\right] - \frac{1}{8 \pi G_2}\int_{\partial \mathfrak{m}} \hspace{-2mm} du \sqrt{h} \left[\frac{2\alpha-1}{2\alpha}\Phi K -\frac{1}{2} n_\mu \nabla^\mu \Phi\right] \, ,
\end{equation}
for arbitrary dilaton potential $U(\Phi)$ and parameter $\alpha$. Note that here for convenience we factor out an arbitrary length scale $L$ such that the potential $U(\Phi)$ is dimensionless.  
The boundary term arises from imposing the (reduced) conformal boundary conditions \eqref{eq:confbcsind},
\begin{equation}\label{eqn: bdry cond alpha=1}
    \left.\Phi^{-\alpha}\sqrt{h}\right|_{\partial \mathfrak{m}} = L \, , \qquad \left.\Phi^{-\alpha} \left(\Phi K +\alpha n^\mu \nabla_\mu \Phi \right)\right|_{\partial \mathfrak{m}} = \kappa \, ,
\end{equation}
for arbitrary constants $L$, $\kappa$, and $\alpha$. From a higher dimensional perspective, $L=\gamma^{-1}L_{d}$, $\kappa=\gamma\mathcal{K}$, and $\alpha=(d-1)/2(d-2)$. In this section, however, we will treat them as arbitrary parameters defining the two-dimensional theory. Further, we will take the topology of $\partial \mathfrak{m}$ to be a circle $S^1$, parametrized by Euclidean time coordinate $u$ whose periodicity we will specify momentarily.

\vspace{2mm}

\noindent \textbf{Euclidean solution and boundary conditions.} For a generic dilaton potential $U(\Phi)$,  the theory \eqref{eqn: Euclidean action alpha=1} admits a classical solution
\begin{equation}\label{eqn: sol for generic potential}
    \frac{ds^2}{L^2} = f(r) d\tau^2 + \frac{dr^2}{f(r)} \, , \qquad f(r,\rh) = \frac{1}{\tilde{\Phi}}\int_{\rh}^{r} dr' U(\Phi(r'))>0 \, , \qquad \Phi(r) = \tilde{\Phi} \,  r \, ,
\end{equation}
for dimensionless time and radial coordinates $(\tau,r)$, and where $\tilde{\Phi}$ is a dimensionless positive definite normalization constant.  
The manifold has the topology of a disk, where the origin and boundary are located at $r=\rh$ and $r=\rb$, respectively, with $f(\rh,\rh)=0$. The positivity of $f(r,\rh)$ 
implies $\rb > \rh$ for $U(\tilde{\Phi}\,\rh)>0$, and $\rb < \rh$ for $U(\tilde{\Phi}\,\rh)<0$. For the Euclidean geometry (\ref{eqn: sol for generic potential}) to be smooth near the origin $r=\rh$ requires the Euclidean time coordinate $\tau$ have periodicity
\beq \tau\sim \tau+\beta\;,\quad \beta =\frac{4\pi}{\left|f'(\rh)\right|}=\frac{4\pi \tilde{\Phi}}{|U(\tilde{\Phi} \,\rh)|}\;,\label{eq:tauperiod}\eeq 
independent of the asymptotic form of potential $U$. Here $f'(r)=\partial_{r}f(r,\rh)$.

At the boundary, the unit normal satisfies $n^\mu \partial_\mu \= \sigma L^{-1}\sqrt{f(\rb)}$,  where the $\sigma\equiv\text{sign}[U(\Phi(\rh))]$ factor ensures the normal vector is outward-pointing. Consequently, the induced metric and the trace of the extrinsic curvature of the boundary are
\begin{equation}\label{eqn: bdry data}
    \left.\frac{ds^2}{L^2}\right|_{\partial \mathfrak{m}} = f(\rb)d\tau^2 = \Phi^{2\alpha} du^2 \, , \qquad \left.K\right|_{\partial \mathfrak{m}} = \sigma\frac{f'(\rb)}{2L\sqrt{f(\rb)}} \, .
\end{equation}
Let us now see the implications of the reduced boundary conditions \eqref{eqn: bdry cond alpha=1}. The first condition leads to Euclidean boundary coordinate $u$ having periodicity
\begin{equation}\label{eqn: beta generic potential}
    u \sim u + \tilde{\beta} \, , \qquad \tilde{\beta} \= \tilde{\Phi}^{-\alpha}\frac{4\pi\sqrt{f(\rb)}}{\rb^{\alpha}|f'(\rh)|} \, .
\end{equation}
Meanwhile, the second boundary condition in \eqref{eqn: bdry cond alpha=1}, upon implementing \eqref{eqn: bdry data}, gives 
\begin{equation}\label{eqn: kappa generic potential}
    \kappa L = \sigma\tilde{\Phi}^{1-\alpha} \frac{\rb f'(\rb)+2\alpha f(\rb)}{2\rb^{\alpha}\sqrt{f(\rb)}} \, .
\end{equation}
For a given $f(r)$, we can express $\rb$ and $\rh$ in terms of $\tilde{\beta}$ and $\kappa L$ by inverting \eqref{eqn: beta generic potential} and \eqref{eqn: kappa generic potential}.

\vspace{2mm}

\noindent \textbf{Conformal canonical ensemble.}  The conformal canonical thermal ensemble is characterized by fixing the following thermodynamic data,
\beq \text{conformal canonical ensemble:} \quad \{\tilde{\beta},\kappa\}\;\;\text{fixed}\;.\label{eq:2Dconfensem}\eeq 
According to the Gibbons-Hawking prescription \cite{Gibbons:1976ue}, the leading (classical) contribution to the canonical thermal partition function is
\beq \mathcal{Z}(\tilde{\beta}, \kappa)\approx \sum_{\{g^{\ast}_{\mu\nu},\Phi^{\ast}\}}e^{-I_{\text{E}}[g^{\ast},\Phi^{\ast}]}\;,\eeq
for on-shell Euclidean action $I_{\text{E}}$ of solutions $\{g^{\ast}_{\mu\nu},\Phi^{\ast}\}$ to the Euclidean equations of motion and conformal boundary conditions (\ref{eqn: bdry data}).
Via the partition function, the conformal energy, entropy, and specific heat at fixed $\kappa$ are defined, respectively, 
\begin{equation}\label{eqn: def of thermo quant}
    E_{\text{conf}} \equiv \left. \partial_{\tilde{\beta}}\right|_{\kappa} I^{\text{on-shell}}_E \, , \qquad \mathcal{S}_{\text{conf}} \equiv \left.\left(\tilde{\beta} \partial_{\tilde{\beta}}-1\right)\right|_{\kappa} I^{\text{on-shell}}_E \, , \qquad C_{\kappa} \equiv \left.-\tilde{\beta}^2 \partial_{\tilde{\beta}}^2 \right|_{\kappa}I^{\text{on-shell}}_E \, .
\end{equation}

The task then is to evaluate the on-shell action $I^{\text{on-shell}}_{E}$.  Substituting the solution \eqref{eqn: sol for generic potential} into  the action \eqref{eqn: Euclidean action alpha=1} gives
\beq 
\begin{split}
I_{E}^{\text{on-shell}}=-\frac{\beta\tilde{\Phi}\sigma}{16\pi G_{2}}\int_{\rh}^{\rb}dr\left[-rf''(r)+f'(r)\right]-\frac{\beta\tilde{\Phi}\sigma}{16\pi G_{2}}\left[\frac{2\alpha-1}{2\alpha}\rb f'(\rb)-f(\rb)\right]\;,
\end{split}
\eeq
where we used that on-shell $R=-f''(r)/L^{2}$. The factor of $\sigma$ depends on whether the system of interest has a black hole ($\sigma=+1$) or cosmological horizon ($\sigma=-1$). Note that for solutions like dS$_{2}$, where the full geometry can have more than one horizon, the boundary divides the spacetime such that only a single horizon is contained in the thermal system of interest. Performing integration by parts in the first term and using $f(\rh)=0$, the on-shell action becomes
\begin{equation}\label{eqn: on-shell action generic potential}
    I^{\text{on-shell}}_E = -\frac{\Phi(\rh)}{4 G_2} +\frac{\tilde{\Phi}\left(\rb f'(\rb) - 2\alpha f(\rb)\right)}{8 G_{2} \alpha f'(\rh)} \, ,
\end{equation}
where we used $\beta=4\pi/|f'(\rh)|$ with $|f'(\rh)|=\sigma f'(\rh)$. 

\vspace{2mm}

\noindent \textbf{Canonical thermodynamics.}  Substituting the on-shell action (\ref{eqn: on-shell action generic potential}) into the definition for the energy and entropy \eqref{eqn: def of thermo quant}, we find\footnote{To compute the energy, first replace $f'(\rh)$ appearing in the denominator of on-shell action (\ref{eqn: on-shell action generic potential}) for conformal temperature $\tilde{\beta}$ (\ref{eqn: beta generic potential}) and a function of $\kappa$, such that $I_{E}$ is linear in $\tilde{\beta}$. }
\begin{equation}\label{eqn: energy and entropy generic potential}
    E_{\text{conf}} =  \sigma\tilde{\Phi}^{1+\alpha}\frac{\rb^\alpha \left(\rb f'(\rb)-2\alpha f(\rb)\right)}{32 \pi G_{2}  \alpha\sqrt{f(\rb)}} \, , \qquad \mathcal{S}_{\text{conf}} = \frac{\Phi(\rh)}{4 G_2} \, .
\end{equation}
Notice the quasi-local conformal energy for a system with a black hole horizon differs from the quasi-local energy for a system with the same $f(\rb)$ but with a cosmological horizon by an overall sign, $\sigma$. A similar effect occurs for the case of Dirichlet boundary conditions \cite{Svesko:2022txo,Anninos:2022hqo,Aguilar-Gutierrez:2024nst}. While there both patches have the same Dirichlet boundary data ($\rb$), here the trace of the extrinsic curvature differs by the same sign $\sigma$, so the two patches do not obey the same conformal boundary conditions.
The conformal entropy, meanwhile, is proportional to the dilaton evaluated at the horizon, i.e., the two-dimensional analog of the Bekenstein-Hawking area-entropy formula.

It is straightforward to check that the on-shell action takes the form
\beq I^{\text{on-shell}}_{E}= \b F_{\text{conf}} = -\mathcal{S}_{\text{conf}}+\tilde{\beta}E_{\text{conf}}\;,\eeq
for conformal temperature $\tilde{\beta}$ (\ref{eqn: beta generic potential}) and canonical free energy $F_{\text{conf}}$.
It is also straightforward to verify the quasi-local first law of conformal thermodynamics is 
\beq \delta E_{\text{conf}} = \tilde{\beta}^{-1}\delta \mathcal{S}_{\text{conf}} - \Pi \delta \kappa \;.
\label{eq:quasilocfirstlaw}\end{equation}
 where $\Pi$ is the variable conjugate to $\kappa$
\beq \Pi\equiv -\left(\frac{\partial E_{\text{conf}}}{\partial\kappa}\right)_{\hspace{-1mm} \mathcal{S}_{\text{conf}}}=-\frac{(\Phi(\rb))^{2\alpha}}{16\pi \alpha G_2}L\;.\eeq
We emphasize the sign in front of the entropy variation is positive independent of the type of horizon in question.  
Finally, for a general $f(r)$, the specific heat at fixed $\kappa$ is given by\footnote{To compute $C_\kappa$, we use the chain-rule to rewrite $\partial_{\tilde{\beta}}$ in terms of $\partial_{\rh}$,  $\partial_{\rb}$, $\partial_{\tilde{\beta}}\rh$, and $\partial_{\tilde{\beta}}\rb$. The latter two can be obtained by differentiating \eqref{eqn: beta generic potential} and \eqref{eqn: kappa generic potential} with respect to $\tilde{\beta}$.}
\begin{equation}\label{eqn: specific heat generic potential}
    C_{\kappa} = \left(\frac{ f''(\rh)}{f'(\rh)} + \frac{ \rb f'(\rh)((2\alpha-1)f'(\rb)-\rb f''(\rb))}{4\alpha^2f(\rb)^2+\rb^2 f'(\rb)^2-2\rb f(\rb)(f'(\rb)+\rb f''(\rb))}\right)^{-1}\frac{\tilde{\Phi}}{4G_2} \, .
\end{equation}

Note that the universal behavior of the conformal entropy (as well as other thermodynamic quantities) in the high-temperature limit, see Eqs. (\ref{eq: intro_entropy}) and (\ref{eq: ndof}), can be simply recovered from this two-dimensional perspective. See Appendix \ref{app: highT}.

\vspace{2mm}

\noindent \textbf{Microcanonical thermodynamics.} Thermal ensembles are characterized by stationarity conditions of an appropriate thermodynamic potential. For example, in standard thermodynamics, the Helmholtz free energy $F$ is stationary at fixed temperature, $\delta F|_{\beta}=0$. This stationarity condition follows from varying $F=E-\beta^{-1}S$ and implementing the first law. Similarly, the conformal free energy $F_{\text{conf}}$ is stationary at fixed conformal temperature $\tilde{\beta}^{-1}$ and fixed $\kappa$:
\beq \delta F_{\text{conf}}|_{\tilde{\beta}^{-1},\kappa}=-\mathcal{S}_{\text{conf}}  \, \delta (\tilde{\beta}^{-1})-\Pi\delta\kappa=0\;,\eeq
where the quasi-local first law (\ref{eq:quasilocfirstlaw}) was implemented. 

Different thermal ensembles may be transformed into one another using an appropriate Legendre
transform of a particular thermodynamic potential. Such is the case for the microcanonical ensemble, where, in standard thermodynamics, the thermodynamic potential is the microcanonical entropy. Similarly, the conformal microcanonical entropy is equal to the (negative) Legendre transform of $\tilde{\beta}F_{\text{conf}}$ with respect to $\tilde{\beta}$, i.e., 
\beq \mathcal{S}_{\text{conf}}=-(\tilde{\beta}F_{\text{conf}}-\tilde{\beta}E_{\text{conf}})\;.\eeq
Moreover, stationarity of $F_{\text{conf}}$ implies stationarity of $\mathcal{S}_{\text{conf}}$, 
\beq 0=\delta F_{\text{conf}}|_{\tilde{\beta}^{-1},\kappa}=-\tilde{\beta}^{-1}\delta \mathcal{S}_{\text{conf}}|_{E_{\text{conf}},\kappa}\;.\eeq
This establishes that the conformal microcanonical ensemble is characterized by fixing both the (quasi-local) conformal energy and $\kappa$, with thermodynamic potential $\mathcal{S}_{\text{conf}}(E_{\text{conf}},\kappa)$.  

\subsection{Reduced (A)dS$_3$} 

Thus far we have formally analyzed the conformal thermodynamics for 2D dilaton theories \eqref{eqn: Euclidean action alpha=1} with generic potential $U(\Phi)$, subject to boundary conditions (\ref{eq:confbcsind}) for arbitrary parameter $\alpha$. Let us now specialize to scenarios when the theory of interest arises from a dimensional reduction of (A)dS$_{3}$. In this case the dilaton potential (\ref{eq:dilapotgenred}) is linear in $\Phi$ and the Euclidean action is 
\begin{equation} \label{eqn: AdS3 action}
    I_E = -\frac{1}{16 \pi G_2}\int_{\mathfrak{m}} d^2 x \sqrt{g} \,\Phi\left[ R \pm \frac{2}{L^2}\right] - \frac{1}{8 \pi G_2}\int_{\partial \mathfrak{m}}du \sqrt{h} \left[\frac{1}{2}\Phi K -\frac{1}{2} n_\mu \nabla^\mu \Phi\right] \, ,
\end{equation}
where `$+$' refers to reduction from AdS$_{3}$ and `$-$' from dS$_{3}$; accordingly,  the (A)dS$_{2}$ length scale $L$ is equivalent to the (A)dS$_{3}$ length scale $\ell_{3}$. Moreover, the parameter $\alpha=1$ such that the boundary conditions (\ref{eqn: bdry cond alpha=1}) are
\begin{equation}\label{eqn: AdS3 bdry cond}
    \left.\Phi^{-1}\sqrt{h}\right|_{\partial \mathfrak{m}} = L \, , \qquad \left.\Phi^{-1} \left(\Phi K + n^\mu \nabla_\mu \Phi \right)\right|_{\partial \mathfrak{m}} = \kappa \, ,
\end{equation}
Let us now effectively describe the conformal thermodynamics for dS$_{3}$ and AdS$_{3}$ in turn.

\vspace{2mm}

\noindent \textbf{dS$_{3}$.} The classical solution is a dilaton on dS$_2$,
\begin{equation}\label{eqn: dS3 classical sol}
    \frac{ds^2}{ L^2} = f(r) d\tau^2 + \frac{dr^2}{f(r)} \, , \qquad f(r) = \rh^2 - r^2 \, , \qquad \Phi(r) = \tilde{\Phi} \,r \, .
\end{equation}
From the three-dimensional perspective $\rh$ denotes the dS$_{3}$ cosmological horizon radius, and positivity of $f(r)$ now  imposes that $\rh>\rb$. The inverse conformal temperature and $\kappa$ is
\begin{equation}
    \tilde{\beta} = \frac{2\pi \sqrt{\rh^2-\rb^2}}{\tilde{\Phi}\rb \rh} \, , \qquad \kappa L = \frac{2\rb^2-\rh^2}{\rb\sqrt{\rh^2-\rb^2}} \, .
\label{eq:invconftempdS3}\end{equation}
Note for $\rh>\rb>0$ the boundary data $\kappa$ is unbounded above and below. Taking $\tfrac{\rh}{\rb}\gg1$ leads to the low conformal temperature regime, $\tilde{\beta}\sim \tfrac{2\pi}{\tilde{\Phi}\rb}$, with $\kappa\ell=-\tfrac{\rh}{\rb} + \mathcal{O}\left(\tfrac{\rb^4}{\rh^4}\right)$. Inverting (\ref{eq:invconftempdS3}) yields
\begin{equation}\label{eqn: dS3 sol to bdry conds}
    \tilde{\Phi}\,\rh = \frac{\pi}{\tilde\beta} \left(\sqrt{\kappa^2 L^2+4}-\kappa L\right) \, , \qquad \tilde{\Phi}\,\rb = \frac{\pi}{\tilde\beta} \sqrt{2-\frac{2\kappa L}{\sqrt{\kappa^2 L^2+4}}} \, .
\end{equation}

The on-shell action for the solution \eqref{eqn: dS3 classical sol} is
\begin{equation}
    I^{\text{on-shell}}_E = - \frac{\tilde{\Phi}\,\rh}{8 G_2}=-\frac{\pi\mathfrak{c}_{\text{dS}}}{6\tilde{\beta}} \, , \qquad \mathfrak{c}_{\text{dS}} \equiv \frac{3\sqrt{\rh^2-\rb^2}}{2\rb G_2} = \frac{3\left(\sqrt{\kappa^2 L^2+4}-\kappa L\right)}{4 G_2} \, .
\end{equation} 
The conformal energy, entropy, and free energy at fixed $\kappa$ precisely reproduce the higher-dimensional 
the higher-dimensional physics reviewed in Section \ref{sec: cct} with the same central charge  (\ref{eq:cardyformdS3}), upon recalling the relation $2\pi \ell_{3}/G_{3}=1/G_{2}$ and that $\kappa=\mathcal{K}$.

\vspace{2mm}

\noindent \textbf{AdS$_{3}$.} The classical solution is given by a dilaton $\Phi$ on an AdS$_2$ background,
\begin{equation}\label{eqn: AdS3 classical sol}
    \frac{ds^2}{L^2} = f(r) d\tau^2 + \frac{dr^2}{f(r)} \, , \qquad f(r) = r^2 - r_h^2 \, , \qquad \Phi(r) = \tilde{\Phi} \,r \, ,
\end{equation}
where $r_h$ is the AdS$_{2}$-black hole horizon.
Substituting this $f(r)$ into \eqref{eqn: beta generic potential} and \eqref{eqn: kappa generic potential} gives (with $\sigma=+1$)
\begin{equation}
    \tilde{\beta} = \frac{2\pi \sqrt{\rb^2-\rh^2}}{\tilde{\Phi}\rb \rh} \, , \qquad \kappa L = \frac{2\rb^2-\rh^2}{\rb\sqrt{\rb^2-\rh^2}} \, .
\end{equation}
For $\rb>\rh>0$, the boundary data $\kappa$ is bounded from below by $\kappa=\tfrac{2}{L}$, the value of the trace of the extrinsic curvature at the conformal boundary of AdS$_3$. Inverting the boundary data as in the dS$_{3}$ example, we find the on-shell action  \eqref{eqn: on-shell action generic potential} reads
\begin{equation}
    I^{\text{on-shell}}_E = - \frac{\tilde{\Phi}\,\rh}{8 G_2}\equiv-\frac{\pi\mathfrak{c}_{\text{AdS}}}{6\tilde{\beta}} \, , \qquad \mathfrak{c}_{\text{AdS}} \equiv \frac{3\sqrt{\rb^2-\rh^2}}{2\rb G_2} = \frac{3\left(\kappa L-\sqrt{\kappa^2 L^2-4}\right)}{4 G_2} \, .
\label{eq:centralchargeAdS}\end{equation}
The resulting thermodynamic quantities match those in $d=3$, with $\mathfrak{c}_{\text{AdS}}$ given in \eqref{eq:cardyformAdS3}. 

 \subsection{Reduced near-extremal theories}
 
The spherical reduction from three dimensions is simple, in the sense that $U(\Phi)$ becomes exactly linear, $U(\Phi) = \pm 2 \Phi$. As seen in Section \ref{sec:spherered}, reducing the theory from higher dimensions gives more complicated forms for the dilaton potential. 
To take the near-extremal limit, we expand the dilaton about $\Phi = \Phi_{0} + \phi$, for $\phi \ll \Phi_{0}$. 
Assuming $U(\Phi_{0})=0$, the dilaton potential has the generic expansion
\begin{equation}\label{eqn: potential expand}
   U(\Phi) = 2 \gamma_1 (\Phi-\Phi_0) + 3\gamma_2\frac{(\Phi-\Phi_0)^2}{\Phi_0} +4\gamma_3\frac{(\Phi-\Phi_0)^3}{\Phi_0^2} + ... \,,
\end{equation}
where $\gamma_i$ for $i=1,2,...$ are real constants characterizing the potential. From our metric ansatz (\ref{eq:dimansatzmain}), we can always choose $\gamma$ and $L_{d}$ such that $\Phi_{0}=1$ and $\gamma_{1}=\pm1$, as we will do moving forward. For the examples above, the full potentials for the near-Nariai and near-extremal RN black hole are\footnote{Recall we have pulled out an overall factor of $L=L_{d}$ of the dilaton potential.} 
\begin{eqnarray} 
   U^{\text{Nariai}}(\Phi) = \frac{2}{\sqrt{\Phi}}-2\sqrt{\Phi} \quad , \quad  U^{\text{RN}}(\Phi) = \frac{2}{\sqrt{\Phi}}-\frac{2}{\Phi^{3/2}} \,,
\label{eq:dSEMpotetnails}\end{eqnarray}
where we chose $L_{d}=r_{\text{N}}$ and $\gamma=1/\sqrt{d-3}$ for the near-Nariai solution, and $L_{4}=Q$ and $\gamma=1$ for the near-extremal RN solution. Upon expanding, we find
\begin{eqnarray}
\begin{cases}
        \text{Near-Nariai:}  \quad \gamma_{1}=-1\,,\quad \gamma_2 = \frac{1}{3}\,, \quad\gamma_3 = -\tfrac{3}{16} \,, \\
        \text{Near-extremal RN black hole:} \quad \gamma_{1}=+1\,,\quad  \gamma_2 = -1\, , \quad \gamma_3 = \frac{15}{16} \,.
        \end{cases}
\label{eq:specialvals}\end{eqnarray}

For other black hole solutions, the expressions for $\gamma_{2}$ and $\gamma_{3}$ tend to become cumbersome functions of $L_{d}$ and $\Lambda_{d}$.
For example, near-extremal AdS$_{4}$-RN, has the dilaton potential
\beq U(\Phi)=\gamma^{2}\left(\frac{2}{\sqrt{\Phi}}+\frac{6L_{4}^{2}\sqrt{\Phi}}{\ell_{4}^{2}}\right)-\frac{2Q^{2}}{L_{4}^{2}\gamma^{2}\Phi^{3/2}}\;, \label{eqn: potential AdSRN}\eeq
for AdS$_{4}$ length $\ell_{4}$. Choosing 
\beq 
\begin{split}
&L_{4}=
\frac{\sqrt{2}Q}{\gamma\sqrt{1+\gamma^{2}}}\;,
\quad \gamma=(1-12(Q/\ell_{4})^{2})^{1/4}\;,
\end{split}
\eeq
fixes $\Phi_{0}=1$ and $\gamma_{1}=+1$, and
\beq
\begin{split} 
&\gamma_{2}=-\frac{(2+\gamma^{2})}{3}\;,\quad 
\gamma_{3}=\frac{3}{16}(3+2\gamma^{2})\;.
\end{split}
\eeq
In the flat space limit, $\ell_{4}\to\infty$, we recover $L_{4}=Q$, $\gamma=1$, and (\ref{eq:specialvals}), but generally, $\gamma_{2}\neq-1$.

Another interesting example to consider is the near-extremal charged black holes with positive cosmological constant, which generally have three distinct horizons and, consequently, three extremal limits: cold, Nariai, and ultracold, cf.  \cite{Castro:2022cuo}. The dilaton potential for the effective dilaton-gravity describing dS$_{4}$-RN is easily obtained from the analytic continuation of \eqref{eqn: potential AdSRN}, with $\ell_4 \to i \ell_4$. Notably, in the  ultracold limit (where the inner, outer black hole and cosmological horizons all coincide), the dilaton potential expansion (\ref{eqn: potential expand}) has $\gamma_1=0$. A detailed analysis of the thermodynamics in this case is left for future work.

% \textcolor{red}{We note that the dilaton potential for dS$_4$-RN system can be obtained via analytic continuation, $\ell_4 \to i \ell_4$, of \eqref{eqn: potential AdSRN}. In such case, there exists a limit where all three horizon coincide, known as the ultracold limit, see \cite{Castro:2022cuo} for more details. The dilaton potential expansion of this limit has $\gamma_1=0$. A detailed analysis of this potential lies beyond the scope of the current paper and is left for future work.}

In this section, we will first prove that truncating the dilaton potential to first order, i.e., JT gravity, with conformal boundary conditions gives a trivial theory. Then we will proceed by solving the theory perturbatively for a generic potential of the form \eqref{eqn: potential expand}. In Sections \ref{subsec:redNN}  and \ref{subsec:redNERN}, we will specialize the solution to the particular $\gamma_i$ for near-Nariai and near-extremal black holes, respectively, and analyze the implications for the thermodynamics.

\subsubsection{JT gravity with reduced CBCs has no dynamics}

Let us suggestively rewrite the full two-dimensional effective action (\ref{eqn: Euclidean action alpha=1}) for $d>3$ as
\beq
\begin{split}
 I_{E}&=-\frac{\chi(\mathfrak{m})}{4G_{2}}-\frac{1}{16 \pi G_2}\int_{\mathfrak{m}} \hspace{-1mm} d^2 x \sqrt{g} \left[(\Phi-1) R + L^{-2}U(\Phi)\right]-\frac{1}{8\pi G_{2}}\int_{\partial\mathfrak{m}} \hspace{-3mm}du\sqrt{h}(\Phi-1) K\\
 &+\frac{1}{16\pi G_{2}\alpha}\int_{\partial\mathfrak{m}}\hspace{-2mm}du\sqrt{h}(\Phi K+\alpha n_{\mu}\nabla^{\mu}\Phi)\;,
\end{split}
\label{eq:totalact}\eeq
where we have added and subtracted the two-dimensional topological term proportional to the Euler characteristic of manifold $\mathfrak{m}$,
\begin{equation}\label{eqn: Euler def}
   4\pi\,\chi(\mathfrak{m}) \equiv \int_{\mathfrak{m}} d^2x \sqrt{g} R + 2\int_{\partial \mathfrak{m}}du \sqrt{h} K  \, . 
\end{equation}
For the present analysis we always take $\mathfrak{m}$ to have disk topology such that $\chi(\mathfrak{m})=+1$.

Now directly implement the boundary conditions (\ref{eqn: bdry cond alpha=1}) at the level of the action (\ref{eq:totalact}). Note with a little massaging, it is straightforward to show using (\ref{eqn: bdry cond alpha=1}) that,
\beq 
\begin{split}
& \int_{\partial\mathfrak{m}} du\sqrt{h}K =\int_{\partial\mathfrak{m}} du L\Phi^{2\alpha-1}\left[\kappa-\alpha \Phi^{-\alpha}n_{\mu}\nabla^{\mu}\Phi\right]\;, \\
& \int_{\partial\mathfrak{m}}du\sqrt{h}(\Phi K+\alpha n_{\mu}\nabla^{\mu}\Phi) =\int_{\partial\mathfrak{m}}du(\kappa L)\Phi^{2\alpha}\;.
\end{split}
\eeq
Consequently, the action (\ref{eq:totalact}) takes the form 
\beq 
\begin{split}
  I_{E}&=-\frac{\chi(\mathfrak{m})}{4G_{2}}-\frac{1}{16 \pi G_2}\int_{\mathfrak{m}} \hspace{-1mm} d^2 x \sqrt{g} \left[(\Phi-1) R + L^{-2}U(\Phi)\right]+\frac{1}{16\pi G_{2}\alpha}\int_{\partial\mathfrak{m}}\hspace{-2mm}du(\kappa L)\Phi^{2\alpha}\\
  &-\frac{1}{8\pi G_{2}}\int_{\partial\mathfrak{m}}\hspace{-3mm}du\frac{(\Phi-1)}{\Phi}\left[\Phi^{2\alpha}\kappa L-\alpha \Phi^{\alpha}n_{\mu}\nabla^{\mu}\Phi\right]\;.
\end{split}
\label{eq:fullactwithbcs}\eeq

Next, set $\Phi = 1 + \phi$ for $\phi\ll 1$. Assuming $U(\Phi=1) = 0$, to zeroth order, the action (\ref{eq:fullactwithbcs}) becomes,
\begin{equation}
    I^{(0)}_E = -\frac{1}{4G_{2}} +\frac{1}{16\pi G_{2}\alpha}\int_{\partial\mathfrak{m}}du(\kappa L) \,.
\label{eq:leadingactionJT}\end{equation}
The second integral is easy to evaluate, after which we obtain, off-shell, 
\begin{equation}\label{eqn: IE Phi0}
    I^{(0)}_E = - \frac{1}{4 G_2} + \frac{\tilde{\beta} \kappa L}{16 \pi \alpha G_2}\,.
\end{equation}
Since the second term is linear in $\tilde{\beta}$, it only shifts the energy by purely a $\kappa L$-dependent value. In the case for reduced Nariai (extremal black hole) case, this is the Nariai (extremal) energy.

To capture leading deviations away from extremality, naively we might expect we need only work at linear order in the dilaton expansion (\ref{eqn: potential expand}), which results in the linear-in-$\phi$ action 
\beq 
\begin{split}
 I_{E}&=I_{E}^{(0)}-\frac{1}{16\pi G_{2}}\int_{\mathfrak{m}}d^{2}x\sqrt{g}\phi\left(R+\frac{2\gamma_{1}}{L^{2}}\right)-\frac{1}{8\pi G_{2}}\int_{\partial \mathfrak{m}}du \kappa L\phi+\frac{1}{8\pi G_{2}}\int_{\partial\mathfrak{m}}du\kappa L\phi\;.
\end{split}
\eeq
Notice the last two boundary terms precisely cancel. Thus, we have at linear order in our small $\phi$ expansion,
 \beq I_{E}=- \frac{1}{4 G_2} + \frac{\tilde{\beta} \kappa L}{16 \pi \alpha G_2}-\frac{1}{16\pi G_{2}}\int_{\mathfrak{m}}d^{2}x\sqrt{g}\phi\left(R+\frac{2\gamma_{1}}{L^{2}}\right)\;.\eeq
This is a purely off-shell statement, and merely a consequence of incorporating the conformal boundary conditions into the action. When we work on-shell and implement the dilaton equation of motion $R+\frac{2\gamma_{1}}{L^{2}}=0$, the last term vanishes. 
Thence, at this level, the linear-order JT-bulk contribution will not generate deviations from the extremal thermodynamics (up to a shift in the conformal energy). Note this conclusion does not apply for (A)dS$_{3}$ as there the dilaton potential is exactly linear. Deviations away from extremality will only be captured if we move to higher-orders in our dilaton expansion. Interestingly, a similar phenomenon is observed when computing the two-sphere partition function (with no boundaries) \cite{Ivo:2025yek}. The de Sitter JT gravity partition function is divergent and in order to reproduce the finiteness of the higher-dimensional $S^{2}\times S^{2}$ partition function, it is essential to include the quadratic term in the dilaton potential.

\subsubsection{Generalized near-extremal thermodynamics} \label{sec: generalized 2d thermo}
In order to get the non-trivial thermodynamics encountered in the higher dimensional cases, we need to take into account higher-order terms in the dilaton potential \eqref{eqn: potential expand}.

The classical solution \eqref{eqn: sol for generic potential} for the potential \eqref{eqn: potential expand} is given by
\begin{equation}\label{eqn: sol for potential expand}
    \frac{ds^2}{L^2} = f(r) d\tau^2 + \frac{dr^2}{f(r)} \, , \quad f(r,\rh) = \gamma_1 (r^2-\rh^2) + \gamma_2\tilde{\phi}(r^3-\rh^3)+\gamma_3\tilde{\phi}^2(r^4-\rh^4)+... \, ,
\end{equation}
where $\tilde{\phi}$ is a positive normalization constant characterizing the dilaton, $\phi(r) = \tilde\phi \, r$. For $\tilde{\phi}\ll 1$, the leading term in the metric corresponds to the (A)dS$_2$ geometry and sub-leading terms describe corrections away from it. 

The positivity of $f(r)$ imposes a restriction on the permissible value of $\rb$ and $\rh$ depending on $\gamma_1$. Motivated by the higher-dimensional perspective, for $\gamma_1=+1$, we define the solution with $\rh>0$ and $\rh<0$ as outer patch and inner patch, respectively. The positivity of $f(r)$ then leads to
\beq
\begin{split}
&\textbf{Outer patch:}\;\;0<\rh<r<\rb\;\; \text{with} \;\;\rh>0\;,\\
&\textbf{Inner patch:}\;\;\rb<r<\rh<0\;\; \text{with}\;\; \rh<0\;.
\end{split}
\eeq
It follows that $\rb$ has the same sign as $\rh$, or equivalently, $\tfrac{\rh}{\rb}>0$. For $\gamma_1=-1$, we define the solution with $\rh>0$ and $\rh<0$ as cosmic patch and black hole patch, respectively. The positivity of $f(r)$ leads to
\beq
\begin{split}
&\textbf{Cosmic patch:}\;\;-\rh + \gamma_2\tilde{\phi}\rh^2+\dots<\rb<r<\rh\;\; \text{with} \;\;\rh>0\;,\\
&\textbf{Black hole patch:}\;\;\rh<r<\rb<-\rh + \gamma_2\tilde{\phi}\rh^2+\dots\;\; \text{with}\;\; \rh<0\;.
\end{split}
\eeq
In this case, $\rb$ can either be positive or negative regardless of the sign of $\rh$.

To compute thermodynamic quantities, we expand the exact boundary conditions \eqref{eqn: beta generic potential} and \eqref{eqn: kappa generic potential} to leading order in $\tilde\phi$ to obtain
\begin{equation}\label{eqn: b k gamma1}
    \begin{cases}
        \tilde{\beta} = 2\pi \sqrt{\gamma_1\left(\frac{\rb^2}{\rh^2}-1\right)}\left(1+\tilde{\phi}\left(\left(\frac{\gamma_2}{2\gamma_1}-\alpha\right)\rb - \frac{\gamma_2 }{2\gamma_1}\left(\frac{3 \rb+2\rh}{\rb+\rh}\right)\rh\right)\right)  \, , \\
        \kappa L = \frac{\frac{\rb}{\rh}}{\sqrt{\gamma_1\left(\frac{\rb^2}{\rh^2}-1\right)}}\left(1+\tilde{\phi}\left(\left(1+\frac{\gamma_2}{\gamma_1}\right)\rb - \left(\frac{\gamma_2 \rh}{2 \gamma_1(\rb+\rh)}+\frac{\alpha \rh}{\rb}\right)\rh\right)\right) \, .
    \end{cases}
\end{equation}
Note that the leading terms for both $\tilde{\beta}$ and $\kappa$ only depend on $\gamma_1$. These terms correspond to the proper size and trace of the extrinsic curvature of the boundary in the exact (A)dS$_2$ geometry. In parallel to the higher dimensional case, it is convenient to define
\begin{equation}
   \delta \tilde{\beta}\equiv\tilde{\beta}- \b_{\text{ex}} \quad , \quad  \b_{\text{ex}} \equiv \frac{2\pi}{\sqrt{\kappa^2L^2-\gamma_1}} \,.
\end{equation}
Consistency with $\tilde{\phi}\ll 1$ implies that $|\delta \tilde{\beta}| \ll1$. Upon inverting \eqref{eqn: b k gamma1}, we obtain
\begin{equation}\label{eqn: rb rh gamma1}
    \rh = \frac{\sqrt{\kappa^2 L^2-\gamma_1}}{\kappa L}\rb = \frac{\gamma_1(\kappa^2L^2-\gamma_1)^2\d\b}{2\pi\left(\left(\gamma_1+\gamma_2-2\alpha \gamma_1\right)\kappa^3L^3+2\alpha\kappa L - \gamma_2(\kappa^2L^2-\gamma_1)^{3/2}\right)}\,.
\end{equation}
For finite $(\kappa^2 L^2 - \gamma_1)$, the first equality implies that the boundary is finitely away from the horizon. The positivity of $f(r)$ implies that $\kappa L > 1$ for $\gamma_1=+1$ and $\kappa L \in \mathbb{R}$ for $\gamma_1=-1$. Thermodynamic quantities are given (to leading order) by 
\begin{equation}\label{eqn: thermo gamma1}
\begin{cases}
    E_{\text{conf}}=\frac{\kappa L }{16 \pi \alpha G_2}+  \tilde{\phi}\rh \left(8\pi \sqrt{\gamma_1} \right)^{-1} \left(\frac{\rb^2}{\rh^2}-1\right)^{-1/2} \, , \\
    \mathcal{S}_{\text{conf}}=\frac{1}{4G_2}+  \frac{\tilde{\phi} \rh}{4} \, ,\\
    C_\kappa = -\left((1+\frac{\gamma_2}{\gamma_1})\frac{\rb^3}{\rh^3}-\frac{2\alpha \rb}{\rh}-\frac{\gamma_2}{\gamma_1}\right)^{-1}\frac{1}{4G_2} \, .
    \end{cases}
\end{equation}
We can use \eqref{eqn: rb rh gamma1} to write down the thermodynamic quantities in terms of the boundary data. This result universally describes the near-extremal limit at finite temperatures. We will see below that by just tuning the $\gamma_i$, we will reproduce the thermodynamics of four-dimensional near-Nariai and near-extremal black holes. Given this, a few general comments are in order:
\begin{itemize}
    \item The leading term in $E_{\text{conf}}$ and $\mathcal{S}_{\text{conf}}$ corresponds to the exact extremal limit and it is independent of the inverse conformal temperature. The first correction is proportional to $\delta\b$.
    \item The specific heat to this order only depends on the ratio $\tfrac{\rb}{\rh}$, which means that it can be written purely as a function of $\kappa L$.
    \item A last but important remark is that even to first order, $\rh$ depends explicitly on $\gamma_2$, which means that to correctly capture $E_{\text{conf}}$ and $\mathcal{S}_{\text{conf}}$, it is necessary to keep the first correction to the dilaton potential away from the exact (A)dS$_2$ potential. The same happens for $C_\kappa$, where $\gamma_2$ appears explicitly. Higher order $\gamma_i$'s do not contribute to leading order.
\end{itemize}

This is again a consequence of the fact that JT gravity does not provide any non-trivial behavior away from the extremal limit, so further corrections are needed in the case of reduced conformal boundary conditions. To contrast, in the Dirichlet problem, JT gravity does capture the leading behavior away from extremality, and it is consistent to truncate $f(r)$, as all $\gamma_i$'s with $i\geq2$ will only contribute to subleading corrections to the thermodynamic quantities.

\textbf{Near zero-temperature extremal limit.} So far, we restricted our analysis to cases where $(\kappa^2 L^2 -\gamma_1)$ was finite. When $\gamma_1= +1$ (the near-AdS$_2$ case), it is also possible to have $(\kappa^2 L^2 -\gamma_1) \to 0$. In this case, the boundary is moving close to the conformal boundary of AdS$_2$ and the extremal conformal temperature goes to zero. Note that this starts looking similar to the near-extremal limits of Section \ref{subsec: near_zero}.

From \eqref{eqn: rb rh gamma1}, one can see that this limit amounts to having $\rh^2 \ll \rb^2$, while keeping $\phi\ll 1$. This means that at least to leading order, the higher powers of $\rh$ in the metric will be subleading and we can approximate $f(r)$ near $r\sim \rb$ in \eqref{eqn: sol for potential expand} as,
\begin{equation} \label{eqn: sol for potential expand 2}
    f(r\sim \rb) = r^2-\rh^2 +\gamma_2 \tilde{\phi} r^3 +\gamma_3 \tilde{\phi}^2r^4+ \dots \, ,
\end{equation}
where the ellipsis represents sub-leading corrections (in $\tilde\phi$ and $\rb/\rh$). We are interested in computing the leading corrections to the thermodynamic quantities away from the extremal limit. Assuming $\gamma_2\neq-1$, as we will see, it is self-consistent to truncate the metric expansion before the $\gamma_3$ term. If $\gamma_2=-1$, it is necessary to include the next order, which we discuss in Section \ref{subsec:redNERN}.

To start the analysis, we will assume that both the terms with $\gamma_2$ and $\gamma_3$ contribute to the near-extremal thermodynamics. This amounts to considering $\tfrac{\rh^2}{\rb^2} \sim \tilde{\phi} \ll1$. The boundary data in this case become
\begin{equation}\label{eqn: b k gamma2!=-1 (2)}
    \begin{cases}
        \tilde{\beta} = \frac{2\pi \rb}{\rh} \, , \\
        \kappa L -1= \frac{\rh^2}{2\rb^2}+(1+\gamma_2)\rb \tilde{\phi}\, ,
    \end{cases}
\end{equation}
to leading order. The limits $\tilde{\phi}\ll 1$ and $|\rh|\ll|\rb|$ are translated to $\tilde{\beta}\to0$ and $\kappa L \to 1$. The sign of $(\kappa L-1)$ depends on $\gamma_2$ and can be negative. This is in contrast to the exact AdS$_2$ result where $\kappa L$ is strictly greater (or equal) than $1$. Moreover, notice that one of the corrections away from $\kappa L-1=0$ is proportional to $1+\gamma_2$, so, as mentioned, in the particular case where $\gamma_2=-1$, further analysis is needed. We will analyze that case in the next subsection. Inverting \eqref{eqn: b k gamma2!=-1 (2)}, we find a unique solution of $\rb$ and $\rh$,
\begin{equation}\label{eqn: rb rh gamma2!=-1 (2)}
    \rh = \frac{2\pi \rb}{\tilde{\beta}} = \frac{1}{\tilde{\phi}}\frac{2\pi(\tilde{\beta}^2(\kappa L -1)-2\pi^2)}{(1+\gamma_2)\tilde{\beta}^3}  \, .
\end{equation}
The condition $\tfrac{\rh^2}{\rb^2} \sim \tilde{\phi}$ requires that the combination $\tilde{\beta}^2(\kappa L-1)$ remains finite. 
The leading non-trivial thermodynamic quantities now become
\begin{equation}\label{eqn: thermo gamma2!=1 (2)}
\begin{cases}
    E_{\text{conf}}=\frac{\kappa L }{16 \pi \alpha G_2}+ \frac{\tilde{\phi}\rh^2}{8\pi \rb G_2} + \frac{(1+\gamma_2)\tilde{\phi}^2\rb^2}{16 \pi  G_2}&=\frac{\kappa L }{16 \pi \alpha G_2}+\frac{(\tilde{\beta}^2(\kappa L-1)+6\pi^2)(\tilde{\beta}^2(\kappa L-1)-2\pi^2)}{16\pi (1+\gamma_2)\tilde{\beta}^4 G_2} \, , \\
    \mathcal{S}_{\text{conf}}=\frac{1}{4G_2}+\frac{\tilde{\phi}\rh}{4 G_2}&= \frac{1}{4G_2}+ \frac{\pi(\tilde{\beta}^2(\kappa L-1)-2\pi^2)}{2  (1+\gamma_2)\tilde{\beta}^3 G_2}\, ,\\
    C_\kappa =\frac{\tilde{\phi}\rh}{4G_2}-\frac{\rh^3 }{4(1+\gamma_2) \rb^3G_2}&= \frac{\pi(\tilde{\beta}^2(\kappa L-1)-6\pi^2)}{2(1+\gamma_2)\tilde{\beta}^3 G_2} \, .
    \end{cases}
\end{equation}

Naively, one would imagine it is possible to take only two limiting cases of the above quantities: (i) $\; \tilde{\phi} \ll \tfrac{\rh^2}{\rb^2} \ll 1$, (ii) $\; \tfrac{\rh^2}{\rb^2} \ll \tilde{\phi} \ll 1$. We will analyze those two cases below, and also show that a third scaling regime is available, namely, (iii) $\; \left|\tfrac{\rh^2}{\rb^2} +2 (1+\gamma_2) \rb\tilde{\phi} \right|\ll\tilde{\phi}\ll 1$. 

We remind the reader that in the following, outer patches will be characterized by $\rh>0$, while inner patches will have $\rh<0$.

\noindent \fbox{Case (i) $\; \tilde{\phi} \ll \tfrac{\rh^2}{\rb^2} \ll 1$}

In this case, it is easy to obtain the boundary data by considering the $\tfrac{\rb}{\rh}\to\infty$ limit of \eqref{eqn: b k gamma1},
\begin{equation}\label{eqn: b k gamma2!=-1 (1)}
    \begin{cases}
        \d\b=\tilde{\beta}-\frac{2\pi}{\sqrt{\kappa^2L^2-1}}&= \frac{2\pi(1+\gamma_2) \rb^4\tilde{\phi}}{\rh^3} \, , \\
        \kappa L - 1 &= \frac{\rh^2}{2\rb^2}\,.
    \end{cases}
\end{equation}
Note that in this limit $\b_{\text{ex}} = \tfrac{\sqrt{2}\pi}{\sqrt{\kappa L -1}}$ to leading order and $\delta\b \to 0$. 
Also, $\kappa L - 1 >0$ in this limit. 
Upon inverting these equations, we find
\begin{equation}\label{eqn: rb rh gamma2!=-1 (1)}
    \rh = \sqrt{2(\kappa L -1)}\rb = \frac{1}{\tilde{\phi}}\frac{(\kappa L-1)^2\delta\tilde{\beta}}{8\pi (1+\gamma_2)}  \, .
\end{equation}
The thermodynamic quantities are given by 
\begin{equation}\label{eqn: thermo gamma2!=1 (1)}
\begin{cases}
    E_{\text{conf}}=\frac{\kappa L }{16 \pi \alpha G_2}+ \frac{\tilde{\phi}\rh^2}{8\pi \rb G_2} &=\frac{\kappa L }{16 \pi \alpha G_2}-\frac{(\kappa L-1)^{5/2}\delta \tilde{\beta}}{2\sqrt{2}\pi^2(1+\gamma_2)G_2} \, , \\
    \mathcal{S}_{\text{conf}}=\frac{1}{4G_2}+\frac{\tilde{\phi}\rh}{4 G_2}&= \frac{1}{4G_2}- \frac{(\kappa L-1)^2\delta\tilde{\beta}}{2\pi(1+\gamma_2)G_2}\, ,\\
    C_\kappa=-\frac{\rh^3 }{4(1+\gamma_2) \rb^3G_2} &=-\frac{(\kappa L-1)^{3/2}}{\sqrt{2}(1+\gamma_2)G_2} \, ,
    \end{cases}
\end{equation}
which coincide with the $\kappa L \to 1$ limit of \eqref{eqn: thermo gamma1} or $\d \b \to 0$ limit of \eqref{eqn: thermo gamma2!=1 (2)}. Since $\tfrac{\rb}{\rh}>0$ for both outer and inner patches, the specific heat is positive for the class of dilaton potentials with $\gamma_2<-1$.

\noindent \fbox{Case (ii) $\; \tfrac{\rh^2}{\rb^2} \ll \tilde{\phi} \ll 1$}

In this case, the leading boundary data can be simply obtained from \eqref{eqn: b k gamma2!=-1 (2)}, neglecting the $\tfrac{\rh^2}{2 \rb^2}$ term in the expression for $\kappa L$,
\begin{equation}\label{eqn: b k gamma2!=-1 (3)}
    \begin{cases}
        \tilde{\beta} = \frac{2\pi \rb}{\rh} \, , \\
        \kappa L -1= (1+\gamma_2)\rb \tilde{\phi}\, ,
    \end{cases}
\end{equation}
for which $\kappa L - 1$ in this case can have either sign. Inverting these equations gives
\begin{equation}\label{eqn: rb rh gamma2!=-1 (3)}
    \rh = \frac{2\pi \rb}{\tilde{\beta}} = \frac{1}{\tilde{\phi}}\frac{2\pi (\kappa L-1)}{(1+\gamma_2)\tilde{\beta}}  \, .
\end{equation}
The thermodynamic quantities in this case are given by
\begin{equation}\label{eqn: thermo gamma2!=1 (3)}
\begin{cases}
    E_{\text{conf}}=\frac{\kappa L}{16 \pi \alpha G_2}+ \frac{(1+\gamma_2)\tilde{\phi}^2\rb^2}{16 \pi  G_2}&=\frac{\kappa L }{16 \pi \alpha G_2}+\frac{(\kappa L -1 )^2}{16 \pi (1+\gamma_2) G_2} +\frac{\pi(\kappa L -1 )}{4 (1+\gamma_2)\tilde{\beta}^2 G_2}  \, , \\
    \mathcal{S}_{\text{conf}}=\frac{1}{4G_2}+\frac{\tilde{\phi}\rh}{4 G_2}&= \frac{1}{4G_2}+  \frac{\pi (\kappa L -1)}{2(1+\gamma_2)\tilde{\beta}G_2}\, ,\\
    C_\kappa =\frac{\tilde{\phi}\rh}{4G_2}&= \frac{\pi (\kappa L -1)}{2(1+\gamma_2)\tilde{\beta}G_2}\, ,
    \end{cases}
\end{equation}
which exhibits linear-in-temperature behavior for the entropy and the specific heat.
We note that \eqref{eqn: thermo gamma2!=1 (3)} is simply the $\tilde{\beta}\gg\tfrac{\sqrt{2}\pi}{\sqrt{|\kappa L-1|}}$ limit of \eqref{eqn: thermo gamma2!=1 (2)}. The expression of the specific heat in terms of $\rh$ implies that the system is thermally stable for the outer patch solutions, regardless of $\gamma_2$.

\noindent \fbox{Case (iii) $\; \left|\tfrac{\rh^2}{\rb^2} +2 (1+\gamma_2) \rb\tilde{\phi} \right|\ll\tilde{\phi}\ll 1$}

In this case, we take the limit $\kappa L\to1$ of \eqref{eqn: b k gamma2!=-1 (2)}, resulting in 
\begin{equation}\label{eqn: b k gamma2!=-1 (4)}
    \begin{cases}
        \tilde{\beta} = \frac{2\pi \rb}{\rh} =2\pi\sqrt{-2(1+\gamma_2)\rb \tilde{\phi}}\, , \\
        \kappa L -1=0\, ,
    \end{cases}
\end{equation}
to leading order. Clearly, this limit only exists for $(1+\gamma_2)\rb<0$. The solution to these equations are given by
\begin{equation}\label{eqn: rb rh gamma2!=-1 (4)}
    \rh = \frac{2\pi\rb}{\b} = -\frac{1}{\tilde{\phi}}\frac{4\pi^3}{(1+\gamma_2)\b^3} \, .
\end{equation}
The thermodynamic quantities in this case are given by 
\begin{equation}\label{eqn: thermo gamma2!=1 (4)}
\begin{cases}
    E_{\text{conf}}=\frac{\kappa L}{16 \pi \alpha G_2}-\frac{3 (1+\gamma_2)\tilde{\phi}^2\rb^2}{16 \pi G_2}&=\frac{\kappa L }{16 \pi \alpha G_2}-\frac{3\pi^3}{4 (1+\gamma_2)\b^4 G_2} \, , \\
    \mathcal{S}_{\text{conf}}=\frac{1}{4G_2}+\frac{\tilde{\phi}\rh}{4 G_2}&= \frac{1}{4G_2} - \frac{\pi^3}{(1+\gamma_2)\b^3 G_2}\, ,\\
    C_\kappa =\frac{3\tilde{\phi}\rh}{4G_2}&= -\frac{3\pi^3 }{(1+\gamma_2)\b^3G_2}\, .
    \end{cases}
\end{equation}
The correction to the extremal entropy exhibits cubic-in-temperature behavior. Similarly to the case (ii), the specific heat is always positive for the outer patch solution regardless of $\gamma_2$.

\begin{center}
\pgfornament[height=5pt, color=black]{83}
\end{center}
\vspace{5pt}

Looking at all these three cases slightly away from extremality, we note that (1) they do not depend on $\alpha$; (2) the factor of $(1+\gamma_2)$ only appears as a common denominator; and (3) even if we were to include higher $\gamma_i$ corrections to the metric, they would not appear in the leading deviations away from the extremal thermodynamic quantities. This suggests (provided $\gamma_2\neq-1$) the near-extremal limits with $\tilde\phi\ll 1$ and $\rh \ll \rb$ are universal.

\subsubsection{Reduced near-Nariai} \label{subsec:redNN}

Let us now specialize the previous generic discussion to the particular case of four-dimensional near-Nariai. Recall that the effective 2D action is (with $\alpha=3/4$)
\begin{equation}\label{eqn: action dS4}
    I_E = - \frac{1}{16 \pi G_2}\int_{\mathfrak{m}}d^2x \sqrt{g} \left[\Phi R + L^{-2}U(\Phi)\right] - \frac{1}{8 \pi G_2}\int_{\partial \mathfrak{m}} du \sqrt{h} \left[\frac{1}{3}\Phi K - \frac{1}{2}n^\mu \nabla_\mu \Phi\right] \, ,
\end{equation}
for the dS dilaton potential (\ref{eq:dSEMpotetnails}), with expansion (\ref{eqn: potential expand}) for $\Phi=1+\phi$ and special values (\ref{eq:specialvals}). Here $L=r_{\text{N}}$ (which comes from setting $L_{4}=r_{\text{N}}$ and $\gamma=1$ in the generic potential (\ref{eq:dilatonpotential2})).  The classical solution is (\ref{eqn: sol for potential expand}) for $\gamma_{1}=-1$ and $\gamma_{2}=1/3$. Zeros of the blackening factor $f(r)$ determine the location of the two horizons to be at $r=\rh$ and $r=-\rh + \frac{1}{3}\tilde{\phi}\rh^2$ for any real $\rh$.

Applying the generic formulae for the thermodynamic variables (\ref{eqn: thermo gamma1}), we find
\begin{equation}\label{eqn: thermo near-Nariai}
\begin{cases}
    E_{\text{conf}}=\frac{\kappa L}{12\pi G_2} - \frac{C_\kappa \delta \tilde{\beta}}{\tilde{\beta}_{\text{N}}(\kappa)^2} \, , \\
    \mathcal{S}_{\text{conf}}=\frac{1}{4G_2} - \frac{C_\kappa \delta \tilde{\beta}}{\tilde{\beta}_{\text{N}}(\kappa)} \, ,\\
    C_\kappa = \frac{3\left(1+\kappa^2 L^2\right)^{3/2}}{2\left(9\kappa L+5\kappa^3 L^3-2(1+\kappa^2 L^2)^{3/2}\right)G_2} \,, 
    \end{cases}
\end{equation}
with $\delta\tilde{\beta}\equiv \tilde{\beta}-\tilde{\beta}_{\text{N}}$ for inverse Nariai conformal temperature $\tilde{\beta}_{\text{N}}(\kappa) = \frac{2\pi}{\sqrt{1+\kappa^2 L^2}}$. Upon identifying
\begin{equation}
        \kappa=\mathcal{K} \,, \quad
        L=r_{\text{N}} \,, \quad
        G_{2}^{-1}=4\pi r_{\text{N}}^{2}G_{4}^{-1} \,,
\end{equation}
we precisely recover the conformal thermodynamics of the near-Nariai solution (\ref{eq:nearnariaithermo4D}).\footnote{To attain the $\delta\tilde{\beta}$ correction to the specific heat, we would need to incorporate the $\gamma_{3}$ contribution to the expansion of the dilaton potential, which would in turn modify the classical solution (\ref{eqn: sol for potential expand}).} Note that the conformal entropy is $\mathcal{S}_{\text{conf}} = \tfrac{1}{4G_2} + \tfrac{\tilde{\phi}\,\rh}{4G_2}$, consistent with the Wald entropy, $\mathcal{S}_{\text{Wald}}=\frac{\Phi_{\text{h}}}{4G_{2}}$, for $\Phi_{\text{h}}=1+\phi (\rh)$. 

Finally, given that the geometry (\ref{eqn: sol for potential expand}) is no longer pure dS$_{2}$, the conformal energies of the black hole and the cosmological patches (with opposite $\kappa$) are not equal and opposite. Similarly, the sum of their conformal entropies will not add to the Nariai entropy, which is what happens for instance in JT dS with Dirichlet boundary conditions, see \cite{Anninos:2022hqo}.

\subsubsection{Reduced near-extremal RN} \label{subsec:redNERN}

Lastly, here we recover the conformal thermodynamics of the four-dimensional electrically charged RN black hole in the fixed charge ensemble from an effective two-dimensional description.

The effective two-dimensional Euclidean action has the form (\ref{eqn: action dS4}), where now the potential is that for reduced Einstein-Maxwell (\ref{eq:dSEMpotetnails}), with expansion (\ref{eqn: potential expand}) for $\Phi=1+\phi$. The classical solution is (\ref{eqn: sol for potential expand}). The special values for $\gamma_i$ are given in (\ref{eq:specialvals}), $\gamma_{1}=+1$, $\gamma_{2}=-1$ and $\gamma_{3}=\frac{15}{16}$. Since $\gamma_{2}=-1$, here we must work at cubic order in the expansion of the dilaton potential to uncover the non-trivial thermodynamics of the near-extremal black holes. We will do this for generic $\gamma_3$ first and then specialize to this particular value. 

In what follows we will further assume the combination
\begin{equation}
  \Gamma_3 \equiv 11+4\alpha^2-12\gamma_3 >0\,.
\end{equation}
This is true not only for the $d=4$ RN solution, where $\Gamma_3=2$, but also for general $d$.\footnote{For the RN black hole in  $d$ dimensions, we obtain $\gamma_3=\tfrac{12 d^2-50d+53}{12(d-2)^2}$ and $\alpha=\tfrac{d-1}{2(d-2)}$, so that $\Gamma_3 = \tfrac{4}{d-2}$. See Appendix E in \cite{Banihashemi:2025qqi} for the expression of the dilaton potential in generic spacetime dimensions.} Similarly to what happens in Section \ref{sec: generalized 2d thermo} when $\gamma_2=-1$, the expansions below will break down when $\Gamma_3=0$ and higher orders in the dilaton potential will be needed. We will not consider such potentials here (nor the cases with $\Gamma_3<0$).

\textbf{Near zero-temperature extremal limit with a $\gamma_2=-1$ potential.}
When $\gamma_2=-1$, we include the next-order correction to the metric near $\rb$ (assuming $\rb \gg \rh$). The metric to this order then becomes,
\begin{equation} \label{eqn: sol for potential expand 22}
    f(r \sim \rb) = r^2-\rh^2 - \tilde{\phi} r^3 +\gamma_3 \tilde{\phi}^2r^4+ \dots \, .
\end{equation}
Motivated by the previous analysis, we can consider the cases where each of the subleading terms dominates, defining two possible cases, 
\beq \text{(i)}\; \tfrac{\rh^2}{\rb^2} \sim \tilde{\phi}\;,\quad \text{(ii)}\; \tfrac{\rh^2}{\rb^2} \sim \tilde{\phi}^2\;.\eeq
As before, outer patches will have $\rh>0$, while for inner patches, $\rh<0$. In both cases, $\tfrac{\rb}{\rh}>0$ as a consequence of the positivity of $f(r)$.

\noindent \fbox{ Case (i) $\tfrac{\rh^2}{\rb^2} \sim \tilde{\phi}$} 

For the first case, the boundary data are 
\begin{equation}\label{eqn: b k gamma2=-1}
    \begin{cases}
        \d\b=\tilde{\beta}-\frac{2\pi}{\sqrt{\kappa^2L^2-1}} = -\frac{\pi \tilde{\phi}\rb^4}{4\rh^3}\left(\Gamma_3\tilde{\phi}\rb+\frac{16\alpha\rh^2}{\rb^2}\right)  \, , \\
        \kappa L-1 =\frac{\rh^2}{2\rb^2}\, ,
    \end{cases}
\end{equation}
to leading order. It follows that $\kappa L$ is strictly larger than $1$. The conditions $\tilde{\phi}\ll1$ and $|\rh|\ll|\rb|$ implies that $\d \b \to 0$ and $\kappa L\to 1$. There are two branches of solutions to \eqref{eqn: b k gamma2=-1} which obey $\tfrac{\rb}{\rh}>0$, 
\begin{equation}\label{eqn: rb rh gamma2=-1 (1)}
    \rh= \sqrt{2(\kappa L-1)}\rb = -\frac{1}{\tilde{\phi}}\frac{(\kappa L -1)\delta \tilde{\beta}}{\pi \alpha\pm\sqrt{\pi^2\alpha^2-\frac{\sqrt{2}\pi \Gamma_3}{32}\frac{\delta\tilde{\beta}}{\sqrt{\kappa L-1}}}} \,.
\end{equation}
Consistency with $\tfrac{\rh^2}{\rb^2} \sim \tilde{\phi}$ requires that $\tfrac{\delta\tilde{\beta}}{\sqrt{\kappa L-1}}$ remains finite. The $\Gamma_3$-dependence implies the contribution from the cubic term in the dilaton potential becomes important. The reality condition for $\rh$ imposes a bound $\d \b \leq \d \b_c$, where the critical value is given by
\begin{equation}
    \d \b_c =  \frac{32\pi \alpha^2 \sqrt{\kappa L-1}}{\sqrt{2}\Gamma_3}\, .
\end{equation}
Note this is different from $\b_{\text{ex}}$. When the bound is saturated, the two branches coincide.

\begin{figure}[t!]
        \centering
         %\subfigure[$l = 2$]{
                \includegraphics[width = 0.65\textwidth]{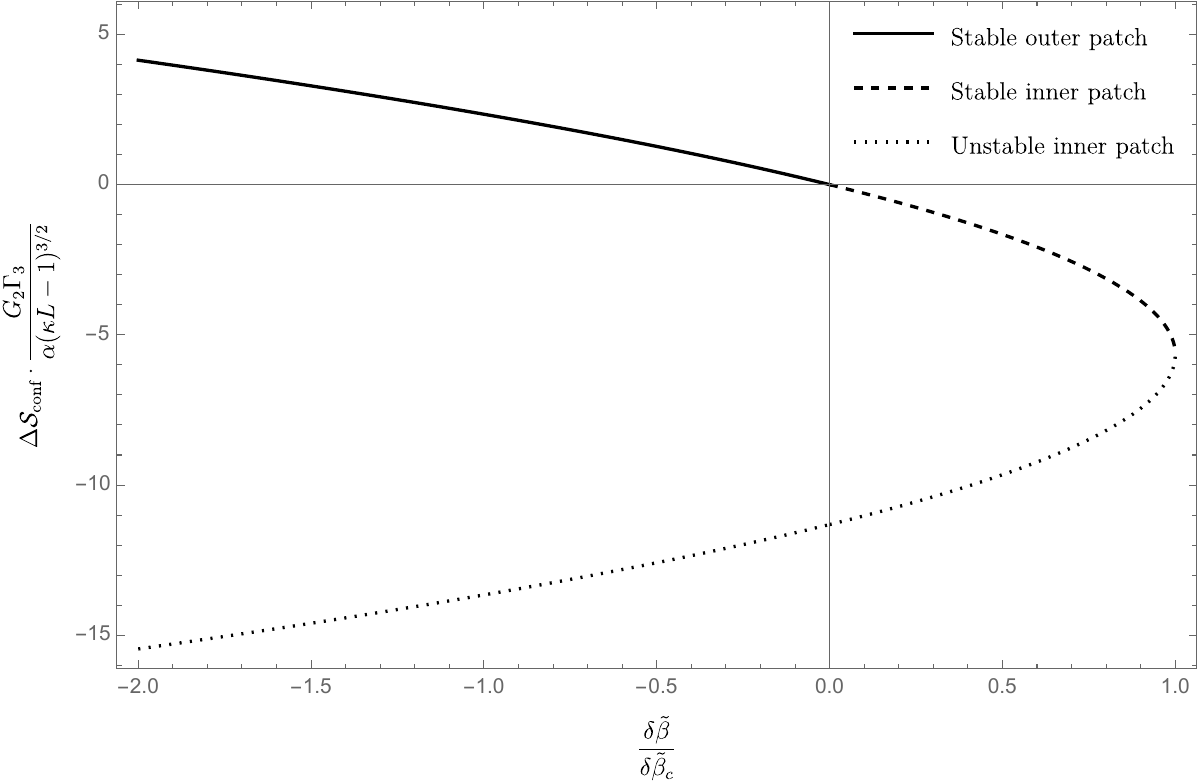}
                \caption{Entropy deviation as a function of $\d \b$ for arbitrary positive parameters $\alpha$ and $\Gamma_3$. The stable outer patch, stable inner patch, and unstable inner patch are marked in solid, dashed, and dotted curves, respectively. Note that outer patches only exist for $\d\b<0$ and no real solutions exist for $\d\b>\d\b_c$.} \label{fig: entropy (i)}
\end{figure}

The thermodynamic quantities in this case are given by
\begin{equation}\label{eqn: thermo gamma2=1 (1)}
\begin{cases}
    E_{\text{conf}}= \frac{\kappa L }{16 \pi \alpha G_2}+\frac{\tilde{\phi}\rh^2 }{8\pi \rb G_2}&= \frac{\kappa L}{16 \pi \alpha G_2}- \frac{\sqrt{2}(\kappa L-1)^{3/2}\d \b}{8\pi^2\alpha \left(1\pm  \sqrt{1-\frac{\d \b}{\d \b_c}}\right)G_2} \, , \\
    \mathcal{S}_{\text{conf}}=\frac{1}{4G_2}+\frac{\tilde{\phi}\rh}{4G_2}&=\frac{1}{4G_2}- \frac{(\kappa L-1)\d \b}{4 \pi \alpha \left(1\pm \sqrt{1-\frac{\d\b}{\d \b_c}}\right)G_2}  \, ,\\
    C_\kappa = \frac{\rh^3}{\Gamma_3 \tilde{\phi}\rb^4+8 \alpha\rb \rh^2 } &
    %=\pm \sqrt{\frac{\pi(\kappa L-1)^{3/2}}{32 \pi \alpha^2\sqrt{\kappa L-1}-\sqrt{2} \Gamma_3\d\b}}\frac{1}{G_2}
    =\pm \frac{(\kappa L-1)^{1/2}}{4 \sqrt{2}\alpha \sqrt{1-\frac{\d\b}{\d\b_c}} G_2}\, .
    \end{cases}
\end{equation}
For a given $\delta\b$, there are two solutions (except at $\delta\b_c$, where there is only one). The specific heat is positive (negative) for the branch of solutions with the upper (lower) sign. When $\d\beta<0$, the branch with positive specific heat corresponds to an outer patch, while for $0<\d\beta<\delta\b_c$, there is an inner patch with positive specific heat. See Fig. \ref{fig: entropy (i)}.

From here we can consider further taking $|\delta \tilde{\beta} |\ll \sqrt{\kappa L-1}\ll1$ of the stable branch, from which we recover \eqref{eqn: thermo gamma1} for $\gamma_2=-1$. In terms of the geometric quantities, this limit corresponds to $\tilde{\phi}^2\ll\tilde{\phi}\ll\tfrac{\rh^2}{\rb^2}\ll1$. 

Alternatively, we can consider $\sqrt{\kappa L-1}\ll|\delta \tilde{\beta} |\ll1 $. The resulting thermodynamic properties are 
\begin{equation}\label{eqn: thermo gamma2=1 (2)}
\begin{cases}
    E_{\text{conf}}=  \frac{\kappa L }{16 \pi \alpha G_2}\pm\frac{(\kappa L-1)^{7/4}\sqrt{|\delta\tilde{\beta}|}}{\pi\sqrt{\sqrt{2}\pi \Gamma_3}G_2}\, , \\
    \mathcal{S}_{\text{conf}}= \frac{1}{4G_2}\pm \frac{\sqrt{2}(\kappa L-1)^{5/4}\sqrt{|\delta \tilde{\beta}|}}{\sqrt{\sqrt{2}\pi \Gamma_3}G_2}  \, ,\\
    C_\kappa = \pm\frac{\pi (\kappa L-1)^{3/4}}{\sqrt{\sqrt{2}\pi  \Gamma_3}\sqrt{|\delta\tilde{\beta}|}G_2} \, .
    \end{cases}
\end{equation}
It can be shown that this limit amounts to $\tilde{\phi}^2\ll\tfrac{\rh^2}{\rb^2}\ll \tilde{\phi}\ll1$.

\vspace{2mm}

\noindent \fbox{ Case (ii) $\tfrac{\rh^2}{\rb^2} \sim \tilde{\phi}^2$} 

The second case is obtained by moving the boundary further away until we have $\tfrac{\rh^2}{\rb^2} \sim \tilde{\phi}^2$. The boundary data are
\begin{equation}
    \begin{cases}
        \tilde{\beta} = \frac{2\pi \rb}{\rh} \, , \\
        \kappa L -1= \frac{\rh^2}{2\rb^2}-\frac{ \Gamma_3 \rb^2 \tilde{\phi}^2}{8}\, .
    \end{cases}
\end{equation}
Note that here $\kappa L -1$ can have either sign. Again, there are two solutions to these equations,
\begin{equation}
    \rh = \frac{2\pi \rb}{\tilde{\beta}} = \pm\frac{4\sqrt{2}\pi\sqrt{2\pi^2-\tilde{\beta}^2(\kappa L-1)}}{\tilde{\beta}^2\sqrt{\Gamma_3}\tilde{\phi}}\,.
\end{equation}
The condition $\tfrac{\rh^2}{\rb}\sim\tilde{\phi}^2$ requires that the combination $\tilde{\beta}^2 (\kappa L-1)$ remains finite. The reality condition for $\rh$ leads to a bound on the boundary data, $\b^2(\kappa L-1)<2\pi^2$. Due to the positive (negative) definiteness of $\rh$, the solution with the upper (lower) sign always corresponds to the outer (inner) patch solution. The thermodynamic quantities are given by
\begin{equation}\label{eqn: thermo gamma2=1 (3)}
\begin{cases}
    E_{\text{conf}}=\frac{\kappa L}{16 \pi \alpha G_2}+ \frac{\tilde{\phi}\rh^2}{8 \pi \rb G_2}- \frac{ \Gamma_3\tilde{\phi}^3\rb^3}{96\pi G_2} &= \frac{\kappa L}{16 \pi \alpha G_2}\pm\frac{\left(4\pi^2 +\tilde{\beta}^2(\kappa L-1)\right)\sqrt{2\pi^2-\tilde{\beta}^2(\kappa L-1)}}{3\sqrt{2}\pi\sqrt{\Gamma_3}\tilde{\beta}^3G_2 }\, , \\
    \mathcal{S}_{\text{conf}}= \frac{1}{4G_2}+ \frac{\tilde{\phi}\rh}{4G_2}  &= \frac{1}{4G_2} \pm \frac{\sqrt{2}\pi\sqrt{2\pi^2-\tilde{\beta}^2(\kappa L-1)}}{\sqrt{\Gamma_3}\tilde{\beta}^2G_2} \, ,\\
    C_\kappa = \frac{\tilde{\phi}\rh}{4 G_2} + \frac{\rh^3}{\Gamma_3\rb^4\tilde{\phi}} &=\pm\frac{\sqrt{2}\pi\left(4\pi^2-\tilde{\beta}^2(\kappa L-1)\right)}{\sqrt{\Gamma_3}\sqrt{2\pi^2-\tilde{\beta}^2(\kappa L-1)}\tilde{\beta}^2 G_2}\, .
    \end{cases}
\end{equation}
The specific heat is positive (negative) for the outer (inner) patch solution. Similarly to the case where $\gamma_2\neq-1$, the dependence on $\gamma_3$ and $\alpha$ only appears as an overall factor in the leading thermodynamic corrections. Note that different scaling behaviors appear depending on whether $\kappa L$ is greater or less than 1.

\noindent \fbox{$\kappa L>1$} 

In this case, $\b$ has an upper bound given by $\tfrac{\sqrt{2}\pi}{\sqrt{\kappa L-1}}$, which coincide with $\b_{\text{ex}}$ in the $\kappa L \to 1$ limit. 

The first limiting case is $\tilde{\phi}^2\ll\tfrac{\rh^2}{\rb^2}\ll\phi \ll 1$. In terms of the boundary data, this amounts to take the inverse conformal temperature near its upper bound, $\d \b =\b - \b_{\text{ex}} \to 0$, while respecting $\sqrt{\kappa L-1}\ll\d\b\ll 1$. Implementing this limit on the thermodynamic quantities \eqref{eqn: thermo gamma2=1 (3)}, we recover \eqref{eqn: thermo gamma2=1 (2)} for both branches of solutions.

Another limit we can take here is $\b^2 (\kappa L-1)\to 0$. This amounts to consider $\sqrt{\kappa L-1}\ll\b^{-1}\ll1$. The resulting thermodynamic quantities are given by
\begin{equation}\label{eqn: thermo gamma2=1 (4)}
\begin{cases}
    E_{\text{conf}}= \frac{\kappa L}{16 \pi \alpha G_2} \pm\frac{4\pi^2}{3\sqrt{\Gamma_3}\b^3 G_2}\, , \\
    \mathcal{S}_{\text{conf}}= \frac{1}{4G_2}\pm\frac{2\pi^2}{\sqrt{\Gamma_3}\b^2 G_2} \, ,\\
    C_\kappa = \pm\frac{4\pi^2}{\sqrt{\Gamma_3}\b^2 G_2}\, .
    \end{cases}
\end{equation}
In this case, the entropy correction scales quadratically in the conformal temperature. 

\noindent \fbox{$\kappa L<1$} 

In this case, there is no upper bound for $\b$, and hence it is possible to take the zero temperature limit, $\b^2 (\kappa L-1)\to-\infty$. The validity of this limit amounts to the condition that $\b^{-1}\ll\sqrt{1-\kappa L}\ll1$, which translate to $\tfrac{\rh^2}{\rb^2}\ll\tilde{\phi}^2\ll\phi \ll 1$ in terms of the two-dimensional variables. By taking this limit, \eqref{eqn: thermo gamma2=1 (3)} becomes
\begin{equation}\label{eqn: thermo gamma2=1 (5)}
\begin{cases}
    E_{\text{conf}}= \frac{\kappa L}{16 \pi \alpha G_2} \pm\sqrt{\frac{(1-\kappa L)^{3}}{\Gamma_3}}\frac{1}{3\sqrt{2}\pi G_2}\pm\sqrt{\frac{1-\kappa L}{\Gamma_3}}\frac{\pi}{\sqrt{2}\tilde{\beta}^2 G_2}\, , \\
    \mathcal{S}_{\text{conf}}= \frac{1}{4G_2}\pm\sqrt{\frac{1-\kappa L}{\Gamma_3}}\frac{\sqrt{2}\pi}{\tilde{\beta}G_2} \, ,\\
    C_\kappa = \pm\sqrt{\frac{1-\kappa L}{\Gamma_3}}\frac{\sqrt{2}\pi }{\tilde{\beta}G_2}\, .
    \end{cases}
\end{equation}
The correction to the conformal entropy is linear-in-temperature. The other limit, $\b^2 (\kappa L-1) \to 0$, gives the same thermodynamics as in the $\kappa L>1$ case, \eqref{eqn: thermo gamma2=1 (4)}.

\textbf{Connecting to the RN near zero-temperature extremal limits.}
It is now straightforward to obtain the thermodynamic behavior near-extremality for the RN electrically charged black hole in four dimensions, from the two-dimensional point of view. Upon identifying the following parameters,
% \begin{equation}
%     \begin{cases}
%         \alpha = \frac{3}{4} \,, \\
%         \gamma_3 = \frac{15}{16} \,,\\
%         L =L_{4}= Q \,, \\
%         \kappa = \k\,, \\
%        % \Phi_0 = Q^2/L_4^2 \,, \\
%         G_{2}^{-1}=4\pi L_{4}^{2}G_{4}^{-1} \,,
%     \end{cases}
% \end{equation}
\begin{equation}
        \alpha = \frac{3}{4} \,, \quad
        \gamma_3 = \frac{15}{16} \,, \quad 
        L =L_{4}= Q \,, \quad 
        \kappa = \k\,, \quad
       % \Phi_0 = Q^2/L_4^2 \,, 
        G_{2}^{-1}=4\pi L_{4}^{2}G_{4}^{-1} \,,
\end{equation}
the scaling limits in Eqs. (\ref{eqn: thermo gamma2=1 (2)}), (\ref{eqn: thermo gamma2=1 (4)}), and (\ref{eqn: thermo gamma2=1 (5)}) reproduce the scaling behavior of the higher-dimensional black hole when $\k Q >1$; see Eqs. \eqref{eq: near-ex RN kq>1 v2}, \eqref{eq: near-ex RN thermo lin-T 2}, and \eqref{eq: near-ex RN thermo lin-T}, respectively. Meanwhile, when $\k Q<1$, the two-dimensional scaling regimes in Eqs. (\ref{eqn: thermo gamma2=1 (4)}) and (\ref{eqn: thermo gamma2=1 (5)}), precisely map to those of the four dimensional analysis, see Eqs. (\ref{eq: near-ex RN thermo lin-T 2}) and (\ref{eq: near-ex RN thermo lin-T}), respectively. For a summary of all the scalings in the near-extremal regime we refer the reader to Table \ref{tab:NElimits}.

\section{Outlook - an effective action for the boundary mode} \label{sec: schwarzian}
Thus far we have focused only on characterizing static solutions. As described in Section \ref{sec:CBCsthermalensem}, however, conformal boundary conditions admit solutions with a non-static conformal factor at the boundary. Restricting to spherically symmetric boundaries, such that the conformal factor only depends on the boundary time, the general equation for this spherically symmetric boundary mode is given by (\ref{eqn: K non-static d-dim}). A similar expression can be found in Lorentzian signature. See also \cite{Strominger:1994xi} for another tractable example of a dynamical boundary in the context of two-dimensional dilaton-gravity.

To fully characterize finite boundaries obeying conformal boundary conditions, it would be desirable to understand universal features of the dynamics for this boundary mode. This would give, among others,  an understanding of the fate of conformal boundaries at late times, and potential microscopic duals, beyond the classical limit. The boundary mode will be relevant for treating perturbative 1-loop quantum effects. Indeed, these effects dominate the thermodynamics of black holes at low-temperatures, such that
the 1-loop quantum effective action acquires temperature dependent logarithmic corrections \cite{Iliesiu:2020qvm,Iliesiu:2022onk}. It would be interesting to study the contributions of the boundary mode to these corrections, including the case of near-extremal black holes in de Sitter \cite{Maulik:2025phe,Blacker:2025zca}.

% arising from zero mode contributions to the Euclidean path integral, cf. \cite{Iliesiu:2020qvm,Iliesiu:2022onk,Maulik:2025phe,Blacker:2025zca}. %From the effective two-dimensional perspective, these zero modes are characterized by the boundary modes. 
% }

Though the boundary dynamics (\ref{eqn: K non-static d-dim}) is amenable to numerical studies, finding analytic solutions becomes an important challenge. A related question is whether there exist effective actions, from which (\ref{eqn: K non-static d-dim}) can be obtained as an equation of motion.

In general, answering both of these questions is hard. There are, however, a collection of cases where this can be achieved  \cite{elsewhere}. In the remainder of this article, we focus on the most tractable example of a spherically symmetric boundary mode, given by the dimensional reduction of AdS$_{3}$, with the finite boundary being close to the AdS$_{3}$ conformal boundary. For the particular case when the boundary is the conformal boundary of AdS$_3$, similar analyses were performed in \cite{Carlip:2005tz,Nguyen:2021pdz}.

\subsection*{Effective description of a dynamical boundary in AdS$_3$}

 Recall the dilaton gravity theory arising from a circular dimensional reduction of AdS$_{3}$ has the action (where we set the AdS$_{3}$ length scale $L=1$)
\beq
    I_E = -\frac{1}{16 \pi G_2}\int_{\mathfrak{m}} \hspace{-2mm} d^2 x \sqrt{g} \left(\Phi R + 2\right) - \frac{1}{8 \pi G_2}\int_{\partial \mathfrak{m}} \hspace{-2mm} du \sqrt{g_{uu}} \left(\frac{1}{2}\Phi K -\frac{1}{2} n_\mu \nabla^\mu \Phi\right) \, ,
\label{eq:actionconf3d}\eeq
 subject to boundary conditions (\ref{eqn: bdry cond alpha=1})
\begin{equation}\label{eqn: bdry cond alpha=12}
    \Phi^{-1}\sqrt{g_{uu}}\;|_{\partial \mathfrak{m}} = 1 \, , \qquad \left(K +\frac{1}{\Phi}n^\mu \nabla_\mu \Phi \right)\biggr|_{\partial \mathfrak{m}} = \kappa \, ,
\end{equation}
for arbitrary (fixed) $\kappa(u)$. Here $u$ denotes a Euclidean boundary time, such that we write $ds^{2}_{\partial M}=g_{uu}du^{2}$ and $\sqrt{h}=\sqrt{g_{uu}}$. The dilaton equation fixes $R=-2$, such that the background is locally AdS$_{2}$, and the geometry describes the hyperbolic disk, where $u$ has periodicity $u\sim u+\beta$.

Non-static solutions can be characterized by introducing a dynamical trajectory near the asymptotic AdS$_{2}$ boundary.  It is convenient to work in Poincar\'e coordinates, where the AdS$_{2}$ line element is 
\beq ds^{2}=\frac{dt^{2}+dz^{2}}{z^{2}}\;,\label{eq:Poinccoord}\eeq
with $t\in[-\infty,\infty]$ and the AdS$_{2}$ boundary is positioned at $z=0$. In these coordinates, the boundary trajectory is parametrized by $(t(u),z(u))$. Our goal is to determine the boundary trajectory. We will do this in two ways. The first approach is to work fully on-shell and impose the boundary conditions (\ref{eqn: bdry cond alpha=12}). The second approach is to work partially on-shell and find the one-dimensional effective action characterizing the dynamics of $t(u)$.

We begin by working totally on-shell. The JT equations of motion fix $R=-2$ while the dilaton has the generic solution 
\beq \Phi=\frac{a+b t+c(t^{2}+z^{2})}{z}\;,\label{eq:Phisolpoincmain}\eeq
for real integration constants $a,b,c$. Next, we introduce a small dimensionless parameter $\epsilon\ll1$ characterizing the proximity of the boundary trajectory to the AdS$_{2}$ boundary, 
such that coordinates $z$ and $t$ have expansions
\beq 
\begin{split}
 &z(u)=z_{0}(u)+\epsilon z_{1}(u)+\epsilon^{2}z_{2}(u)+...\;,\\&t(u)=t_{0}(u)+\epsilon t_{1}(u)+\epsilon^{2}t_{2}(u)+...\;,
\end{split}
\label{eq:pertexpanszt}\eeq
for small $\epsilon$. We now impose the reduced boundary conditions (\ref{eqn: bdry cond alpha=12}). 

First, note that the trace of the extrinsic curvature $K$ for the boundary trajectory is 
\beq K=\frac{t'(t'^{2}+z'^{2}+zz'')-zz't''}{(t'^{2}+z'^{2})^{3/2}}\;.\eeq
Substituting the expansions (\ref{eq:pertexpanszt}) into the second boundary condition (\ref{eqn: bdry cond alpha=12}) then yields
\begin{equation}\label{eq:diffeqkapp}
    2+F(u)\epsilon^2 = \kappa (u) \, , \quad F(u) \equiv \frac{z_1}{t_0'^3}\left(\frac{(b+2c t_0)t_0'^2z_1'}{a+bt_0+c t_0^2}-z_1't_0''+t_0'z_1''\right)-\frac{2c z_1^2}{a+bt_0+c t_0^2}-\frac{z_1'^2}{t_0'^2} \,.
\end{equation}
Note that near the AdS$_{2}$ boundary, $\kappa(u)=2+\epsilon^{2}\kappa_{2}$ since $\kappa$ coincides with the trace of the extrinsic curvature in AdS$_{3}$ ($\kappa=2$) in the limit $\epsilon\to0$. Consistency requires $z_{0}(u)=0$, and one can rearrange $F(u)$ to solve for $t_{0}(u)$ or $z_{1}(u)$ in terms of $\kappa_{2}$. 

Meanwhile, using $z(u)=\epsilon z_{1}(u)$ and the dilaton solution (\ref{eq:Phisolpoincmain}), at leading order in $\epsilon$ the first boundary condition yields the following differential equation 
\beq 0=a+bt_{0}(u)+ct_{0}(u)^{2}-t_{0}'(u)\;.\label{eq:diffeqt0}\eeq
The solution for $t_{0}$ is easily found to be
\beq t_{0}(u)=\frac{-b+\sqrt{-b^{2}+4ac}}{2c}\tan\left[\frac{1}{2}\sqrt{-b^{2}+4ac}(u+c_{1})\right]\;,\eeq
where $c_{1}$ is an integration constant. It is straightforward to verify that the solution $t_{0}(u)$ yields a constant Schwarzian, 
\beq S(t_{0}(u),u)\equiv\left(\frac{t_{0}''}{t_{0}'}\right)'-\frac{1}{2}\frac{t_{0}''^{2}}{t_{0}'^{2}}=2ac-\frac{b^{2}}{2}\;.\eeq
This indicates, near the AdS$_{2}$ boundary, the trajectory $t(u)$ follows from an effective Schwarzian action. We will return to this point shortly. 

Substituting the solution for $t_{0}(u)$ from (\ref{eq:diffeqt0}) back into the condition (\ref{eq:diffeqkapp}) yields a differential equation for $z_{1}(u)$ coupled to $t_0(u)$. Another way of writing this differential equation is in terms of $\left.\Phi\right|_{\partial \mathfrak{m}}$. Denote $\left.\Phi\right|_{\partial \mathfrak{m}}=\Phi_r(u)/\epsilon$. To leading order in $\epsilon$ then
\begin{equation}
    \Phi_r(u) = \frac{a+bt_0(u)+c t_0(u)^2}{z_1(u)} \, .
\end{equation}
Plugging this into $F(u)$ in \eqref{eq:diffeqkapp} for $\kappa(u)=2+\epsilon^{2}\kappa_{2}$ gives the equation of motion for $\Phi_{r}(u)$,
\begin{equation}
    \frac{\Phi_r''}{\Phi_r}-\frac{\Phi_r'^2}{\Phi_r^2}+\kappa_2 \Phi_r^2 = 0 \, .
\label{eq:eomPhir3D}\end{equation}

\noindent \textbf{Effective Schwarzian action.} Let us now recover the above dynamics from an effective action. We work partially on-shell, and first implement the dilaton equations of motion such that $R=-2$, leaving only the boundary term in action (\ref{eq:actionconf3d}). 
Implementing the boundary conditions (\ref{eqn: bdry cond alpha=12}) brings the boundary action to the form
\beq
\begin{split}
I_{\text{bdy}}&=-\frac{1}{16\pi G_{2}}\int_{\partial \mathfrak{m}}du\Phi^{2}\left(2K-\kappa\right)\;.
\end{split}
\label{eq:actionbdry3D}\eeq
Using that $\Phi|_{\partial \mathfrak{m}}=\Phi_{r}(u)/\epsilon$ for small $\epsilon$, then the first boundary condition yields
\beq \label{eqn: z expansion 3d} z=\frac{\epsilon}{\Phi_{r}(u)}\sqrt{t'^{2}+z'^{2}}\approx \frac{\epsilon t'}{\Phi_{r}(u)}+\mathcal{O}(\epsilon^{3})\;.\eeq
Thus, the boundary curve is determined by the ratio of functions $t'/\Phi_{r}$. Substituting this into the extrinsic curvature and expanding to leading order in $\epsilon$ gives
\beq 
\begin{split} \label{eqn: K expansion 3d}
K&=
1+\frac{\epsilon^{2}}{\Phi_{r}^{2}}S(t(u),u)-\frac{\epsilon^{2}}{2\Phi_{r}^{4}}\left(2\Phi_{r}\Phi_{r}''-3(\Phi'_{r})^{2}\right) \,,
\end{split}
\eeq
for Schwarzian $S$. The action (\ref{eq:actionbdry3D}) thus becomes 
\beq \label{eq: brane dynamics}
I_{\text{bdy}}
=-\frac{1}{16\pi G_{2}}\int_{\partial \mathfrak{m}}du\left[2S(t(u),u) -\kappa_{2}\Phi_{r}^{2}-\frac{1}{\Phi_{r}^{2}}(2\Phi_{r}\Phi''_{r}-3(\Phi'_{r})^{2})\right] \,,
\eeq
where we also implemented $\kappa=2+\epsilon^{2}\kappa_{2}$.
Note that, at least to leading order in $\epsilon$, the dynamics of $t(u)$ and $\Phi_r(u)$ are decoupled. In fact, the first term in the action is the standard Schwarzian action for $t(u)$ that appears in the Dirichlet case. 
Varying the action with respect to $\Phi_{r}$ precisely recovers the equation of motion (\ref{eq:eomPhir3D}). 

To connect back to the AdS$_{3}$ gravity theory, it is convenient to reintroduce $\Phi_{r}(u)\equiv e^{\bomega(u)}$, such that the action becomes
\beq 
\begin{split}
I_{\text{bdy}}
=-\frac{1}{16\pi G_{2}}\int_{\partial \mathfrak{m}}du\left[\bomega'(u)^{2}-2\bomega''(u)-\kappa_{2}e^{2\bomega(u)}+2S(t(u),u)\right]\;,
\end{split}\label{eqn: action 3d 0th}
\eeq
and the equation of motion (\ref{eq:eomPhir3D}) becomes
\beq \bomega''(u)+\kappa_{2}e^{2\bomega (u)}=0\;.\eeq
This is recognized to be the one-dimensional Liouville equation, precisely coinciding with equation of motion for non-static conformal factor (\ref{eqn: 0th non-static eqn}). 
When $\kappa_{2}$ is fixed to be a constant, the generic solution for $\bomega$ is 
\beq \bomega(u)=\frac{1}{2}\log\left[\frac{c_{1}}{\kappa_{2}}\text{Sech}^{2}(\sqrt{c_{1}(u+c_{2})^{2}})\right]\;,\eeq
for integration constants $c_{1}$ and $c_{2}$. By a translation in $u$ we can always fix $c_{2}=0$. As in Section \ref{sec:CBCsthermalensem}, we do not expect $c_1$ to spoil the uniqueness of conformal boundary conditions.

It is worth rewriting the effective action \eqref{eq: brane dynamics} in a third suggestive way. As in \cite{Bagrets:2016cdf}, let us define $\Gamma (u) \equiv \int^u \Phi_r(\tilde{u}) d\tilde u$. Then the effective action becomes, 
\beq 
I_{\text{bdy}} =-\frac{1}{8\pi G_{2}}\int_{\partial \mathfrak{m}}du\left[S(t(u),u) - \frac{\kappa_2}{2} \Gamma'(u)^{2} - S\left(\Gamma(u),u\right) \right]\;.
\eeq
Note that $\Gamma(u)$ has non-trivial dynamics even in the case of vanishing $\kappa_2$. In that case, this action reduces exactly to the effective action described in \cite{Anninos:2018svg}, see Eq. (4.14). 
The relative sign between Schwarzian actions appears when coupling a Schwarzian theory to one-dimensional quantum gravity, in which the size of the thermal circle is allowed to fluctuate \cite{Anninos:2021ydw}. This resembles the physical problem studied in this section. 

Turning on $\kappa_2>0$, adds a kinetic term for $\Gamma(u)$, that has a definite sign. This does not break the $SL(2,\mathbb{R})$ symmetry of the problem. Similar actions were discussed in e.g., \cite{Anninos:2018svg,Nayak:2019evx, Jensen:2019cmr}. Given the general boundary dynamics \eqref{eqn: K non-static d-dim} is greatly simplified when the bulk spacetime is AdS$_3$ and the finite boundary is close to the AdS$_{3}$ boundary, this case seems to be the most manageable example to go beyond the classical limit. In this sense, it would be desirable to obtain microscopic theories which exhibit effective actions with such a relative sign, either in the context of deformations to SYK quantum mechanics or in a matrix model picture.

Finally, we do not expect $t(u)$ and $\Gamma(u)$ to be decoupled to next order in $\epsilon$. In fact, the effective action at this order should reproduce \eqref{eq: brane dynamics second order} as an equation of motion. Understanding the structure of such effective actions is an important challenge to discover the potential microscopic interpretation of conformal boundaries, including quantum corrections away from the classical solutions explored. We look forward to address this intriguing question on the near horizon. 

\noindent\section*{Acknowledgments}

We are grateful to Batoul Banihashemi, Christopher Herzog, Diego Hofman, Gloria Odak, Edgar Shaghoulian, Eva Silverstein, David Vegh, and especially Dionysios Anninos, for useful discussions. The work of DAG is funded by UKRI Stephen Hawking Fellowship EP/W005530/1 ``Quantum Emergence of an Expanding Universe". CM is funded
by STFC under grant number ST/X508470/1. AS is partially funded by the Royal
Society under the grant ``Concrete Calculables in Quantum de Sitter''. DAG and AS are further supported by STFC consolidated grant ST/X000753/1. DAG and CM thank Yukawa Institute for Theoretical Physics at Kyoto University, where part of this work was completed during YITP workshop: Quantum Gravity and Information in Expanding Universe (YITP-W-24-19).

\appendix

\section{Variational problem and generalized boundary terms} \label{app:varprobCBC}

Consider $d$-dimensional general relativity, characterized by the Lorentzian action, 
\beq I_{\text{EH}}=\frac{1}{16\pi G}\int_{\mathcal{V}}d^{d}x\sqrt{-g}(R-2\Lambda)\;,\label{eq:EHact}\eeq
where $R$ is the Ricci scalar of a $d$-dimensional spacetime region $\mathcal{V}$, and $\Lambda$ is the cosmological constant. Varying the action (\ref{eq:EHact}) under an arbitrary off-shell metric variation
gives 
\beq
\begin{split}
\hspace{-3mm} 16\pi G\delta I_{\text{EH}}&=\int_{\mathcal{V}}\hspace{-1mm} d^{d}x\sqrt{-g}(G_{\alpha\beta}+\Lambda g_{\alpha\beta})\delta g^{\alpha\beta}+\oint_{\partial\mathcal{V}}\hspace{-2mm} d^{d-1}y\sqrt{|h|}\epsilon n_{\rho}(g^{\sigma\nu}\delta\Gamma^{\rho}_{\;\nu\sigma}-g^{\sigma\rho}\delta\Gamma^{\nu}_{\;\nu\sigma})\;,
\end{split}
\label{eq:bdrytermvarEH}\eeq
where $n_{\mu}$ is the unit normal to the boundary $\partial\mathcal{V}$ of region $\mathcal{V}$ with $\epsilon=n^{2}=\pm1$, where $\epsilon=+1$ or $\epsilon=-1$ when $\partial\mathcal{V}$ is timelike or spacelike, respectively. Further, $h_{ab}$ denotes the induced metric on $\partial\mathcal{V}$ with coordinates $y^{a}$. On-shell, $G_{\alpha\beta}+\Lambda g_{\alpha\beta}=0$, such that the total variation results in a boundary term (\ref{eq:bdrytermvarEH}). After some further massaging, the on-shell variation of the Einstein-Hilbert action is 
\beq
\begin{split}
\hspace{-3mm} 16\pi G\delta I_{\text{EH}}&=\oint_{\partial\mathcal{V}}d^{d-1}y\sqrt{|h|}\epsilon (K_{\mu\nu}-h_{\mu\nu}K)\delta h^{\mu\nu}-2\delta\left(\oint_{\partial\mathcal{V}}d^{d-1}y\sqrt{|h|}\epsilon K\right)\;.
\end{split}
\label{eq:bdrytermGRuse}\eeq
Here, $K_{\mu\nu}=\nabla_{\mu}n_{\nu}-\epsilon a_{\mu}n_{\nu}$ is the extrinsic curvature of the hypersurface $\partial\mathcal{V}$ with `acceleration' $a_{\mu}\equiv n^{\rho}\nabla_{\rho}n_{\mu}$, and $K=g^{\mu\nu}\nabla_{\mu}n_{\nu}$ is the trace of the extrinsic curvature.\footnote{The on-shell variation also includes $\oint_{\partial\mathcal{V}}d^{d-1}y\sqrt{|h|}\epsilon D_{\mu}\delta w^{\mu}$ where $\delta w^{\mu}\equiv \delta n^{\mu}+g^{\mu\nu}\delta n_{\nu}$ and $D_{\mu}$ denotes is covariant derivative parallel to $n$. Such a contribution is only relevant when the region $\mathcal{V}$ has a codimension-2 corner. To have a well-posed variational problem under, say, Dirichlet boundary conditions, then requires one include a Hayward corner term \cite{Hayward:1993my} (see also Eq. (106) of Section 8 of \cite{Hawking:1980gf}). Since we work with spacetimes without corners, we neglect the necessary corner terms needed to make the variational problem well-posed.}

 Let us now evaluate (\ref{eq:bdrytermGRuse}) for various boundary conditions imposed at $\partial\mathcal{V}$. 

 \vspace{2mm}

 \noindent \textbf{Dirichlet boundary conditions.} These boundary conditions fix the induced metric $h_{\mu\nu}$ on $\partial\mathcal{V}$ such that its variation vanishes, $\delta h_{\mu\nu}|_{\partial\mathcal{V}}=0$. Thus, to have a well-posed variational problem with Dirichlet boundary conditions on $\partial\mathcal{V}$,  the Einstein-Hilbert action (\ref{eq:EHact}) must be supplemented with the Gibbons-Hawking-York (GHY) boundary term \cite{York:1972sj,Gibbons:1976ue}
\beq I_{\text{D}}=I_{\text{EH}}+I_{\text{GHY}}=\frac{1}{16\pi G}\int_{\mathcal{V}}d^{d}x\sqrt{-g}(R-2\Lambda)+\frac{\epsilon}{8\pi G}\oint_{\partial\mathcal{V}}d^{d-1}y\sqrt{|h|}K\;,\eeq
where $I_{\text{D}}$ denotes the action for which the variational problem of Einstein gravity is well-posed assuming Dirichlet boundary conditions.

\vspace{2mm}

\noindent \textbf{Neumann boundary conditions.} By Neumann boundary conditions one often means keeping the normal derivative of the metric fixed at the boundary.
Alternatively, following \cite{Krishnan:2016mcj,Krishnan:2016tqj}, rather than fixing the normal derivative of the metric at the boundary, by Neumann boundary conditions we mean fixing the functional derivative of the action with respect to the boundary metric, i.e., the conjugate momenta to $h^{\mu\nu}$. In standard classical particle mechanics, these two descriptions of Neumann boundary conditions are equivalent since the former corresponds to holding the velocity fixed,  while the
latter refers to holding the momentum fixed. Advantageously, with this latter notion of Neumann boundary conditions, one is able to derive a covariant boundary term, which must be added to the Einstein-Hilbert action plus GHY contribution such that the variational problem is well-posed.

To wit, rearrange  (\ref{eq:bdrytermGRuse})  such that
\beq \delta\left(I_{\text{EH}}+\frac{2}{16\pi G}\oint_{\partial\mathcal{V}}d^{d-1}y\sqrt{|h|}\epsilon K\right)=\frac{1}{16\pi G}\oint_{\partial\mathcal{V}}d^{d-1}y\sqrt{|h|}\epsilon(K_{\mu\nu}-h_{\mu\nu}K)\delta h^{\mu\nu}\;.\label{eq:varnocorns}\eeq
The term in parentheses on the left hand side is the familiar Einstein-Hilbert action together with the usual GHY term. First, notice that the right hand side will vanish when the boundary is umbilic, i.e., $K_{\mu\nu}=h_{\mu\nu}K$. We will not assume the boundary is umbilic. Then, using $\delta h^{\mu\nu}=-h^{\mu\alpha}h^{\beta\nu}\delta h_{\alpha\beta}$ and $n^{\mu}h_{\mu\nu}=n^{\nu}h_{\mu\nu}=0$, the right hand side becomes
\beq \frac{1}{16\pi G}\oint_{\partial\mathcal{V}}d^{d-1}y\sqrt{|h|}\epsilon(K_{\mu\nu}-h_{\mu\nu}K)\delta h^{\mu\nu}=\oint_{\partial\mathcal{V}}d^{d-1}y\pi^{\mu\nu}\delta h_{\mu\nu}\;,\label{eq:rhspi}\eeq
where 
\beq \pi^{\mu\nu}\equiv-\frac{\sqrt{|h|}\epsilon}{16\pi G}(K^{\mu\nu}-h^{\mu\nu}K)=\frac{\delta}{\delta h_{\mu\nu}}(I_{\text{EH}}+I_{\text{GHY}})\;.\label{eq:conmomNeu}\eeq
This is the usual conjugate momenta to $h_{\mu\nu}$ in the ADM formalism (where $\epsilon=-1$). Performing integration by parts yields
\beq \delta\left(I_{\text{EH}}+\frac{2}{16\pi G}\oint_{\partial\mathcal{V}}d^{d-1}y\sqrt{|h|}\epsilon K-\oint_{\partial\mathcal{V}}d^{d-1}y\pi^{\mu\nu}h_{\mu\nu}\right)=-\oint_{\partial\mathcal{V}}d^{d-1}yh_{\mu\nu}\delta \pi^{\mu\nu}\;.\eeq
Thus, keeping $\pi^{\mu\nu}$ (\ref{eq:conmomNeu}) fixed, it follows the Einstein-Hilbert action must be supplemented by an additional boundary term for the variational problem with Neumann boundary conditions to be well-posed. Specifically,
\beq I_{\text{N}}=I_{\text{D}}-\oint_{\partial\mathcal{V}}d^{d-1}y\pi^{\mu\nu}h_{\mu\nu}\;.\eeq
Using $\pi^{\mu\nu}h_{\mu\nu}=\frac{\epsilon\sqrt{|h|}}{16\pi G}(d-2)K$, the GHY and Neumann boundary terms combine to yield
\beq I_{\text{N}}=I_{\text{EH}}-\frac{(d-4)}{16\pi G}\oint_{\partial\mathcal{V}}d^{d-1}y\sqrt{|h|}\epsilon K\;.\eeq
Observe that in $d=2$, the GHY and Neumann boundary terms coincide, while in $d=4$ the boundary term exactly vanishes. The boundary term in $d=3$ was uncovered in \cite{Banados:1998ys,Miskovic:2006tm,Detournay:2014fva}.

\vspace{2mm}

\noindent \textbf{Conformal boundary conditions.} These boundary conditions fix the conformal class of the induced metric $h_{\mu\nu}$, denoted by $[h_{\mu\nu}]$, and the trace of extrinsic curvature $K$ at the boundary \cite{Anderson:2006lqb}. Fixing the conformal class of the boundary metric amounts to fixing $h_{\mu\nu}$ up to a Weyl transformation. To this end, it is useful to introduce the conformal metric  $\hat{h}_{\mu\nu}\equiv h^{-\frac{1}{(d-1)}}h_{\mu\nu}$.\footnote{The conformal metric is invariant under Weyl transformations, i.e., for $h_{\mu\nu}\to\Omega^{2}h_{\mu\nu}$, it follows $\hat{h}_{\mu\nu}\to h^{-\frac{1}{(d-1)}}(\Omega^{2(d-1)})^{-\frac{1}{(d-1)}}\Omega^{2}h_{\mu\nu}=\hat{h}_{\mu\nu}$. In the literature often the conformal class, i.e., the equivalence class of metrics equivalent up to a Weyl transformation, is also called the conformal metric.} 
Using $\delta h=h h^{\mu\nu}\delta h_{\mu\nu}$, it follows
\beq \delta \hat{h}_{\mu\nu}=-\frac{1}{(d-1)}h^{-\frac{1}{(d-1)}}h_{\mu\nu} h^{\alpha\beta}\delta h_{\alpha\beta}+h^{-\frac{1}{(d-1)}}\delta h_{\mu\nu}\;.\label{eq:varhhat}\eeq
Observe $h^{\mu\nu}\delta\hat{h}_{\mu\nu}=0$, while\footnote{The first line follows from
 \beq 
 \begin{split}
\sqrt{h}(K^{\mu\nu}-h^{\mu\nu}K)\delta h_{\mu\nu}&=-\frac{(d-2)}{(d-1)}\sqrt{h}Kh^{\mu\nu}\delta h_{\mu\nu}+\sqrt{h}\left(K^{\mu\nu}\delta h_{\mu\nu}-\frac{1}{(d-1)}K h^{\mu\nu}\delta h_{\mu\nu}\right)\\
&=-\frac{2(d-2)}{(d-1)}K\delta\sqrt{h}+h^{\frac{1}{(d-1)}}\sqrt{h}K^{\mu\nu}\delta\hat{h}_{\mu\nu}\;,
\end{split}
\eeq
where we used (\ref{eq:varhhat}). Then add zero $-\frac{1}{(d-1)}Kh^{\mu\nu}\delta\hat{h}_{\mu\nu}$ such that the conjugate momenta to the conformal metric $\hat{h}_{\mu\nu}$ is proportional to the traceless part of the extrinsic curvature.}
\beq 
\begin{split}
\pi^{\mu\nu}\delta h_{\mu\nu}&=\frac{\epsilon}{16\pi G}\biggr\{\frac{2(d-2)}{(d-1)}K\delta\sqrt{h}+h^{\frac{1}{(d-1)}}\sqrt{h}\left(K^{\mu\nu}-\frac{1}{(d-1)}Kh^{\mu\nu}\right)\delta\hat{h}_{\mu\nu}\biggr\}\\
&=-\frac{\epsilon}{16\pi G}\delta\left(-\frac{2(d-2)}{(d-1)}K\sqrt{h}\right)+\pi_{K}\delta K+\hat{\pi}^{\mu\nu}\delta\hat{h}_{\mu\nu}
\end{split}
\label{eq:pidelh}\eeq
where in the second line we used integration by parts and introduced conjugate momenta\footnote{Comparing to Eqs. (2.10) and (2.11) of \cite{Odak:2021axr}, our formulae for the momenta are different by an overall sign. They also set $16\pi G=1$ and do not include $\sqrt{h}$ in their definition for $\hat{\pi}^{\mu\nu}$.}
\beq \pi_{K}\equiv-\frac{\epsilon}{16\pi G}\frac{2(d-2)}{(d-1)}\sqrt{h}\;,\quad \hat{\pi}^{\mu\nu}=-\frac{\epsilon\sqrt{h}}{16\pi G}h^{\frac{1}{(d-1)}}\left(K^{\mu\nu}-\frac{1}{(d-1)}K h^{\mu\nu}\right)\;.\eeq
Return to variation (\ref{eq:varnocorns}) with the relation (\ref{eq:rhspi}). Rearranging and invoking (\ref{eq:pidelh}) gives
\beq
\begin{split}
&\delta\left(I_{\text{EH}}+\frac{2}{16\pi G}\oint_{\partial\mathcal{V}}d^{d-1}y\sqrt{|h|}\epsilon K-\frac{1}{16\pi G}\oint_{\partial\mathcal{V}}d^{d-1}y\sqrt{|h|}\epsilon\frac{2(d-2)}{(d-1)}K\right)\\
&=\oint_{\partial\mathcal{V}}d^{d-1}y \pi_{K}\delta K+\oint_{\partial\mathcal{V}}d^{d-1}y\hat{\pi}^{\mu\nu}\delta\hat{h}_{\mu\nu}\;.
\end{split}
\eeq
Thus, keeping the conformal data $(K,\hat{h}_{\mu\nu})$ fixed, the Einstein-Hilbert action must be supplemented by an additional boundary term for the variational problem to well-posed, 
\beq I_{\text{C}}=I_{\text{EH}}+\frac{2}{(d-1)16\pi G}\oint_{\partial\mathcal{V}}d^{d-1}y\sqrt{|h|}\epsilon K\;.\eeq
The same boundary term is uncovered in \cite{Odak:2021axr}. In $d=3$, this boundary term also coincides with the one in \cite{Anastasiou:2020zwc}, obtained from a different perspective.

\vspace{2mm}

\noindent \textbf{Generalized boundary conditions.} Thus far, to have a well-posed variational problem, the Einstein-Hilbert action is supplemented by a boundary term of the form
\beq I_{d}=I_{\text{EH}}+\frac{\Theta_{d}}{8\pi G}\oint_{\partial\mathcal{V}}d^{d-1}y\sqrt{|h|}\epsilon K\;,\eeq
with constant $\Theta_{d}$ depending on the choice of boundary conditions; specifically,
\[ \Theta_{d} =  \begin{cases} 
      1\,, & \text{Dirichlet}\\
      -\frac{(d-4)}{2}\,, & \text{Neumann} \\
      \frac{1}{(d-1)}\,, & \text{Conformal} 
   \end{cases}
\]
Notice in $d=2$, $\Theta_{2}$ coincides for the three boundary conditions. More generally, one can consider the generalized boundary term \cite{Liu:2024ymn} (see also \cite{Parvizi:2025shq,Parvizi:2025wsg})
\beq I_{d,p}=I_{\text{EH}}+\frac{\Theta_{d,p}}{8\pi G}\oint_{\partial\mathcal{V}}d^{d-1}y\sqrt{|h|}\epsilon K\;,\quad \Theta_{d,p}=\frac{2p(d-1)-1}{(d-1)(2p-1)}\;,\label{eq:genbdrycondsapp}\eeq
to generalize the conformal and Neumann boundary conditions; $p=0$ coincides with the term needed to impose conformal boundary conditions, the singular limits $p\to\pm$ recover the usual Dirichlet boundary terms, and $p=\frac{(d-3)}{2(d-1)}$ gives the same term needed when assuming Neumann boundary conditions.\footnote{Also notice for $p=1/2(d-1)$ the boundary term vanishes, while for $p=1/2$ the boundary term diverges. The former condition implies (for vanishing cosmological constant $\Lambda=0$) that the solutions exhibit scale invariance. The latter condition arises because $p=1/2$ requires fixing $\sqrt{h}K$, the integrand of the boundary term. See the discussion below Eq. (2.13) of \cite{Liu:2024ymn}.} For any finite $p$, this more general boundary term is such that Einstein gravity has a well-posed variational problem for the one-parameter family of boundary conditions consisting of fixing either (i) $(h^{p}K,\hat{h}_{\mu\nu})$ or (ii) $(h^{p}K,\hat{K}_{\mu\nu})$, where $\hat{K}_{\mu\nu}=K_{\mu\nu}-\frac{1}{(d-1)}h_{\mu\nu}K$ is the traceless part of extrinsic curvature.

\subsection{Adding Maxwell}

Let us now consider Einstein-Maxwell-$\Lambda$ gravity, characterized by the sum of Einstein-Hilbert action (\ref{eq:EHact}) and the Maxwell action
\beq I_{\text{Max}}=-\frac{1}{64\pi G\mu_{0}}\int_{\mathcal{V}}d^{d}x\sqrt{-g}F^{2}\;,\label{eq:maxact}\eeq
where $F^{2}\equiv F_{\mu\nu}F^{\mu\nu}$ for Maxwell field strength $F_{\mu\nu}\equiv\nabla_{\mu}A_{\nu}-\nabla_{\nu}A_{\mu}=2\partial_{[\mu}A_{\nu]}$ with $U(1)$ gauge field $A_{\mu}$, and $\mu_{0}$ is the electromagnetic coupling. A total variation of the action  (\ref{eq:maxact}) yields 
\beq
\begin{split} 
-64\pi G\mu_{0}\delta I_{\text{Max}}
&=\int d^{d}x\sqrt{-g}\left[\left(2F^{\alpha}_{\;\mu}F_{\alpha\nu}-\frac{1}{2}g_{\mu\nu}F^{2}\right)\delta g^{\mu\nu}-4(\nabla_{\alpha}F^{\alpha\beta})\delta A_{\beta}\right]\\
&+4\epsilon\oint_{\partial\mathcal{V}}d^{d-1}y\sqrt{|h|}n_{\alpha}F^{\alpha\beta}\delta A_{\beta}\;.
\end{split}
\eeq
Combined with the variation of the Einstein-Hilbert action, the bulk term vanishes on-shell, i.e., when
\beq G_{\alpha\beta}+\Lambda g_{\alpha\beta}=8\pi G T^{\text{Max}}_{\alpha\beta}\;,\quad T^{\text{Max}}_{\alpha\beta}=-\frac{2}{\sqrt{-g}}\frac{\delta I_{\text{Max}}}{\delta g^{\alpha\beta}}=\frac{1}{16\pi G\mu_{0}}\left(F^{\mu}_{\;\alpha}F_{\mu\beta}-\frac{1}{4}g_{\alpha\beta}F^{2}\right)\;,\eeq
and 
\beq \nabla_{\alpha}F^{\alpha\beta}=0\;.\eeq

Thus,  the on-shell variation of the Einstein-Maxwell action leaves the boundary term,
\beq
\begin{split} 
\delta I_{\text{Max}}&=-\frac{\epsilon}{16\pi G\mu_{0}}\oint_{\partial\mathcal{V}}d^{d-1}y\sqrt{|h|}n_{\alpha}F^{\alpha\beta}\delta A_{\beta}\;,
\end{split}
\label{eq:remainderbdry}\eeq
accompanied by the variation (\ref{eq:bdrytermGRuse}). If we assume the gauge field obeys Dirichlet boundary conditions, i.e., $\delta A_{\mu}|_{\partial\mathcal{V}}=0$, the boundary term vanishes. Consequently, independent of the boundary conditions imposed on the metric, the Maxwell action does not need to be supplemented by a boundary term if the gauge field is fixed.

Alternatively, suppose Dirichlet boundary conditions are not imposed on $A_{\mu}$. Then write
\beq 
\begin{split} \delta I_{\text{Max}}&=-\delta I_{A}+\frac{\epsilon}{16\pi G\mu_{0}}\oint_{\partial\mathcal{V}}d^{d-1}y\delta\left(\sqrt{|h|}n_{\alpha}F^{\alpha\beta}\right)A_{\beta}\;,
\end{split}
\eeq
with
\beq I_{A}=\frac{\epsilon}{16\pi G\mu_{0}}\oint_{\partial\mathcal{V}}d^{d-1}y\sqrt{|h|}n_{\alpha}F^{\alpha\beta}A_{\beta}\;.\label{eq:IAbdryterm}\eeq
If one further imposes any boundary conditions on the metric and fixes $\delta(\sqrt{h} n_{\alpha}F^{\alpha\beta})|_{\partial\mathcal{V}}=0$, then the boundary term (\ref{eq:IAbdryterm}) is the necessary for the Maxwell action  to be well-posed (see Eq. (24) of \cite{Hawking:1995ap}). In the context of black hole thermodynamics, this amounts to working in a (canonical) thermal ensemble of fixed (electric) charge.

%%%%%%%%%%%%%%%%%%%%%%%%%%%%%%%%%%%%%%%%%%%%%%%%%

\section{Spherical dimensional reduction} \label{app:reductionscoords}

Here we provide details of spherical reductions of the $d$-dimensional Einstein--Hilbert action with an arbitrary GHY-like boundary term. Our approach is a straightforward extension of \cite{Svesko:2022txo}, and a similar treatment is given in Appendix E of \cite{Banihashemi:2025qqi}. 

The starting point is  the Lorentzian Einstein--Hilbert action with  cosmological constant $\Lambda_{d}$ in $d$-spacetime dimensions,
\beq I_{d}=\frac{1}{16\pi G_{d}}\int_{\mathcal{M}} \hspace{-2mm}d^{d}X\sqrt{-G}[\mathcal{R}-2\Lambda_{d}]+\frac{\Theta_{d}}{8\pi G_{d}}\int _{\partial \mathcal{M}}\hspace{-3mm}d^{d-1}Y\sqrt{-H}\mathcal{K}\;. \label{eq:ddimaction}\eeq
Here  $G_{MN}$ is the $d$-dimensional metric, $\mathcal{R}$ the Ricci scalar, and we supplement the action with an appropriate boundary term where $H_{MN}$ is the induced metric of the boundary $\partial \mathcal{M}$ with $\mathcal{K}$ being the trace of its extrinsic curvature.

Consider the metric ansatz
\beq ds^{2}_{d}=G_{MN}dX^{M}dX^{N}=g_{\mu\nu}(x)dx^{\mu}dx^{\nu}+L^{2}_{d}\Phi^{2/(d-2)}(x)d\Omega_{d-2}^{2}\;.\label{eq:dimansatzapp}\eeq
Here $M,N=0,1,...,d-1$; $\mu,\nu=0,1$, $\Phi(x)$ is the dilaton, and $L_{d}$ is some bulk length scale, e.g., the bulk (A)dS$_{d}$ radius $\ell_{d}$. A standard calculation (see \emph{e.g.} \cite{Grumiller:2001ea,Narayan:2020pyj}) shows the $d$-dimensional Ricci scalar decomposes as
\beq \mathcal{R}=R+\frac{(d-3)(d-2)}{L^{2}_{d}\Phi^{2/(d-2)}}+\frac{(d-3)}{(d-2)}\frac{1}{\Phi^{2}}(\nabla\Phi)^{2}-\frac{2}{\Phi}\Box\Phi\;,\label{eq:Riccscalred}\eeq
where $R$ is the Ricci scalar and $\nabla_{\mu}$ denotes the covariant derivative with respect to the two-dimensional metric $g_{\mu\nu}$. Assuming the codimension-1 boundary $\partial\mathcal{M}$ is at a $x^{\mu}$-hypersurface, such that the induced metric obeys $ds^{2}_{d-1}=H_{AB}dY^{A}dY^{B}=h_{ab}(y)dy^{a}dy^{b}+L_{d}^{2}\Phi^{2/(d-2)}(x)d\Omega_{d-2}^{2}$, then the $(d-1)$-dimensional extrinsic curvature $\mathcal{K}$ reduces to 
\beq \mathcal{K}=K+\frac{1}{\Phi}n^{\mu}\nabla_{\mu}\Phi\;,\label{eq:Kred}\eeq
where $K$ is the extrinsic curvature of the 1-dimensional boundary and $n_{\mu}$ is the unit normal vector used to define the $(d-1)$-dimensional boundary metric. Further
\beq \int_{\mathcal{M}}\hspace{-2mm}d^{d}X\sqrt{-G}=L_{d}^{d-2}\Omega_{d-2}\int_{\mathfrak{m}}\hspace{-2mm}d^{2}x\sqrt{-g}\,\Phi\label{eq:detred}\;,\eeq
where $\mathfrak{m}$ is the two-dimensional Lorentzian manifold endowed with metric $g_{\mu\nu}$. Substituting the dimensionally reduced scalar curvatures (\ref{eq:Riccscalred}) and (\ref{eq:Kred}), and (\ref{eq:detred}) into the $d$-dimensional Einstein--Hilbert action results in
\beq 
\begin{split}
I_{\text{EH}}&
=\frac{1}{16\pi G_{2}}\int_{\mathfrak{m}}\hspace{-2mm}d^{2}x\sqrt{-g}\left[\Phi(R-2\Lambda_{d})+\frac{(d-3)(d-2)}{L^{2}_{d}}\Phi^{\frac{(d-4)}{(d-2)}}+\frac{(d-3)}{(d-2)}\frac{1}{\Phi}(\nabla\Phi)^{2}\right]\\
&+\frac{1}{8\pi G_{2}}\int_{\partial \mathfrak{m}}\hspace{-2mm}dy\sqrt{-h}(-n^{\mu}\nabla^{\mu}\Phi)
\end{split}
\eeq
where we performed integration by parts to eliminate a $\Box\Phi$ contribution, and we defined the (dimensionless) two-dimensional Newton's constant $G_{2}$
\begin{equation}
    \frac{1}{G_2} \equiv \frac{L_d^{d-2}\Omega_{d-2}}{G_d}\,.
\end{equation} 
The generalized GHY term, meanwhile, reduces to\footnote{For dimensional reduction of boundary terms of other theories, see \cite{Saskowski:2024tat}.}
\beq \frac{\Theta_{d}}{8\pi G_{d}}\int _{\partial\mathcal{M}}\hspace{-3mm}d^{d-1}Y\sqrt{-H}\mathcal{K}=\frac{\Theta_{d}}{8\pi G_{2}}\int_{\partial \mathfrak{m}}\hspace{-2mm}dy\sqrt{-h}(\Phi K+n^{\mu}\nabla_{\mu}\Phi)\;.\eeq
Combined, spherical reduction of the bulk action (\ref{eq:ddimaction}) results in the two-dimensional dilaton theory of gravity
\beq
\begin{split}
I_{d}&=\frac{1}{16\pi G_{2}}\int_{\mathfrak{m}}\hspace{-2mm}d^{2}x\sqrt{-g}\biggr(\Phi (R-2\Lambda_{d})+\frac{(d-3)(d-2)}{L^{2}_{d}}\Phi^{\frac{(d-4)}{(d-2)}}+\frac{(d-3)}{(d-2)}\frac{(\nabla\Phi)^{2}}{\Phi}\biggr)\\
&+\frac{1}{8\pi G_{2}}\int_{\partial \mathfrak{m}}\hspace{-3mm}dy\sqrt{-h}[\Theta_{d}\Phi K+(\Theta_{d}-1)n^{\mu}\nabla_{\mu}\Phi]\;,
\end{split}
\label{eq:redactgen}\eeq
Notice when $\Theta_{d}=1$ (Dirichlet boundary conditions), the boundary term simplifies, but is otherwise more complicated.

\subsection*{Eliminating the kinetic term}

When $d=3$, a dramatic simplification occurs such that the bulk kinetic term is eliminated. The resulting bulk two-dimensional action is that of standard JT gravity, while the  boundary term is dependent on the assumed boundary conditions of the higher-dimensional theory from whence it came. 
 For $d>3$ the reduced action (\ref{eq:redactgen}) takes a more general form, however, the kinetic two-dimensional kinetic term can be eliminated via an appropriate Weyl rescaling \cite{Svesko:2022txo}.  To this end,  recall how the Ricci scalar and trace of the extrinsic curvature  transform under the Weyl rescaling $\bar{g}_{\mu\nu}=\omega^{2}g_{\mu\nu}$ in a two-dimensional spacetime\footnote{To see how the conformal transformation of the extrinsic curvature arises, first note the induced metric transforms as $\bar{h}_{\mu\nu}=\omega^{2}h_{\mu\nu}$, such that $\bar{n}_{\mu}=\omega n_{\mu}$. Then, 
$$\bar{\nabla}_{\mu}\bar{n}_{\nu}=\omega \nabla_{\mu}n_{\nu}-(n_{\nu}\partial_{\mu}\omega-n_{\rho}g_{\mu\nu}g^{\rho\delta}\partial_{\delta}\omega)$$
where we used $\bar{\Gamma}^{\rho}_{\;\mu\nu}=\Gamma^{\rho}_{\;\mu\nu}+\omega^{-1}(\delta^{\rho}_{\mu}\partial_{\nu}\omega+\delta^{\rho}_{\nu}\partial_{\mu}\omega-g_{\mu\nu}g^{\rho\delta}\partial_{\delta}\omega)$. Consequently, $\bar{K}=\bar{g}^{\mu\nu}\bar{\nabla}_{\mu}\bar{n}_{\nu}=\omega^{-2}g^{\mu\nu}[ \omega \nabla_{\mu}n_{\nu}-(n_{\nu}\partial_{\mu}\omega-n_{\rho}g_{\mu\nu}g^{\rho\delta}\partial_{\delta}\omega)]=\omega^{-1}K+\omega^{-2}n_{\mu}\nabla^{\mu}\omega$.}
\beq \bar{R}=\omega^{-2}R-2\omega^{-3}\Box\omega+2\omega^{-4}(\nabla\omega)^{2}\;,\quad \bar{K}=\omega^{-1}K+\omega^{-2}n_{\mu}\nabla^{\mu}\omega\;.\label{eq:weylrescal2d}\eeq
Then, rescaling $g_{\mu\nu}\to \omega^{2}g_{\mu\nu}$, the reduced action (\ref{eq:redactgen}) becomes
\beq 
\begin{split}
I_{d}&=\frac{1}{16\pi G_{2}}\int_{\mathfrak{m}} \hspace{-1mm}d^{2}x\sqrt{-g}\biggr[\Phi R+2\omega^{-1}(\nabla^{\mu}\Phi)(\nabla_{\mu}\omega)+\frac{(d-3)}{(d-2)}\frac{1}{\Phi}(\nabla\Phi)^{2}-2\Lambda_{d}\Phi\omega^{2}\\
&+\frac{(d-3)(d-2)\omega^{2}}{L_{d}^{2}}\Phi^{\frac{(d-4)}{(d-2)}}\biggr]+\frac{1}{8\pi G_{2}}\int_{\partial \mathfrak{m}}\hspace{-3mm}dy\sqrt{-h}[\Theta_{d}\Phi K+(\Theta_{d}-1)(\omega^{-1}\Phi n_{\mu}\nabla^{\mu}\omega+n_{\mu}\nabla^{\mu}\Phi)]\;.
\end{split}
\eeq
where we used $n_{\mu}\nabla^{\mu}\Phi\to \omega^{-1}n^{\mu}\nabla_{\mu}\Phi$ and performed integration by parts to eliminate the $\Box\omega$ contribution. The kinetic term in the bulk contribution is eliminated by choosing $\omega=\gamma\Phi^{\eta}$ for $\eta=-(d-3)/2(d-2)$ and $\gamma$ some constant, resulting in 
\beq
\begin{split}
 I_{d}&=\frac{1}{16\pi G_{2}}\int_{\mathfrak{m}}\hspace{-2mm}d^{2}x\sqrt{-g}[\Phi R+U(\Phi)]+\frac{1}{8\pi G_{2}}\int_{\partial \mathfrak{m}}\hspace{-3mm}dy\sqrt{-h}\left[\Theta_{d}\Phi K+\frac{(\Theta_{d}-1)(d-1)}{2(d-2)}n_{\mu}\nabla^{\mu}\Phi\right]\;,
\end{split}
\label{eq:redactnariai}\eeq
with dilaton potential 
\beq U(\Phi)=\gamma^{2}\left(\frac{(d-3)(d-2)}{L_{d}^{2}}\Phi^{-1/(d-2)}-2\Lambda_{d}\Phi^{1/(d-2)}\right)\;.\eeq

\subsection*{Reduced boundary conditions}

For consistency, boundary conditions imposed on the bulk theory lead to boundary conditions on the dimensionally reduced theory. For example, consider the metric ansatz (\ref{eq:dimansatzapp}). Imposing Dirichlet boundary conditions in the bulk, $\delta h_{AB}|_{\partial\mathcal{M}}=0$, leads to applying both $\delta h_{ab}|_{\partial \mathfrak{m}}=0$ and $\delta\Phi|_{\partial \mathfrak{m}}=0$ in the two-dimensional dilaton theory. Clearly, these boundary conditions do not change under a Weyl rescaling of the two-dimensional metric. 

For conformal boundary conditions, where $(\hat{H}_{MN},\mathcal{K})$ are fixed at the boundary, note for the ansatz  (\ref{eq:dimansatzapp}) the conformal metric $\hat{H}_{MN}=H^{-1/(d-1)}H_{MN}$ has
\beq \hat{H}_{ab}=H^{-\frac{1}{(d-1)}}H_{ab}=(L_{d}^{2(d-2)}\Phi^{2})^{-\frac{1}{(d-1)}}h^{-\frac{1}{(d-1)}}h_{ab}=(L_{d}^{2(d-2)}\Phi^{2})^{-\frac{1}{(d-1)}}h^{\frac{(d-2)}{(d-1)}}\hat{h}_{ab}\;,\eeq
where we used $H_{ab}=h_{ab}$, and $\hat{h}_{ab}=h^{-1}h_{ab}=1$ is the one-dimensional conformal metric. Under a Weyl scaling, $g_{\mu\nu}\to \gamma^{2}\Phi^{2\eta}g_{\mu\nu}$, then 
\beq
\begin{split}
\hat{H}_{ab}&=(L_{d}^{2(d-2)}\gamma^{2}\Phi^{2(\eta+1)})^{-\frac{1}{(d-1)}}h^{\frac{(d-2)}{(d-1)}}\gamma^{2}\Phi^{2\eta}\hat{h}_{ab}\\
&=\Phi^{\frac{2}{(d-1)}(\eta(d-2)-1)}h^{\frac{(d-2)}{(d-1)}}\gamma^{\frac{2(d-2)}{(d-1)}}L_{d}^{\frac{-2(d-2)}{(d-1)}}\;.
\end{split}
\eeq
Thus, fixing the component $\hat{H}_{ab}$ at the boundary amounts to, at the level of the effective two-dimensional theory, fixing 
\beq \gamma\Phi^{\frac{\eta(d-2)-1}{(d-2)}}\sqrt{h}\biggr|_{\partial \mathfrak{m}}\equiv L_{d}\;.\label{eq:Hredapp}\eeq

Meanwhile combining  $\mathcal{K}$ (\ref{eq:Kred}) with the Weyl rescaling (\ref{eq:weylrescal2d}) gives\footnote{Up to the factor of $\gamma$, our reduced expression (\ref{eq:mathcalKredapp}) matches Eq. (E.35) of \cite{Banihashemi:2025qqi} (upon sending their $d\to d-1$).}
\beq \mathcal{K}=\frac{\Phi^{-(\eta+1)}}{\gamma}\left(\Phi K+(\eta+1)n^{\mu}\nabla_{\mu}\Phi\right)\;.\label{eq:mathcalKredapp}\eeq
Thence, fixing $\mathcal{K}$ in the bulk is equivalent to fixing 
\beq \gamma^{-1}\Phi^{-(\eta+1)}\left(\Phi K+(\eta+1)n^{\mu}\nabla_{\mu}\Phi\right)\biggr|_{\partial \mathfrak{m}}=\mathcal{K}\label{eq:mathcalredapp}\eeq
in the two-dimensional dilaton theory.\footnote{If the bulk obeys the one-parameter family of boundary conditions \cite{Liu:2024ymn} $(H^{p}\mathcal{K},\hat{H}_{MN})$, then the reduced boundary condition (\ref{eq:mathcalKredapp}) becomes $\Phi^{(\eta+1)(2p-1)}\left(\Phi K+(\eta+1)n^{\mu}\nabla_{\mu}\Phi\right)|_{\partial \mathfrak{m}}=\text{fixed}$.}

\subsection{Spherical reduction of dyonic RN-AdS$_{4}$}

Let us now consider the spherical reduction of Einstein-Maxwell-$\Lambda$ theory. We need only append the above by separately describing the reduction of the Maxwell term (with a boundary term)
\beq I_{\text{Max}}=-\frac{1}{64\pi G_{d}\mu_{0}}\int_{\mathcal{M}}\hspace{-2mm} d^{d}X\sqrt{-G}\mathcal{F}^{2}+\frac{1}{16\pi G_{d}\mu_{0}}\int_{\partial\mathcal{M}}\hspace{-3mm} d^{d-1}Y\sqrt{-H}n_{M}\mathcal{F}^{MN}\mathcal{A}_{N}\;,\eeq
where $\mathcal{F}_{MN}=\partial_{M}\mathcal{A}_{N}-\partial_{N}\mathcal{A}_{M}$, for $d$-dimensional gauge field $\mathcal{A}_{M}$.  Restricting the metric ansatz (\ref{eq:dimansatzapp}) (with $g_{\mu\nu}\to \omega^{2} g_{\mu\nu}$) to $d=4$, the bulk term reduces to 
\beq\int_{\mathcal{M}}\hspace{-1mm} d^{4}X\sqrt{-G}\mathcal{F}^{2}=L_{4}^{2}\int_{\mathfrak{m}}\hspace{-2mm} d^{2}x\sqrt{-g}\omega^{2}\,\Phi\hspace{-1mm}\int\hspace{-1mm} d\Omega_{2}\left(\omega^{-4}F^{\mu\nu}F_{\mu\nu}+\frac{2}{L^{4}_{4}\Phi^{2}\sin^{2}(\theta)}\mathcal{F}_{\theta\phi}\mathcal{F}_{\theta\phi}\right)\;,\label{eq:redbulkactmax1}\eeq
where $F_{\mu\nu}$ denotes the two-dimensional Maxwell field strength. Meanwhile, the boundary term reduces to 
\beq
\begin{split} 
\int_{\partial\mathcal{M}}\hspace{-2mm}d^{3}Y\sqrt{-H}n_{M}\mathcal{F}^{MN}\mathcal{A}_{N}&=L^{2}_{4}\int_{\partial \mathfrak{m}} \hspace{-2mm} dy\sqrt{-h}\omega^{-2}\,\Phi \hspace{-1mm}\int d\Omega_{2}\; n_{\mu}F^{\mu\nu}A_{\nu}\;,
\end{split}
\label{eq:redbdryact1}\eeq
where recall $n_{\mu}\to \omega n_{\mu}$ under a conformal rescaling of $g_{\mu\nu}$, and we take $n_{M}$ to point only in the $r-t$ plane.

We focus on the dimensional reduction of the dyonic RN-AdS$_{4}$ black hole,  
\beq ds^{2}=-fdt^{2}+f^{-1}dr^{2}+r^{2}d\Omega_{2}^{2}\;,\quad f(r)=1-\frac{2m}{r}+\frac{Q_{e}^{2}+Q_{m}^{2}}{r^{2}}+\frac{r^{2}}{\ell_{4}^{2}}\;,\label{eq:RNAdSmet4}\eeq
with gauge field and field strength
\beq 
\begin{split}
&\mathcal{A}=\mathcal{A}_{M}dX^{M}=A_{\mu}dx^{\mu}+\mathcal{A}_{\phi}d\phi=-\frac{Q_{e}}{r}dt+Q_{m}\cos(\theta)d\phi\;,\\
&F_{rt}=\frac{Q_{e}}{r^{2}}\;,\quad \mathcal{F}_{\theta\phi}=-Q_{m}\sin(\theta)\;.
\end{split}
\label{eq:gaugeAF}\eeq
Here $m$, $Q_{e}$, and $Q_{m}$ are mass, electric and magnetic charge parameters of the black hole, and $\ell_{4}$ is the AdS$_{4}$ radius. Substituting this solution into the bulk and boundary actions (\ref{eq:redbulkactmax1}) and (\ref{eq:redbdryact1}) and evaluating the angular integral gives\footnote{Setting $G_{4}=1$ and $\mu_{0}=1/4$, the reduced boundary term coincides with Eq. (6.47) of \cite{Brown:2018bms} upon $\omega^{2}=\Phi^{-1/2}/2$ and subsequently rescaling $\Phi\to 2\Phi$.}
\beq
\begin{split} 
&\int_{\mathcal{M}}\hspace{-1mm} d^{4}X\sqrt{-G}\mathcal{F}^{2}=4\pi L_{4}^{2}\int_{\mathfrak{m}}\hspace{-2mm} d^{2}x\sqrt{-g}\left(\omega^{-2}\Phi F^{\mu\nu}F_{\mu\nu}+\frac{2\omega^{2}q_{m}^{2}}{L^{4}_{4}\Phi}\right)\;,\\
& \int_{\partial\mathcal{M}}\hspace{-2mm}d^{3}Y\sqrt{-H}n_{M}\mathcal{F}^{MN}\mathcal{A}_{N}=4\pi L^{2}_{4}\int_{\partial \mathfrak{m}} \hspace{-2mm} dy\sqrt{-h}\omega^{-2}\,\Phi n_{\mu}F^{\mu\nu}A_{\nu}\;.
\end{split}
\label{eq:redbdrybulkmaxapp}\eeq
At this level the $F_{\mu\nu}$ is the two-dimensional Maxwell field strength, inherited from the higher-dimensional metric. Notably, we have not imposed the on-shell condition for $F_{\mu\nu}$ (\ref{eq:gaugeAF}), however, we did so for $\mathcal{F}_{\theta\phi}$ because ultimately we take the four-dimensional bulk theory to be in an ensemble where the magnetic charge is fixed. 

There are two possible reduced boundary conditions. First, the reduced Dirichlet boundary condition $\delta \mathcal{A}_{M}|_{\partial \mathcal{M}}=0$ trivially implies the Dirichlet boundary conditions of the 2D gauge field
\beq \delta A_{\mu}|_{\partial \mathfrak{m}}=0\;.\eeq
Meanwhile, the boundary condition $\delta (\sqrt{-H}n_{M}\mathcal{F}^{MN})|_{\partial \mathcal{M}}=0$ implies
\beq \sqrt{-h}n_{\mu}F^{\mu\nu}\Phi\omega^{-2}|_{\partial \mathfrak{m}}=\text{const}\;,\eeq
for $G_{\mu\nu}=\omega^{2}g_{\mu\nu}$, $n_{\mu}\to \omega n_{\mu}$ and $H=L_{4}^{4}\Phi^{2} h$. 

\subsection*{Extremal RN}

The RN black hole (\ref{eq:RNAdSmet4}) has inner and outer black hole horizons, the real, positive roots $r_{\pm}$ of $f(r_{\pm})=0$. The extremal RN black hole occurs when $r_{+}=r_{-}=r_{\text{ex}}$, where $f(r_{\text{ex}})=\partial_{r}f(r_{\text{ex}})=0$. In the extremal limit the mass $m$ and charge $Q^{2}=Q_{e}^{2}+Q_{m}^{2}$ are
\beq 
\begin{split}
&Q_{\text{ex}}^{2}=r_{\text{ex}}^{2}\left(1+\frac{3r_{\text{ex}}^{2}}{\ell_{4}^{2}}\right)\;,\\    
&m_{\text{ex}}=r_{\text{ex}}\left(1+\frac{2r_{\text{ex}}^{2}}{\ell_{4}^{2}}\right)\;.
\end{split}
\eeq
In this limit the Hawking temperature of the RN black hole vanishes. 

The near-horizon geometry of the extremal RN-AdS$_{4}$ black hole can be found by introducing dimensionless radial and time coordinates $(\rho,\hat{t})$ \cite{Bardeen:1999px,Hartman:2008pb}
\beq r=r_{\text{ex}}+\epsilon \ell_{2}\rho\;,\quad t=\frac{\hat{t}\ell_{2}}{\epsilon}\;,\eeq
for small dimensionless positive parameter $\epsilon$ and 
\beq \ell_{2}^{2}=\frac{r_{\text{ex}}^{2}}{1+6r_{\text{ex}}^{2}/\ell_{4}^{2}}=\frac{\ell_{4}^{2}}{6}\left[1-\left(1+\frac{12Q_{\text{ex}}^{2}}{\ell_{4}^{2}}\right)^{-1/2}\right]\;.\eeq
In the asymptotically flat limit, $\ell_{4}\to\infty$, then $\ell_{2}^{2}=Q_{\text{ex}}^{2}$. The near-horizon geometry is then
\beq ds^{2}\approx \ell_{2}^{2}\left(-\rho^{2}d\hat{t}^{2}+\frac{d\rho^{2}}{\rho^{2}}\right)+r_{\text{ex}}^{2}d\Omega_{2}^{2}\;,\eeq
which has the form of $\text{AdS}_{2}\times S^{2}$ with AdS$_{2}$ radius $\ell_{2}$.

\section{Details on boundary dynamics} \label{app: brane dynamics}

Here we give an example of a perturbative solution to \eqref{eqn: K non-static d-dim} beyond the leading correction away from the static boundary solution. We show that the existence of these solutions does not spoil the uniqueness property of conformal boundary conditions. For concreteness, consider Euclidean AdS$_3$ space in  global coordinates,
$
    f(r) = \frac{r^2+\ell^2}{\ell^2} \, .
$
For constant  $\k$, the differential equation \eqref{eqn: K non-static d-dim} admits a static solution, 
\begin{equation}
    \bomega(u) = \omega_s = - \frac{1}{2}\log \left(-2 + \frac{\k \ell}{2}\left(\k \ell + \sqrt{\k^2\ell^2-4}\right)\right) \, ,
\end{equation}
where $\omega_s \in \mathbb{R}$ for $\k \ell>2$. Taking $\k \ell \to 2$, the Weyl factor goes to infinity, corresponding to the finite boundary being pushed to the AdS$_3$ boundary. As the bulk geometry does not contain a conical singularity, the periodicity of the boundary coordinate $u$ is a free parameter.

Let $\delta \bomega(u) \equiv \bomega(u) - \omega_s$ be a difference between the exact solution and the static one. We also impose conditions that $\delta\bomega(u_0)=\upsilon$ and $\partial_u\delta\bomega(u_0)=0$ for arbitrary real constants $\upsilon$ and $u_0$. For $\delta \bomega(u)$ to describe a small fluctuation around the static solution, we treat $\upsilon$ as a small parameter and write $\delta\bomega(u)$ perturbatively in $\upsilon$. At the linearized level, we obtain a solution \eqref{eqn: lin non-static sol} with frequency 
\begin{equation}
    \Omega^2 =2-\frac{\k \ell}{2}\left(\k \ell -\sqrt{\k^2\ell^2-4}\right).
\end{equation}
For $\k \ell>2$, the frequency approaches zero as $\k\ell \to 2^{+}$. Beyond the linear order, we find 
\begin{eqnarray}
    \delta \bomega(u) &=& \upsilon \cos\left(\Omega \frac{\Delta u}{\r}\right) + \upsilon^2 \left(-a_1 + a_1 \cos\left(\Omega \frac{\Delta u}{\r}\right)\right)  \\
    &&+ \upsilon^3 \left(-(a_2+a_3) + a_2 \cos\left(\Omega \frac{\Delta u}{\r}\right) +a_3 \cos\left(3\Omega \frac{\Delta u}{\r}\right) + a_4 \frac{\Delta u}{\r}\sin\left(\Omega \frac{\Delta u}{\r}\right)\right) + \mathcal{O}(\upsilon^4) \, ,\nonumber
\end{eqnarray}
where we denote $\Delta u \equiv u-u_0$ and
\begin{equation}
    a_1 = \frac{\Omega^2}{2} \,, \quad
    a_2 = \frac{1}{48}\left(-1+24\Omega^4\right) \,, \quad
    a_3 = \frac{1}{48} \,, \quad
    a_4 = \frac{\Omega}{4}\left(-1+2\Omega^2\right) \,.
\end{equation}
As expected from solving the second-order-differential equation, the solution is uniquely fixed in terms of the two constants of integration, $\upsilon$ and $u_0$. Moreover, we can deduce from the solution that the linearized solution \eqref{eqn: lin non-static sol} is reliable as long as $\upsilon \Omega^2 \ll1$.

At order $\mathcal{O} (\upsilon^3)$, the solution contains a secular term, $\tfrac{\Delta u}{\r}\sin\left(\Omega\tfrac{\Delta u}{\r}\right)$, which grows linearly in $u$. According to the Poincar\'e-Lindstet method, one may absorb this term into a shift of the frequency. Concretely, we combine the secular term and the zeroth-order term into
\begin{equation}
    \upsilon \cos \left(\left(\Omega - \upsilon^2 a_4 \right)\frac{\Delta u}{\r}\right)+\mathcal{O}\left(\upsilon^3\right) \, .
\end{equation}
The periodic condition for the Weyl factor, $\delta\bomega(u+\beta) =\delta\bomega(u)$, now becomes
\begin{equation}
    \tilde{\beta}\left(\Omega - \upsilon^2 a_4  + \mathcal{O}\left(\upsilon^3\right)\right) = 2 \pi n \, ,\qquad n \in \mathbb{N} \, ,
\end{equation}
which fixes the constant $\upsilon$ to be
\begin{equation}
    \upsilon^2 = \frac{\left(\tilde{\beta}-\frac{2\pi n}{\Omega}\right)\Omega}{\tilde{\beta}^2a_4} \, .
\end{equation}
The validity of the perturbation theory then enforces that $\left|\tilde{\beta} - \tfrac{2\pi n}{\Omega}\right|\ll\tfrac{4\pi^2 n^2 }{\Omega^6}$. 

This analysis shows that the apparent family of solutions close to the static one is uniquely fixed in terms of the boundary data. Hence, the existence of non-static solutions does not spoil the uniqueness properties of the Euclidean problem, as mentioned in Section \ref{sec2: gravity}.

\section{High-temperature universality from 2D} \label{app: highT}

Here we derive conformal thermodynamics in the high temperature limit from the perspective of the effective two-dimensional dilaton gravity theory. Recall the class of dilaton potential obtained from dimensionally reducing $d$-dimensional Einstein gravity is given by \eqref{eq:dilapotgenred},
\beq U(\Phi)=\gamma^{2}\left(\frac{(d-3)(d-2)L^2}{L_{d}^{2}}\Phi^{-1/(d-2)}-2\Lambda_{d}L^2\Phi^{1/(d-2)}\right)\;,\eeq
where we factor out an overall $L^{-2}$ as in the reduced action (\ref{eqn: Euclidean action alpha=1}) and set $\alpha = \tfrac{d-1}{2(d-2)}$ in the action  and boundary conditions (\ref{eqn: bdry cond alpha=1}). In the limit $\Phi \gg 1$, the potential is dominated by the last term,
\begin{equation}\label{eqn: poten large phi}
    U(\Phi) \to -2\Lambda_dL^2 \Phi^{1/(d-2)} \, , \qquad \text{as}\quad \Phi\to\infty,
\end{equation}
where we set the parameter $\gamma =1$ without loss of generality. Consequently, the solution (\ref{eqn: sol for generic potential}) is,
\begin{equation}
    \frac{ds^2}{L^2} = f(r)dt^2 +\frac{dr^2}{f(r)} \,, \quad f(r) = \pm(d-2)^2 \left(\rh^{\frac{d-1}{d-2}}-r^{\frac{d-1}{d-2}}\right) \tilde{\Phi}^{-\frac{d-3}{d-2}} \,, \quad \Phi(r)=\tilde{\Phi}\,r\,,
\end{equation}
where $L=L_{d}$, satisfying $L^2= \tfrac{(d-1)(d-2)}{|\Lambda_d|}$. Here the upper (lower) sign refers a positive (negative) cosmological constant $\Lambda_{d}$. To compute thermodynamic quantities, we invert \eqref{eqn: beta generic potential} and \eqref{eqn: kappa generic potential},
\begin{equation}\label{eqn: sol large phi}
    \left(\frac{\rh}{\rb}\right)^{\frac{d-1}{d-2}}=\frac{2 \sqrt{\kappa^2L^2\pm (d-1)^2}}{\sqrt{\kappa^2L^2\pm(d-1)^2}+\kappa L}\, , \qquad \left(\tilde{\Phi} \rh\right)^{\frac{1}{d-2}}= \pm\frac{4\pi\left(\sqrt{\kappa^2 L^2 \pm(d-1)^2}-\kappa L\right)}{(d-1)^2\b}\,.
\end{equation}
Plugging \eqref{eqn: poten large phi} into \eqref{eqn: energy and entropy generic potential} and \eqref{eqn: specific heat generic potential}, we obtain the conformal thermodynamic quantities. To leading order in the large dilaton expansion, we have
\begin{equation}\label{eqn: thermo large phi}
\begin{cases}
 E_{\text{conf}} = \frac{(d-2)(\tilde{\Phi}\rh)^{\frac{d-1}{d-2}}}{16 \pi G_2} \left|\left(\frac{\rh}{\rb}\right)^{\frac{d-1}{d-2}}-1\right|^{-1/2}\,, \\
 %\frac{(d-2)(\tilde{\Phi}\rh)^{\frac{d-1}{d-2}}}{16 \pi G_2 \sqrt{\pm\left(\left(\frac{\rh}{\rb}\right)^{\frac{d-1}{d-2}}-1\right)}} \dami{} \,, \\
    \mathcal{S}_{\text{conf}}=\frac{\tilde{\Phi} \rh}{4 G_2}\,, \\
    C_\kappa = \frac{(d-2)\tilde{\Phi} \rh}{4 G_2} \,.
\end{cases}
\end{equation}
Inserting $\tilde{\Phi} \rh$ from \eqref{eqn: sol large phi}, we obtain that the conformal entropy in terms of the boundary data is
\begin{equation}
     \mathcal{S}_{\text{conf}}=\frac{(\pm1)^{d-2}(4\pi)^{d-2}\left(\sqrt{\kappa^2 L^2 \pm(d-1)^2}-\kappa L\right)^{d-2}}{(d-1)^{2(d-2)} 4 G_2} \frac{1}{\b^{d-2}} \,,
\end{equation}
where the `$+$' (`$-$') sign refers to positive (negative) cosmological constant. Comparing with the higher-dimensional high-temperature expansion \eqref{eq: intro_entropy}-\eqref{eq: ndof}, we find a precise agreement, upon identifying $\kappa = \k$ and $G_2^{-1}=L^{d-2} \Omega_{d-2}G_d^{-1}$, and $L=\ell$.

Lastly, we emphasize that, to obtain \eqref{eqn: thermo large phi}, we only assumed the large dilaton limit of the potential is governed by \eqref{eqn: poten large phi}. Contributions from terms that grow slower than \eqref{eqn: poten large phi}, e.g., adding a fixed electric charge, will only affect the sub-leading corrections of \eqref{eqn: thermo large phi}. This is the two-dimensional analog of the universality statement of conformal thermodynamics at high-temperatures.

\bibliography{bdryrefs}

\providecommand{\href}[2]{#2}\begingroup\raggedright\begin{thebibliography}{100}

\bibitem{Turiaci:2024cad}
G.~J. Turiaci, \emph{{Les Houches lectures on two-dimensional gravity and
  holography}},  \href{https://arxiv.org/abs/2412.09537}{{\tt 2412.09537}}.

\bibitem{Cavaglia:1998xj}
M.~Cavaglia, \emph{{Geometrodynamical formulation of two-dimensional dilaton
  gravity}}, \href{http://dx.doi.org/10.1103/PhysRevD.59.084011}{\emph{Phys.
  Rev. D} {\bf 59} (1999) 084011},
  [\href{https://arxiv.org/abs/hep-th/9811059}{{\tt hep-th/9811059}}].

\bibitem{Grumiller:2007ju}
D.~Grumiller and R.~McNees, \emph{{Thermodynamics of black holes in two (and
  higher) dimensions}},
  \href{http://dx.doi.org/10.1088/1126-6708/2007/04/074}{\emph{JHEP} {\bf 04}
  (2007) 074}, [\href{https://arxiv.org/abs/hep-th/0703230}{{\tt
  hep-th/0703230}}].

\bibitem{Anninos:2017hhn}
D.~Anninos and D.~M. Hofman, \emph{{Infrared Realization of dS$_2$ in
  AdS$_2$}}, \href{http://dx.doi.org/10.1088/1361-6382/aab143}{\emph{Class.
  Quant. Grav.} {\bf 35} (2018) 085003},
  [\href{https://arxiv.org/abs/1703.04622}{{\tt 1703.04622}}].

\bibitem{Witten:2020ert}
E.~Witten, \emph{{Deformations of JT Gravity and Phase Transitions}},
  \href{https://arxiv.org/abs/2006.03494}{{\tt 2006.03494}}.

\bibitem{Jackiw:1984je}
R.~Jackiw, \emph{{Lower Dimensional Gravity}},
  \href{http://dx.doi.org/10.1016/0550-3213(85)90448-1}{\emph{Nucl. Phys. B}
  {\bf 252} (1985) 343--356}.

\bibitem{Teitelboim:1983ux}
C.~Teitelboim, \emph{{Gravitation and Hamiltonian Structure in Two Space-Time
  Dimensions}},
  \href{http://dx.doi.org/10.1016/0370-2693(83)90012-6}{\emph{Phys. Lett. B}
  {\bf 126} (1983) 41--45}.

\bibitem{Maldacena:2016upp}
J.~Maldacena, D.~Stanford and Z.~Yang, \emph{{Conformal symmetry and its
  breaking in two dimensional Nearly Anti-de-Sitter space}},
  \href{http://dx.doi.org/10.1093/ptep/ptw124}{\emph{PTEP} {\bf 2016} (2016)
  12C104}, [\href{https://arxiv.org/abs/1606.01857}{{\tt 1606.01857}}].

\bibitem{Sachdev:1992fk}
S.~Sachdev and J.~Ye, \emph{{Gapless spin fluid ground state in a random,
  quantum Heisenberg magnet}},
  \href{http://dx.doi.org/10.1103/PhysRevLett.70.3339}{\emph{Phys. Rev. Lett.}
  {\bf 70} (1993) 3339}, [\href{https://arxiv.org/abs/cond-mat/9212030}{{\tt
  cond-mat/9212030}}].

\bibitem{Kit_SYK}
A.~Kitaev, \emph{{A simple model of quantum holography}}, .

\bibitem{Saad:2019lba}
P.~Saad, S.~H. Shenker and D.~Stanford, \emph{{JT gravity as a matrix
  integral}},  \href{https://arxiv.org/abs/1903.11115}{{\tt 1903.11115}}.

\bibitem{Johnson:2019eik}
C.~V. Johnson, \emph{{Nonperturbative Jackiw-Teitelboim gravity}},
  \href{http://dx.doi.org/10.1103/PhysRevD.101.106023}{\emph{Phys. Rev. D} {\bf
  101} (2020) 106023}, [\href{https://arxiv.org/abs/1912.03637}{{\tt
  1912.03637}}].

\bibitem{Jiang:2019pam}
J.~Jiang and Z.~Yang, \emph{{Thermodynamics and Many Body Chaos for generalized
  large q SYK models}},
  \href{http://dx.doi.org/10.1007/JHEP08(2019)019}{\emph{JHEP} {\bf 08} (2019)
  019}, [\href{https://arxiv.org/abs/1905.00811}{{\tt 1905.00811}}].

\bibitem{Anninos:2020cwo}
D.~Anninos and D.~A. Galante, \emph{{Constructing AdS$_{2}$ flow geometries}},
  \href{http://dx.doi.org/10.1007/JHEP02(2021)045}{\emph{JHEP} {\bf 02} (2021)
  045}, [\href{https://arxiv.org/abs/2011.01944}{{\tt 2011.01944}}].

\bibitem{Anninos:2022qgy}
D.~Anninos, D.~A. Galante and S.~U. Sheorey, \emph{{Renormalisation Group Flows
  of the SYK Model}},  \href{https://arxiv.org/abs/2212.04944}{{\tt
  2212.04944}}.

\bibitem{Chapman:2024pdw}
S.~Chapman, S.~Demulder, D.~A. Galante, S.~U. Sheorey and O.~Shoval,
  \emph{{Krylov complexity and chaos in deformed Sachdev-Ye-Kitaev models}},
  \href{http://dx.doi.org/10.1103/PhysRevB.111.035141}{\emph{Phys. Rev. B} {\bf
  111} (2025) 035141}, [\href{https://arxiv.org/abs/2407.09604}{{\tt
  2407.09604}}].

\bibitem{Maxfield:2020ale}
H.~Maxfield and G.~J. Turiaci, \emph{{The path integral of 3D gravity near
  extremality; or, JT gravity with defects as a matrix integral}},
  \href{https://arxiv.org/abs/2006.11317}{{\tt 2006.11317}}.

\bibitem{Witten:2020wvy}
E.~Witten, \emph{{Matrix Models and Deformations of JT Gravity}},
  \href{http://dx.doi.org/10.1098/rspa.2020.0582}{\emph{Proc. Roy. Soc. Lond.
  A} {\bf 476} (2020) 20200582}, [\href{https://arxiv.org/abs/2006.13414}{{\tt
  2006.13414}}].

\bibitem{Turiaci:2020fjj}
G.~J. Turiaci, M.~Usatyuk and W.~W. Weng, \emph{{2D dilaton-gravity,
  deformations of the minimal string, and matrix models}},
  \href{http://dx.doi.org/10.1088/1361-6382/ac25df}{\emph{Class. Quant. Grav.}
  {\bf 38} (2021) 204001}, [\href{https://arxiv.org/abs/2011.06038}{{\tt
  2011.06038}}].

\bibitem{Eberhardt:2023rzz}
L.~Eberhardt and G.~J. Turiaci, \emph{{2D dilaton gravity and the
  Weil-Petersson volumes with conical defects}},
  \href{https://arxiv.org/abs/2304.14948}{{\tt 2304.14948}}.

\bibitem{Kruthoff:2024gxc}
J.~Kruthoff and A.~Levine, \emph{{Semi-classical dilaton gravity and the very
  blunt defect expansion}},  \href{https://arxiv.org/abs/2402.10162}{{\tt
  2402.10162}}.

\bibitem{Anninos:2018svg}
D.~Anninos, D.~A. Galante and D.~M. Hofman, \emph{{De Sitter horizons \&
  holographic liquids}},
  \href{http://dx.doi.org/10.1007/JHEP07(2019)038}{\emph{JHEP} {\bf 07} (2019)
  038}, [\href{https://arxiv.org/abs/1811.08153}{{\tt 1811.08153}}].

\bibitem{Anninos:2022hqo}
D.~Anninos and E.~Harris, \emph{{Interpolating geometries and the stretched
  dS$_{2}$ horizon}},
  \href{http://dx.doi.org/10.1007/JHEP11(2022)166}{\emph{JHEP} {\bf 11} (2022)
  166}, [\href{https://arxiv.org/abs/2209.06144}{{\tt 2209.06144}}].

\bibitem{Chapman:2021eyy}
S.~Chapman, D.~A. Galante and E.~D. Kramer, \emph{{Holographic complexity and
  de Sitter space}},
  \href{http://dx.doi.org/10.1007/JHEP02(2022)198}{\emph{JHEP} {\bf 02} (2022)
  198}, [\href{https://arxiv.org/abs/2110.05522}{{\tt 2110.05522}}].

\bibitem{Maldacena:2019cbz}
J.~Maldacena, G.~J. Turiaci and Z.~Yang, \emph{{Two dimensional Nearly de
  Sitter gravity}},
  \href{http://dx.doi.org/10.1007/JHEP01(2021)139}{\emph{JHEP} {\bf 01} (2021)
  139}, [\href{https://arxiv.org/abs/1904.01911}{{\tt 1904.01911}}].

\bibitem{Cotler:2019nbi}
J.~Cotler, K.~Jensen and A.~Maloney, \emph{{Low-dimensional de Sitter quantum
  gravity}}, \href{http://dx.doi.org/10.1007/JHEP06(2020)048}{\emph{JHEP} {\bf
  06} (2020) 048}, [\href{https://arxiv.org/abs/1905.03780}{{\tt 1905.03780}}].

\bibitem{Nanda:2023wne}
K.~K. Nanda, S.~K. Sake and S.~P. Trivedi, \emph{{JT gravity in de Sitter space
  and the problem of time}},
  \href{http://dx.doi.org/10.1007/JHEP02(2024)145}{\emph{JHEP} {\bf 02} (2024)
  145}, [\href{https://arxiv.org/abs/2307.15900}{{\tt 2307.15900}}].

\bibitem{Cotler:2024xzz}
J.~Cotler and K.~Jensen, \emph{{Non-perturbative de Sitter Jackiw-Teitelboim
  gravity}}, \href{http://dx.doi.org/10.1007/JHEP12(2024)016}{\emph{JHEP} {\bf
  12} (2024) 016}, [\href{https://arxiv.org/abs/2401.01925}{{\tt 2401.01925}}].

\bibitem{Anninos:2024iwf}
D.~Anninos, C.~Baracco and B.~M\"uhlmann, \emph{{Remarks on 2D quantum
  cosmology}},
  \href{http://dx.doi.org/10.1088/1475-7516/2024/10/031}{\emph{JCAP} {\bf 10}
  (2024) 031}, [\href{https://arxiv.org/abs/2406.15271}{{\tt 2406.15271}}].

\bibitem{Blommaert:2024ydx}
A.~Blommaert, T.~G. Mertens and J.~Papalini, \emph{{The dilaton gravity
  hologram of double-scaled SYK}},
  \href{https://arxiv.org/abs/2404.03535}{{\tt 2404.03535}}.

\bibitem{Berkooz:2024lgq}
M.~Berkooz and O.~Mamroud, \emph{{A cordial introduction to double scaled
  SYK}}, \href{http://dx.doi.org/10.1088/1361-6633/ada889}{\emph{Rept. Prog.
  Phys.} {\bf 88} (2025) 036001}, [\href{https://arxiv.org/abs/2407.09396}{{\tt
  2407.09396}}].

\bibitem{Blommaert:2024whf}
A.~Blommaert, A.~Levine, T.~G. Mertens, J.~Papalini and K.~Parmentier,
  \emph{{An entropic puzzle in periodic dilaton gravity and DSSYK}},
  \href{https://arxiv.org/abs/2411.16922}{{\tt 2411.16922}}.

\bibitem{Collier:2025pbm}
S.~Collier, L.~Eberhardt and B.~M\"uhlmann, \emph{{The complex Liouville
  string: the gravitational path integral}},
  \href{https://arxiv.org/abs/2501.10265}{{\tt 2501.10265}}.

\bibitem{Goel:2020yxl}
A.~Goel, L.~V. Iliesiu, J.~Kruthoff and Z.~Yang, \emph{{Classifying boundary
  conditions in JT gravity: from energy-branes to $\alpha$-branes}},
  \href{http://dx.doi.org/10.1007/JHEP04(2021)069}{\emph{JHEP} {\bf 04} (2021)
  069}, [\href{https://arxiv.org/abs/2010.12592}{{\tt 2010.12592}}].

\bibitem{Godet:2020xpk}
V.~Godet and C.~Marteau, \emph{{New boundary conditions for AdS$_{2}$}},
  \href{http://dx.doi.org/10.1007/JHEP12(2020)020}{\emph{JHEP} {\bf 12} (2020)
  020}, [\href{https://arxiv.org/abs/2005.08999}{{\tt 2005.08999}}].

\bibitem{York:1986it}
J.~W. York, Jr., \emph{{Black hole thermodynamics and the Euclidean Einstein
  action}}, \href{http://dx.doi.org/10.1103/PhysRevD.33.2092}{\emph{Phys. Rev.
  D} {\bf 33} (1986) 2092--2099}.

\bibitem{Whiting:1988qr}
B.~F. Whiting and J.~W. York, Jr., \emph{{Action Principle and Partition
  Function for the Gravitational Field in Black Hole Topologies}},
  \href{http://dx.doi.org/10.1103/PhysRevLett.61.1336}{\emph{Phys. Rev. Lett.}
  {\bf 61} (1988) 1336}.

\bibitem{Braden:1990hw}
H.~W. Braden, J.~D. Brown, B.~F. Whiting and J.~W. York, Jr., \emph{{Charged
  black hole in a grand canonical ensemble}},
  \href{http://dx.doi.org/10.1103/PhysRevD.42.3376}{\emph{Phys. Rev. D} {\bf
  42} (1990) 3376--3385}.

\bibitem{Brown:1992bq}
J.~D. Brown and J.~W. York, Jr., \emph{{The Microcanonical functional integral.
  1. The Gravitational field}},
  \href{http://dx.doi.org/10.1103/PhysRevD.47.1420}{\emph{Phys. Rev. D} {\bf
  47} (1993) 1420--1431}, [\href{https://arxiv.org/abs/gr-qc/9209014}{{\tt
  gr-qc/9209014}}].

\bibitem{Brown:1994gs}
J.~D. Brown, J.~Creighton and R.~B. Mann, \emph{{Temperature, energy and heat
  capacity of asymptotically anti-de Sitter black holes}},
  \href{http://dx.doi.org/10.1103/PhysRevD.50.6394}{\emph{Phys. Rev. D} {\bf
  50} (1994) 6394--6403}, [\href{https://arxiv.org/abs/gr-qc/9405007}{{\tt
  gr-qc/9405007}}].

\bibitem{Anninos:2011af}
D.~Anninos, S.~A. Hartnoll and D.~M. Hofman, \emph{{Static Patch Solipsism:
  Conformal Symmetry of the de Sitter Worldline}},
  \href{http://dx.doi.org/10.1088/0264-9381/29/7/075002}{\emph{Class. Quant.
  Grav.} {\bf 29} (2012) 075002}, [\href{https://arxiv.org/abs/1109.4942}{{\tt
  1109.4942}}].

\bibitem{Anninos:2012qw}
D.~Anninos, \emph{{De Sitter Musings}},
  \href{http://dx.doi.org/10.1142/S0217751X1230013X}{\emph{Int. J. Mod. Phys.
  A} {\bf 27} (2012) 1230013}, [\href{https://arxiv.org/abs/1205.3855}{{\tt
  1205.3855}}].

\bibitem{Anninos:2022ujl}
D.~Anninos, D.~A. Galante and B.~M\"uhlmann, \emph{{Finite features of quantum
  de Sitter space}},
  \href{http://dx.doi.org/10.1088/1361-6382/acaba5}{\emph{Class. Quant. Grav.}
  {\bf 40} (2023) 025009}, [\href{https://arxiv.org/abs/2206.14146}{{\tt
  2206.14146}}].

\bibitem{Galante:2023uyf}
D.~A. Galante, \emph{{Modave lectures on de Sitter space \& holography}},
  \href{http://dx.doi.org/10.22323/1.435.0003}{\emph{PoS} {\bf Modave2022}
  (2023) 003}, [\href{https://arxiv.org/abs/2306.10141}{{\tt 2306.10141}}].

\bibitem{Shyam:2021ciy}
V.~Shyam, \emph{{$ \mathrm{T}\overline{\mathrm{T}} $ +
  \ensuremath{\Lambda}$_{2}$ deformed CFT on the stretched dS$_{3}$ horizon}},
  \href{http://dx.doi.org/10.1007/JHEP04(2022)052}{\emph{JHEP} {\bf 04} (2022)
  052}, [\href{https://arxiv.org/abs/2106.10227}{{\tt 2106.10227}}].

\bibitem{Coleman:2021nor}
E.~Coleman, E.~A. Mazenc, V.~Shyam, E.~Silverstein, R.~M. Soni, G.~Torroba
  et~al., \emph{{De Sitter microstates from T$ \overline{T} $ +
  \ensuremath{\Lambda}$_{2}$ and the Hawking-Page transition}},
  \href{http://dx.doi.org/10.1007/JHEP07(2022)140}{\emph{JHEP} {\bf 07} (2022)
  140}, [\href{https://arxiv.org/abs/2110.14670}{{\tt 2110.14670}}].

\bibitem{Silverstein:2022dfj}
E.~Silverstein, \emph{{Black hole to cosmic horizon microstates in string/M
  theory: timelike boundaries and internal averaging}},
  \href{https://arxiv.org/abs/2212.00588}{{\tt 2212.00588}}.

\bibitem{Silverstein:2024xnr}
E.~Silverstein and G.~Torroba, \emph{{Timelike-bounded dS$_{4}$ holography from
  a solvable sector of the T$^{2}$ deformation}},
  \href{http://dx.doi.org/10.1007/JHEP03(2025)156}{\emph{JHEP} {\bf 03} (2025)
  156}, [\href{https://arxiv.org/abs/2409.08709}{{\tt 2409.08709}}].

\bibitem{Banihashemi:2022jys}
B.~Banihashemi and T.~Jacobson, \emph{{Thermodynamic ensembles with
  cosmological horizons}},
  \href{http://dx.doi.org/10.1007/JHEP07(2022)042}{\emph{JHEP} {\bf 07} (2022)
  042}, [\href{https://arxiv.org/abs/2204.05324}{{\tt 2204.05324}}].

\bibitem{Banihashemi:2022htw}
B.~Banihashemi, T.~Jacobson, A.~Svesko and M.~Visser, \emph{{The minus sign in
  the first law of de Sitter horizons}},
  \href{http://dx.doi.org/10.1007/JHEP01(2023)054}{\emph{JHEP} {\bf 01} (2023)
  054}, [\href{https://arxiv.org/abs/2208.11706}{{\tt 2208.11706}}].

\bibitem{Gross:2019ach}
D.~J. Gross, J.~Kruthoff, A.~Rolph and E.~Shaghoulian, \emph{{$T\overline{T}$
  in AdS$_2$ and Quantum Mechanics}},
  \href{http://dx.doi.org/10.1103/PhysRevD.101.026011}{\emph{Phys. Rev. D} {\bf
  101} (2020) 026011}, [\href{https://arxiv.org/abs/1907.04873}{{\tt
  1907.04873}}].

\bibitem{Svesko:2022txo}
A.~Svesko, E.~Verheijden, E.~P. Verlinde and M.~R. Visser, \emph{{Quasi-local
  energy and microcanonical entropy in two-dimensional nearly de Sitter
  gravity}}, \href{http://dx.doi.org/10.1007/JHEP08(2022)075}{\emph{JHEP} {\bf
  08} (2022) 075}, [\href{https://arxiv.org/abs/2203.00700}{{\tt 2203.00700}}].

\bibitem{Batra:2024qju}
G.~Batra, \emph{{Timelike boundaries in de Sitter JT gravity and the Gao-Wald
  theorem}}, \href{http://dx.doi.org/10.1007/JHEP01(2025)044}{\emph{JHEP} {\bf
  01} (2025) 044}, [\href{https://arxiv.org/abs/2407.08913}{{\tt 2407.08913}}].

\bibitem{Aguilar-Gutierrez:2024nst}
S.~E. Aguilar-Gutierrez, A.~Svesko and M.~R. Visser,
  \emph{{$\text{T}\overline{\text{T}}$ deformations from AdS$_2$ to dS$_2$}},
  \href{https://arxiv.org/abs/2410.18257}{{\tt 2410.18257}}.

\bibitem{Batra:2024kjl}
G.~Batra, G.~B. De~Luca, E.~Silverstein, G.~Torroba and S.~Yang,
  \emph{{Bulk-local dS$_3$ holography: the Matter with $T\bar T+\Lambda_2$}},
  \href{https://arxiv.org/abs/2403.01040}{{\tt 2403.01040}}.

\bibitem{Hayward:1990zm}
G.~Hayward, \emph{{Euclidean action and the thermodynamics of manifolds without
  boundary}}, \href{http://dx.doi.org/10.1103/PhysRevD.41.3248}{\emph{Phys.
  Rev. D} {\bf 41} (1990) 3248--3251}.

\bibitem{Wang:2001gt}
B.~B. Wang and C.~G. Huang, \emph{{Thermodynamics of de Sitter space-time in
  York's formalism}},
  \href{http://dx.doi.org/10.1142/S0217732301004637}{\emph{Mod. Phys. Lett. A}
  {\bf 16} (2001) 1487--1492}.

\bibitem{Draper:2022ofa}
P.~Draper and S.~Farkas, \emph{{Euclidean de Sitter black holes and
  microcanonical equilibrium}},
  \href{http://dx.doi.org/10.1103/PhysRevD.105.126021}{\emph{Phys. Rev. D} {\bf
  105} (2022) 126021}, [\href{https://arxiv.org/abs/2203.01871}{{\tt
  2203.01871}}].

\bibitem{Anderson:2006lqb}
M.~T. Anderson, \emph{{On boundary value problems for Einstein metrics}},
  \href{http://dx.doi.org/10.2140/gt.2008.12.2009}{\emph{Geom. Topol.} {\bf 12}
  (2008) 2009--2045}, [\href{https://arxiv.org/abs/math/0612647}{{\tt
  math/0612647}}].

\bibitem{An:2021fcq}
Z.~An and M.~T. Anderson, \emph{{The initial boundary value problem and
  quasi-local Hamiltonians in General Relativity}},
  \href{https://arxiv.org/abs/2103.15673}{{\tt 2103.15673}}.

\bibitem{anderson2010extension}
M.~T. Anderson, \emph{Extension of symmetries on einstein manifolds with
  boundary}, {\emph{Selecta Mathematica} {\bf 16} (2010) 343--375}.

\bibitem{Witten:2018lgb}
E.~Witten, \emph{{A note on boundary conditions in Euclidean gravity}},
  \href{http://dx.doi.org/10.1142/S0129055X21400043}{\emph{Rev. Math. Phys.}
  {\bf 33} (2021) 2140004}, [\href{https://arxiv.org/abs/1805.11559}{{\tt
  1805.11559}}].

\bibitem{An:2025gvr}
Z.~An and M.~T. Anderson, \emph{{Well-posed geometric boundary data in General
  Relativity, II: Dirichlet boundary data}},
  \href{https://arxiv.org/abs/2505.07128}{{\tt 2505.07128}}.

\bibitem{Anninos:2023epi}
D.~Anninos, D.~A. Galante and C.~Maneerat, \emph{{Gravitational
  observatories}}, \href{http://dx.doi.org/10.1007/JHEP12(2023)024}{\emph{JHEP}
  {\bf 12} (2023) 024}, [\href{https://arxiv.org/abs/2310.08648}{{\tt
  2310.08648}}].

\bibitem{Anninos:2024xhc}
D.~Anninos, R.~Arias, D.~A. Galante and C.~Maneerat, \emph{{Gravitational
  Observatories in AdS$_4$}},  \href{https://arxiv.org/abs/2412.16305}{{\tt
  2412.16305}}.

\bibitem{Banihashemi:2025qqi}
B.~Banihashemi, E.~Shaghoulian and S.~Shashi, \emph{{Thermal effective actions
  from conformal boundary conditions in gravity}},
  \href{https://arxiv.org/abs/2503.17471}{{\tt 2503.17471}}.

\bibitem{edgar}
K.~Allameh and E.~Shaghoulian, \emph{{to appear}}, .

\bibitem{An:2025rlw}
Z.~An and M.~T. Anderson, \emph{{Well-posed geometric boundary data in General
  Relativity, I: Conformal-mean curvature boundary data}},
  \href{https://arxiv.org/abs/2503.12599}{{\tt 2503.12599}}.

\bibitem{Anninos:2024wpy}
D.~Anninos, D.~A. Galante and C.~Maneerat, \emph{{Cosmological observatories}},
  \href{http://dx.doi.org/10.1088/1361-6382/ad5824}{\emph{Class. Quant. Grav.}
  {\bf 41} (2024) 165009}, [\href{https://arxiv.org/abs/2402.04305}{{\tt
  2402.04305}}].

\bibitem{Liu:2024ymn}
X.~Liu, J.~E. Santos and T.~Wiseman, \emph{{New Well-Posed boundary conditions
  for semi-classical Euclidean gravity}},
  \href{http://dx.doi.org/10.1007/JHEP06(2024)044}{\emph{JHEP} {\bf 06} (2024)
  044}, [\href{https://arxiv.org/abs/2402.04308}{{\tt 2402.04308}}].

\bibitem{Bredberg:2011xw}
I.~Bredberg and A.~Strominger, \emph{{Black Holes as Incompressible Fluids on
  the Sphere}}, \href{http://dx.doi.org/10.1007/JHEP05(2012)043}{\emph{JHEP}
  {\bf 05} (2012) 043}, [\href{https://arxiv.org/abs/1106.3084}{{\tt
  1106.3084}}].

\bibitem{Anninos:2011zn}
D.~Anninos, T.~Anous, I.~Bredberg and G.~S. Ng, \emph{{Incompressible Fluids of
  the de Sitter Horizon and Beyond}},
  \href{http://dx.doi.org/10.1007/JHEP05(2012)107}{\emph{JHEP} {\bf 05} (2012)
  107}, [\href{https://arxiv.org/abs/1110.3792}{{\tt 1110.3792}}].

\bibitem{Odak:2021axr}
G.~Odak and S.~Speziale, \emph{{Brown-York charges with mixed boundary
  conditions}}, \href{http://dx.doi.org/10.1007/JHEP11(2021)224}{\emph{JHEP}
  {\bf 11} (2021) 224}, [\href{https://arxiv.org/abs/2109.02883}{{\tt
  2109.02883}}].

\bibitem{Brown:1992br}
J.~D. Brown and J.~W. York, Jr., \emph{{Quasilocal energy and conserved charges
  derived from the gravitational action}},
  \href{http://dx.doi.org/10.1103/PhysRevD.47.1407}{\emph{Phys. Rev. D} {\bf
  47} (1993) 1407--1419}, [\href{https://arxiv.org/abs/gr-qc/9209012}{{\tt
  gr-qc/9209012}}].

\bibitem{Gibbons:1976ue}
G.~W. Gibbons and S.~W. Hawking, \emph{{Action Integrals and Partition
  Functions in Quantum Gravity}},
  \href{http://dx.doi.org/10.1103/PhysRevD.15.2752}{\emph{Phys. Rev. D} {\bf
  15} (1977) 2752--2756}.

\bibitem{Banihashemi:2024yye}
B.~Banihashemi, E.~Shaghoulian and S.~Shashi, \emph{{Flat space gravity at
  finite cutoff}},
  \href{http://dx.doi.org/10.1088/1361-6382/ada2d7}{\emph{Class. Quant. Grav.}
  {\bf 42} (2025) 035010}, [\href{https://arxiv.org/abs/2409.07643}{{\tt
  2409.07643}}].

\bibitem{Anninos:2021ydw}
D.~Anninos, D.~M. Hofman and S.~Vitouladitis, \emph{{One-dimensional Quantum
  Gravity and the Schwarzian theory}},
  \href{http://dx.doi.org/10.1007/JHEP03(2022)121}{\emph{JHEP} {\bf 03} (2022)
  121}, [\href{https://arxiv.org/abs/2112.03793}{{\tt 2112.03793}}].

\bibitem{Capoferri:2024sgo}
M.~Capoferri, S.~Murro and G.~Schmid, \emph{{On boundary conditions for
  linearised Einstein\textquoteright{}s equations}},
  \href{http://dx.doi.org/10.1016/j.aml.2024.109210}{\emph{Appl. Math. Lett.}
  {\bf 158} (2024) 109210}, [\href{https://arxiv.org/abs/2407.07576}{{\tt
  2407.07576}}].

\bibitem{Faulkner:2009wj}
T.~Faulkner, H.~Liu, J.~McGreevy and D.~Vegh, \emph{{Emergent quantum
  criticality, Fermi surfaces, and AdS(2)}},
  \href{http://dx.doi.org/10.1103/PhysRevD.83.125002}{\emph{Phys. Rev. D} {\bf
  83} (2011) 125002}, [\href{https://arxiv.org/abs/0907.2694}{{\tt
  0907.2694}}].

\bibitem{Faulkner:2010tq}
T.~Faulkner and J.~Polchinski, \emph{{Semi-Holographic Fermi Liquids}},
  \href{http://dx.doi.org/10.1007/JHEP06(2011)012}{\emph{JHEP} {\bf 06} (2011)
  012}, [\href{https://arxiv.org/abs/1001.5049}{{\tt 1001.5049}}].

\bibitem{Nickel:2010pr}
D.~Nickel and D.~T. Son, \emph{{Deconstructing holographic liquids}},
  \href{http://dx.doi.org/10.1088/1367-2630/13/7/075010}{\emph{New J. Phys.}
  {\bf 13} (2011) 075010}, [\href{https://arxiv.org/abs/1009.3094}{{\tt
  1009.3094}}].

\bibitem{Bobev:2017uzs}
N.~Bobev and P.~M. Crichigno, \emph{{Universal RG Flows Across Dimensions and
  Holography}}, \href{http://dx.doi.org/10.1007/JHEP12(2017)065}{\emph{JHEP}
  {\bf 12} (2017) 065}, [\href{https://arxiv.org/abs/1708.05052}{{\tt
  1708.05052}}].

\bibitem{Brown:2018bms}
A.~R. Brown, H.~Gharibyan, H.~W. Lin, L.~Susskind, L.~Thorlacius and Y.~Zhao,
  \emph{{Complexity of Jackiw-Teitelboim gravity}},
  \href{http://dx.doi.org/10.1103/PhysRevD.99.046016}{\emph{Phys. Rev. D} {\bf
  99} (2019) 046016}, [\href{https://arxiv.org/abs/1810.08741}{{\tt
  1810.08741}}].

\bibitem{Castro:2022cuo}
A.~Castro, F.~Mariani and C.~Toldo, \emph{{Near-extremal limits of de Sitter
  black holes}}, \href{http://dx.doi.org/10.1007/JHEP07(2023)131}{\emph{JHEP}
  {\bf 07} (2023) 131}, [\href{https://arxiv.org/abs/2212.14356}{{\tt
  2212.14356}}].

\bibitem{Ivo:2025yek}
V.~Ivo, J.~Maldacena and Z.~Sun, \emph{{Physical instabilities and the phase of
  the Euclidean path integral}},  \href{https://arxiv.org/abs/2504.00920}{{\tt
  2504.00920}}.

\bibitem{Strominger:1994xi}
A.~Strominger and L.~Thorlacius, \emph{{Conformally invariant boundary
  conditions for dilaton gravity}},
  \href{http://dx.doi.org/10.1103/PhysRevD.50.5177}{\emph{Phys. Rev. D} {\bf
  50} (1994) 5177--5187}, [\href{https://arxiv.org/abs/hep-th/9405084}{{\tt
  hep-th/9405084}}].

\bibitem{Iliesiu:2020qvm}
L.~V. Iliesiu and G.~J. Turiaci, \emph{{The statistical mechanics of
  near-extremal black holes}},
  \href{http://dx.doi.org/10.1007/JHEP05(2021)145}{\emph{JHEP} {\bf 05} (2021)
  145}, [\href{https://arxiv.org/abs/2003.02860}{{\tt 2003.02860}}].

\bibitem{Iliesiu:2022onk}
L.~V. Iliesiu, S.~Murthy and G.~J. Turiaci, \emph{{Revisiting the logarithmic
  corrections to the black hole entropy}},
  \href{http://dx.doi.org/10.1007/JHEP07(2025)058}{\emph{JHEP} {\bf 07} (2025)
  058}, [\href{https://arxiv.org/abs/2209.13608}{{\tt 2209.13608}}].

\bibitem{Maulik:2025phe}
S.~Maulik, A.~Mitra, D.~Mukherjee and A.~Ray, \emph{{Logarithmic corrections to
  near-extremal entropy of charged de Sitter black holes}},
  \href{https://arxiv.org/abs/2503.08617}{{\tt 2503.08617}}.

\bibitem{Blacker:2025zca}
M.~J. Blacker, A.~Castro, W.~Sybesma and C.~Toldo, \emph{{Quantum corrections
  to the path integral of near extremal de Sitter black holes}},
  \href{https://arxiv.org/abs/2503.14623}{{\tt 2503.14623}}.

\bibitem{elsewhere}
C.~Maneerat, \emph{{to appear}}, .

\bibitem{Carlip:2005tz}
S.~Carlip, \emph{{Dynamics of asymptotic diffeomorphisms in (2+1)-dimensional
  gravity}}, \href{http://dx.doi.org/10.1088/0264-9381/22/14/014}{\emph{Class.
  Quant. Grav.} {\bf 22} (2005) 3055--3060},
  [\href{https://arxiv.org/abs/gr-qc/0501033}{{\tt gr-qc/0501033}}].

\bibitem{Nguyen:2021pdz}
K.~Nguyen, \emph{{Holographic boundary actions in AdS$_{3}$/CFT$_{2}$
  revisited}}, \href{http://dx.doi.org/10.1007/JHEP10(2021)218}{\emph{JHEP}
  {\bf 10} (2021) 218}, [\href{https://arxiv.org/abs/2108.01095}{{\tt
  2108.01095}}].

\bibitem{Bagrets:2016cdf}
D.~Bagrets, A.~Altland and A.~Kamenev,
  \emph{{Sachdev\textendash{}Ye\textendash{}Kitaev model as Liouville quantum
  mechanics}},
  \href{http://dx.doi.org/10.1016/j.nuclphysb.2016.08.002}{\emph{Nucl. Phys. B}
  {\bf 911} (2016) 191--205}, [\href{https://arxiv.org/abs/1607.00694}{{\tt
  1607.00694}}].

\bibitem{Nayak:2019evx}
P.~Nayak, J.~Sonner and M.~Vielma, \emph{{Extended Eigenstate Thermalization
  and the role of FZZT branes in the Schwarzian theory}},
  \href{http://dx.doi.org/10.1007/JHEP03(2020)168}{\emph{JHEP} {\bf 03} (2020)
  168}, [\href{https://arxiv.org/abs/1907.10061}{{\tt 1907.10061}}].

\bibitem{Jensen:2019cmr}
K.~Jensen, \emph{{Scrambling in nearly thermalized states at large central
  charge}},  \href{https://arxiv.org/abs/1906.05852}{{\tt 1906.05852}}.

\bibitem{Hayward:1993my}
G.~Hayward, \emph{{Gravitational action for space-times with nonsmooth
  boundaries}}, \href{http://dx.doi.org/10.1103/PhysRevD.47.3275}{\emph{Phys.
  Rev. D} {\bf 47} (1993) 3275--3280}.

\bibitem{Hawking:1980gf}
S.~W. Hawking, \emph{{THE PATH INTEGRAL APPROACH TO QUANTUM GRAVITY}},
  pp.~746--789.
\newblock 1980.

\bibitem{York:1972sj}
J.~W. York, Jr., \emph{{Role of conformal three geometry in the dynamics of
  gravitation}},
  \href{http://dx.doi.org/10.1103/PhysRevLett.28.1082}{\emph{Phys. Rev. Lett.}
  {\bf 28} (1972) 1082--1085}.

\bibitem{Krishnan:2016mcj}
C.~Krishnan and A.~Raju, \emph{{A Neumann Boundary Term for Gravity}},
  \href{http://dx.doi.org/10.1142/S0217732317500778}{\emph{Mod. Phys. Lett. A}
  {\bf 32} (2017) 1750077}, [\href{https://arxiv.org/abs/1605.01603}{{\tt
  1605.01603}}].

\bibitem{Krishnan:2016tqj}
C.~Krishnan, K.~V.~P. Kumar and A.~Raju, \emph{{An alternative path integral
  for quantum gravity}},
  \href{http://dx.doi.org/10.1007/JHEP10(2016)043}{\emph{JHEP} {\bf 10} (2016)
  043}, [\href{https://arxiv.org/abs/1609.04719}{{\tt 1609.04719}}].

\bibitem{Banados:1998ys}
M.~Banados and F.~Mendez, \emph{{A Note on covariant action integrals in
  three-dimensions}},
  \href{http://dx.doi.org/10.1103/PhysRevD.58.104014}{\emph{Phys. Rev. D} {\bf
  58} (1998) 104014}, [\href{https://arxiv.org/abs/hep-th/9806065}{{\tt
  hep-th/9806065}}].

\bibitem{Miskovic:2006tm}
O.~Miskovic and R.~Olea, \emph{{On boundary conditions in three-dimensional AdS
  gravity}},
  \href{http://dx.doi.org/10.1016/j.physletb.2006.07.045}{\emph{Phys. Lett. B}
  {\bf 640} (2006) 101--107}, [\href{https://arxiv.org/abs/hep-th/0603092}{{\tt
  hep-th/0603092}}].

\bibitem{Detournay:2014fva}
S.~Detournay, D.~Grumiller, F.~Sch\"oller and J.~Sim\'on, \emph{{Variational
  principle and one-point functions in three-dimensional flat space Einstein
  gravity}}, \href{http://dx.doi.org/10.1103/PhysRevD.89.084061}{\emph{Phys.
  Rev. D} {\bf 89} (2014) 084061}, [\href{https://arxiv.org/abs/1402.3687}{{\tt
  1402.3687}}].

\bibitem{Anastasiou:2020zwc}
G.~Anastasiou, O.~Miskovic, R.~Olea and I.~Papadimitriou, \emph{{Counterterms,
  Kounterterms, and the variational problem in AdS gravity}},
  \href{http://dx.doi.org/10.1007/JHEP08(2020)061}{\emph{JHEP} {\bf 08} (2020)
  061}, [\href{https://arxiv.org/abs/2003.06425}{{\tt 2003.06425}}].

\bibitem{Parvizi:2025shq}
A.~Parvizi, M.~M. Sheikh-Jabbari and V.~Taghiloo, \emph{{Freelance Holography,
  Part I: Setting Boundary Conditions Free in Gauge/Gravity Correspondence}},
  \href{https://arxiv.org/abs/2503.09371}{{\tt 2503.09371}}.

\bibitem{Parvizi:2025wsg}
A.~Parvizi, M.~M. Sheikh-Jabbari and V.~Taghiloo, \emph{{Freelance Holography,
  Part II: Moving Boundary in Gauge/Gravity Correspondence}},
  \href{https://arxiv.org/abs/2503.09372}{{\tt 2503.09372}}.

\bibitem{Hawking:1995ap}
S.~W. Hawking and S.~F. Ross, \emph{{Duality between electric and magnetic
  black holes}}, \href{http://dx.doi.org/10.1103/PhysRevD.52.5865}{\emph{Phys.
  Rev. D} {\bf 52} (1995) 5865--5876},
  [\href{https://arxiv.org/abs/hep-th/9504019}{{\tt hep-th/9504019}}].

\bibitem{Grumiller:2001ea}
D.~Grumiller, \emph{{Quantum dilaton gravity in two-dimensions with matter}},
  other thesis, 5, 2001.

\bibitem{Narayan:2020pyj}
K.~Narayan, \emph{{Aspects of two-dimensional dilaton gravity, dimensional
  reduction, and holography}},
  \href{http://dx.doi.org/10.1103/PhysRevD.104.026007}{\emph{Phys. Rev. D} {\bf
  104} (2021) 026007}, [\href{https://arxiv.org/abs/2010.12955}{{\tt
  2010.12955}}].

\bibitem{Saskowski:2024tat}
R.~J. Saskowski, \emph{{The fate of boundary terms in dimensional reductions}},
   \href{https://arxiv.org/abs/2405.10363}{{\tt 2405.10363}}.

\bibitem{Bardeen:1999px}
J.~M. Bardeen and G.~T. Horowitz, \emph{{The Extreme Kerr throat geometry: A
  Vacuum analog of AdS(2) x S**2}},
  \href{http://dx.doi.org/10.1103/PhysRevD.60.104030}{\emph{Phys. Rev. D} {\bf
  60} (1999) 104030}, [\href{https://arxiv.org/abs/hep-th/9905099}{{\tt
  hep-th/9905099}}].

\bibitem{Hartman:2008pb}
T.~Hartman, K.~Murata, T.~Nishioka and A.~Strominger, \emph{{CFT Duals for
  Extreme Black Holes}},
  \href{http://dx.doi.org/10.1088/1126-6708/2009/04/019}{\emph{JHEP} {\bf 04}
  (2009) 019}, [\href{https://arxiv.org/abs/0811.4393}{{\tt 0811.4393}}].

\end{thebibliography}\endgroup

\end{document}